\documentclass[a4paper,clock]{article}
\usepackage[dvips]{epsfig}
\usepackage{amssymb}
\usepackage{bm}
\usepackage{dcolumn}

\catcode`\@=11
\def\marginnote#1{}
\newcount\hour
\newcount\minute
\newtoks\amorpm
\hour=\time\divide\hour by60
\minute=\time{\multiply\hour by60 \global\advance\minute
by-\hour}\edef\standardtime{{\ifnum\hour<12
\global\amorpm={am}%
        \else\global\amorpm={pm}\advance\hour by-12 \fi
        \ifnum\hour=0 \hour=12 \fi
        \number\hour:\ifnum\minute<10
0\fi\number\minute\the\amorpm}}
\edef\militarytime{\number\hour:\ifnum\minute<10
0\fi\number\minute}

\def\draftlabel#1{{\@bsphack\if@filesw {\let\thepage\relax
   \xdef\@gtempa{\write\@auxout{\string
      \newlabel{#1}{{\@currentlabel}{\thepage}}}}}\@gtempa
   \if@nobreak \ifvmode\nobreak\fi\fi\fi\@esphack}
        \gdef\@eqnlabel{#1}}
\def\@eqnlabel{}
\def\@vacuum{}
\def\draftmarginnote#1{\marginpar{\raggedright\scriptsize\tt#1}}
\def\draft{\oddsidemargin -.5truein
        \def\@oddfoot{\sl preliminary draft \hfil
        \rm\thepage\hfil\sl\today\quad\militarytime}
        \let\@evenfoot\@oddfoot \overfullrule 3pt
        \let\label=\draftlabel
        \let\marginnote=\draftmarginnote

\def\@eqnnum{(\theequation)\rlap{\kern\marginparsep\tt\@eqnlabel}%
\global\let\@eqnlabel\@vacuum}  }

\def\numberbysection{\@addtoreset{equation}{section}
        \def\theequation{\thesection.\arabic{equation}}}

\def\underline#1{\relax\ifmmode\@@underline#1\else
 $\@@underline{\hbox{#1}}$\relax\fi}

\catcode`@=12
\relax

\numberbysection

\advance \topmargin by -\headheight
\advance \topmargin by -\headsep

\evensidemargin \oddsidemargin
\def\br{\begin{eqnarray}}
\def\er{\end{eqnarray}}
\def\be{\begin{equation}}
\def\ee{\end{equation}}

\def\({\left(}
\def\){\right)}

\relax

\def\a{\alpha}

\def\b{\beta}

\def\d{\delta}
\def\D{\Delta}

\def\g{\gamma}
\def\G{\Gamma}

\def\l{\lambda}
\def\L{\Lambda}

\def\pa{\partial}

\def\tp0{\Theta_{+}^{(0)}}
\def\tm0{\Theta_{-}^{(0)}}

\def\vp{\varphi}

\def\f#1#2#3 {f^{#1#2}_{#3}}

\def\win1{{\sf w_{1+\infty}}}

\def\Win1{{\sf W_{1+\infty}}}

\def\rlx{\relax\leavevmode}
\def\inbar{\vrule height1.5ex width.4pt depth0pt}
\def\IZ{\rlx\hbox{\sf Z\kern-.4em Z}}
\def\IR{\rlx\hbox{\rm I\kern-.18em R}}
\def\IC{\rlx\hbox{\,$\inbar\kern-.3em{\rm C}$}}
\def\IN{\rlx\hbox{\rm I\kern-.18em N}}
\def\IO{\rlx\hbox{\,$\inbar\kern-.3em{\rm O}$}}
\def\IP{\rlx\hbox{\rm I\kern-.18em P}}
\def\IQ{\rlx\hbox{\,$\inbar\kern-.3em{\rm Q}$}}
\def\IF{\rlx\hbox{\rm I\kern-.18em F}}
\def\IG{\rlx\hbox{\,$\inbar\kern-.3em{\rm G}$}}
\def\IH{\rlx\hbox{\rm I\kern-.18em H}}
\def\II{\rlx\hbox{\rm I\kern-.18em I}}
\def\IK{\rlx\hbox{\rm I\kern-.18em K}}
\def\IL{\rlx\hbox{\rm I\kern-.18em L}}
\def\one{\hbox{{1}\kern-.25em\hbox{l}}}
\def\0#1{\relax\ifmmode\mathaccent"7017{#1}%
B        \else\accent23#1\relax\fi}

\def\PRE#1#2#3{{\sl Phys. Rev.} {\bf E#1} (#2) #3}
\def\PRA#1#2#3{{\sl Phys. Rev.} {\bf A#1} (#2) #3}

\def\JMP#1#2#3{{\sl J. Math. Phys.} {\bf #1} (#2) #3}

\def\PR#1#2#3{{\sl Phys. Reports} {\bf #1} (#2) #3}

\def\JPA#1#2#3{{\sl J. Physics} {\bf A#1} (#2) #3}
\def\JPAMT#1#2#3{{\sl J. Physics A: Math. Theor.} {\bf A#1} (#2) #3}
\def\JPB#1#2#3{{\sl J. Physics B: At. Opt. Phys.} {\bf B#1} (#2) #3}

\def\PHSD#1#2#3{{\sl Physica} {\bf D#1} (#2) #3}

\def\JHEP#1#2#3{{\sl JHEP} {\bf #1} (#2) #3}

\def\JCP#1#2#3{{\sl Journal of Computational Physics} {\bf #1} (#2) #3}
\def\SAM#1#2#3{{\sl Stud. Appl. Math.} {\bf #1} (#2) #3}
\def\MAA#1#2#3{{\sl Methods and  Applications of Analysis} {\bf #1} (#2) #3}
\def\OL#1#2#3{{\sl Optics Letters} {\bf #1} (#2) #3}
\def\Nonl#1#2#3{{\sl Nonlinearity} {\bf #1} (#2) #3}
\def\CJP#1#2#3{{\sl Canadian Journal of Physics} {\bf #1} (#2) #3}
\def\OC#1#2#3{{\sl Optics Communications} {\bf #1} (#2) #3}
\def\JPAMG#1#2#3{{\sl J. Physics A: Math. Gen.} {\bf A#1} (#2) #3}

\begin{document}

\begin{titlepage}

\vspace{.2in}
\begin{center}
{\large\bf Quasi-integrability in the modified defocusing 
 non-linear Schr\"odinger model and dark solitons}
\end{center}

\vspace{.2in}

\begin{center}

H. Blas$^{a}$ and M. Zambrano$^{b}$

\par \vskip .2in \noindent

$^{a}$ Instituto de F\'{\i}sica\\
Universidade Federal de Mato Grosso\\
Av. Fernando Correa, $N^{0}$ \, 2367\\
Bairro Boa Esperan\c ca, Cep 78060-900, Cuiab\'a - MT - Brazil \\

$^{b}$ Instituto de Ci\^encias Matem\'aticas e de Computa\c c\~ao; ICMC/USP;\\ 
Universidade de S\~ao Paulo,\\
Caixa Postal 668, CEP 13560-970, S\~ao Carlos-SP, Brazil\\
\normalsize
\end{center}

\vspace{.3in}

\begin{abstract}
\vspace{.3in}

The concept of quasi-integrability has been examined in the context of deformations of the defocusing non-linear Schr\"odinger model (NLS). Our results show that the quasi-integrability concept, recently discussed in the context of deformations of the sine-Gordon, Bullough-Dodd and focusing NLS models, holds for the modified defocusing NLS model with dark soliton solutions and it exhibits the new feature of an infinite sequence of alternating conserved and asymptotically conserved charges. For the special case of two dark soliton solutions, where the field components are eigenstates of a space-reflection symmetry, the first four and the sequence of even order charges are exactly conserved in the scattering process of the solitons. Such results are obtained through analytical and numerical methods, and employ adaptations of algebraic techniques used in integrable field theories. We perform extensive numerical simulations and consider the scattering of dark solitons for the cubic-quintic NLS model with potential  $V =\eta I^2 - \frac{\epsilon}{6} I^3 $ and the saturable type potential satisfying $V'[I] =2 \eta I - \frac{\epsilon I^q}{1+ I^q},\,q \in \IZ_{+}$, with a deformation parameter $\epsilon \in \IR$ and $I=|\psi|^2$. The issue of the renormalization of the charges and anomalies, and their (quasi)conservation laws are properly addressed. The saturable NLS supports elastic scattering of two soliton solutions for a wide range of values of $\{\eta, \epsilon,  q\}$. Our results may find potential applications in several areas of non-linear science, such as the Bose-Einstein condensation. 
\end{abstract}

\end{titlepage}

\section{Introduction}
 
The theory of soliton collisions  in the integrable models is well known in the current literature. In particular, the solitons emerge with their velocities and shapes completely unaltered after collision between them, the only noticed outcome being their phase shifts. However, certain non-linear field theory models with relevant physical applications possess soliton-like solutions and it is difficult to know {\sl a priori} if they are in fact true solitons. The so-called quasi-integrability concept has recently been put forward in the context  of certain deformations of the integrable sine-Gordon (SG), Bullough-Dodd (BD) and focusing non-linear Schr\"{o}dinger (NLS) models \cite{jhep1, jhep2}. The idea is that many non-integrable theories posses solitary wave solutions that behave similar to solitons, i.e. the scattering of such solitons basically preserve their shapes and velocities. So, certain deformations of the SG,  BD  and focusing NLS models were considered as quasi-integrable theories possessing infinite number of charges that are asymptotically
conserved. This means that during the collision  of two soliton-like solutions such
charges do vary in time and sometimes significantly, however their values remain the same in the  past (before the collision) and in the future (after the collision). The deformations of the topological SG and BD models studied in \cite{jhep1} furnish kink-type solitary waves presenting soliton-like behaviour. Whereas, the deformed focusing NLS model considered in \cite{jhep2} supports bright solitary waves presenting soliton-like behaviour. The both numerical and analytical techniques were used to describe the concept of quasi-integrability.

Regarding the analytical calculations of solitary wave collisions in deformed NLS models some results have been obtained  for the
cases of small perturbations of the integrable NLS model \cite{keener, malomed, ablowitz}. For nearly integrable models it has been
implemented a perturbation theory allied to the the inverse scattering transform (IST) method
\cite{kivshar}. These methods have been applied mainly to the focusing NLS solitons that decay at infinity, i.e. the bright solitons. On the other hand, the integrable defocusing NLS model supports dark solitons which are intensity dips sitting on a continuous wave background (cwb) with a phase change across their intensity minimum. In order to apply to the dark soliton perturbation and the non-vanishing boundary conditions (nvbc), inherent to this type of solution,  various improvements  have been made to the calculations and methods \cite{chen, lashkin, ao} based on the so-called complete set of squared Jost solution (eigenstates of the linearised NLS operator). The
implementation of direct methods, however, consider small perturbations around the integrable NLS model \cite{ablowitz, yu}. Moreover, in \cite{ablowitz} it has been argued that the squared Jost functions associated with the dark soliton might be an insufficient basis, so rendering problematic  the issue of the  completeness of the basis for the solution space of a deformed NLS equation. The solitary wave collisions of certain deformed NLS models,  beyond small perturbations, have been considered in the recent literature; e.g. in  \cite{jpa2} it has been computed the spatial shifts of fast dark-dark  soliton collisions and in \cite{theocharis} some properties of slow dark soliton collisions have been studied.

It would be interesting to understand the issue of quasi-integrability in the other variant of the NLS model, i.e. the  deformed defocusing NLS model, since such theories also appear in many areas of non-linear science, condensed matter physics, plasma physics  and,  in particular, in the study of Bose-Einstein condensates. In this context, our aim is to predict the results of solitary wave collisions and test the quasi-integrability concept in deformed defocusing NLS models with nvbc. However, the non-vanishing boundary condition inherent to the dark solitons, which may change when the deformations
are present, introduces some complications when applying the techniques
developed for bright solitons in order to study the quasi-integrability concept. In the analytic treatments the nvbc introduces the issue of the renormalization of the charges  and anomalies in the expressions of the quasi-conservation laws which must be properly addressed. This is closely related  to the fact that the dark soliton solution is a composite object, i.e. it incorporates the continuous wave background ({\sl cwb}) plus the dark soliton itself. In addition, to simulate the time dependence of field configurations for computing
dark solitons properties the choice of the efficient and accurate numerical methods, the so-called time-splitting cosine pseudo-spectral finite difference (TSCP) and the time-splitting finite difference with transformation (TSFD-T) methods  \cite{bao}  will be made in order to control the highly oscillatory phase background. In fact, these methods allowed us to improve in several orders of magnitude the accuracy in the computation of the charges and anomalies presented in \cite{jhep2}.

The deformed focusing NLS with bright soliton solutions  and the structures responsible for the phenomenon of quasi-integrability have been discussed in \cite{jhep2}. In this context it has been shown that this model possesses an infinite number of asymptotically conserved charges. An explanation found so far  for this behaviour of the
charges is that some special soliton type solutions are eigenstates of a
space-time parity transformation.  On the other hand, the  deformed defocusing NLS with dark soliton solutions also presents the above remarkable properties, however in this case we can say even more. As we will show, there are some soliton like solutions which present a special space-reflection symmetry for any time, so this property implies the exact conservation of the sequence of the even order charges, which in the absence of this special symmetry would be conserved only asymptotically, as in the models studied in \cite{jhep1, jhep2}. Here  we will show the connection between the space-reflection parity and  the exactly conserved charges which involves an
interplay between the space-reflection parity and internal transformations in the affine Kac-Moody  
algebra underlying the anomalous Lax equation. However, it seems to be that such parity property is a sufficient but not a necessary condition in order to have the sequence of exactly conserved charges. In fact, as we will show by numerical simulations, there are certain soliton like configurations without this symmetry which also exhibit such conserved charges. So, these properties constitute the distinguishing new features associated to the deformed defocusing NLS with dark soliton solutions, as compared to the previous quasi-integrable models \cite{jhep1, jhep2}.

The paper is organized as follows. In the next section we discuss the deformed defocusing NLS model with  non-vanishing boundary conditions. In section \ref{quasi}, we discuss the concept of quasi-integrability  for deformed defocusing NLS, based on an anomalous zero curvature condition. In  \ref{stp} and \ref{srs} we discuss the relationships between the space-time parity and asymptotically conserved charges, and the space-reflection parity and the exactly conserved charges, respectively. In section \ref{renor} we perform the renormalizations of the charges and anomalies relevant for dark soliton solutions.  In sections \ref{spacetime} and \ref{space1}  we discuss the space-time and space-reflection symmetries of dark soliton solutions of the integrable NLS model, respectively. In \ref{solitary1} we show  the vanishing of the anomalies associated to solitary waves.  In \ref{svanish1}, by using general symmetry considerations, we show that the anomalies $\b_r^{(n)}$ vanish when evaluated on a general solitary wave of the deformed NLS model. In \ref{svanish2} we show, by direct computation, the vanishing of the first non-trivial anomaly $\b_r^{(4)}$ when evaluated on a general solitary wave of the deformed NLS. In section \ref{simul} we present the results of our numerical simulations which allowed us to compute and study various properties of dark soliton scattering of the full equations of motion for two deformations of the NLS model and several values of the deformation parameter $\epsilon$. In \ref{first} we present the results of the simulations of the first deformation with  $V = \eta |\psi|^4 -\frac{\epsilon}{6} |\psi|^6$, and in \ref{second} the second deformation with $V'[I] = 2 \eta I - \epsilon \frac{I^q}{1+ I^q}$. In \ref{conclu} we present some conclusions and discussions. The appendix \ref{solitary} presents the eqs. of motion in the $R$ and $\vp$ parametrizations, as well as some useful identities  and \ref{apen2} presents relevant expressions corresponding to the gauge transformation.

\section{Deformations of defocusing NLS}

\label{defor}
We will consider non-relativistic models of a complex scalar field in $(1+1)-$dimensions with Lagrangian density given by
\br
\label{lag}
{\cal L} = \frac{i}{2}\( \bar{\psi} \pa_t  \psi -\psi \pa_t \bar{\psi}\)- \pa_{x} \bar{\psi} \pa_x\psi - V(|\psi|^2),
\er 
where $\bar{\psi}$ stands for the complex conjugation of $\psi$ and $V: \IR_{+} \rightarrow \IR $ . The Euler-Lagrange equation following from (\ref{lag}) becomes
\br
\label{nlsd}
i \frac{\pa}{\pa t} \psi(x,t) + \frac{\pa^2}{\pa x^2} \psi(x,t) -  \frac{\pa V[|\psi(x,t)|^2]}{\pa |\psi(x,t)|^2} \psi(x,t) =  0.\er 

The model (\ref{nlsd}) defines the deformed NLS model and it supports dark soliton type solutions in analytical form for some special functions $V[I]$. The potential  $V[I] = \eta I^2,\,(\eta>0)$, corresponds to the integrable NLS model and supports N-dark soliton solutions. The potential  $V[I] = \eta  I^2-\epsilon I^3/6$ defines the non-integrable cubic-quintic NLS model (CQNLS) which possesses analytical bright and dark type solitons   \cite{cowan, crosta}. In \cite{sombra, cowan} the bright solitary waves of the cubic-quintic focusing NLS have been regarded as quasi-solitons presenting partially inelastic collisions in certain region of parameter space. Among the models with saturable non-linearities \cite{kivshar}, the case $V[I] = \frac{1}{2} \rho_s (I+\frac{\rho_s^2}{I+ \rho_s})$ also exhibits analytical dark solitons \cite{kroli}. The deformed NLS model with $V'[I] = 2 \eta I - \epsilon \frac{I^q}{1+ I^q}$ will deserve a careful consideration below. It passes  the Painlev\'e test for arbitrary positive integers  $q\in \IZ_{+}$ and $\epsilon =1$ \, \cite{enns}. However, its Lax pair formulation and analytical solutions for a general set $\{ \eta, \epsilon, q\}$, to the best of our knowledge, are not known in the literature.     

The qualitative properties of travelling waves of the NLS model for general non-linearities and non-vanishing boundary conditions have been studied in \cite{chiron}. Exact analytical dark-soliton solutions of the deformed  NLS model with arbitrary potential $V$ are not available, and one can resort to numerical simulations to obtain such solutions. However, dark-soliton solutions can be presented in an implicit form as a curve in a complex plane (for the complex variable $Z= A + i B$, where $\psi \equiv (A + i B) e^{(kx + w t)}$) (see \cite{kivshar} and references therein) or in a small-amplitude approximation where the deformed NLS model is reduced to the Korteweg-de Vries equation \cite{bass}.     

Among the possible deformations of the NLS model the case $ V = \frac{ 2\eta}{2+ \epsilon} \( |\psi|^2\)^{2+\epsilon}, \epsilon \in \IR$, for $\eta<0$ has recently been considered in \cite{jhep2} in order to study  the concept of quasi-integrability for bright soliton collisions. An analytical solution with vanishing boundary condition (bright soliton) for this potential is well known in the literature.  However, an analytical dark soliton solution, as far as we know,  is not known in the literature.
 
In this paper we will study analytically  and numerically  some deformations of the NLS model of the type (\ref{nlsd}). The {\bf first} deformation we will consider in our study is defined by the non-integrable cubic-quintic NLS model
\br   
\label{cqnls2}
i \frac{\pa}{\pa t} \psi(x,t) +  \frac{\pa^2}{\pa x^2} \psi(x,t) -\( 2\eta |\psi(x,t)|^2 - \frac{\epsilon}{2} |\psi(x,t)|^4\) \psi(x,t) =0,\,\,\,\, \eta > 0,\,\,\epsilon \in \IR
\er
where we have considered $V'[I] = 2 \eta I - \frac{\epsilon}{2} I^2$ in (\ref{nlsd}) (here and in the next sections $V'[y] \equiv \frac{d V[y]}{d y},\, V''[y] \equiv \frac{d^2 V[y]}{d y^2}$). This deformation possesses analytical  solitary waves  as we will present below.

The {\bf second} deformation considers a saturable non-linearity and it is defined by the equation  
\br   
\label{enns1}
i \frac{\pa}{\pa t} \psi(x,t) +  \frac{\pa^2}{\pa x^2} \psi(x,t) -\( 2\, \eta |\psi(x,t)|^2 -  \frac{ \epsilon \, |\psi(x,t)|^{2 q}}{1+ |\psi(x,t)|^{2 q}}\) \psi(x,t) =0,\,\,\,\, \eta > 0,\,\,\,\, q \in \IZ_{+},\,\,\epsilon \in \IR.
\er
We do not know any analytical solutions for a general set of parameters of this model, however dark solitary wave solutions will be obtained numerically.
 
The both deformations reproduce the integrable defocusing NLS model in the limit $\epsilon \rightarrow 0 $.

\section{Quasi-integrability of deformed NLS}
\label{quasi}
Next we undertake an analytical study of the properties of the deformed NLS model (\ref{nlsd}), in order to do this  we will use the well known techniques from the integrable field theories. We follow the developments and notations  put forward in \cite{jhep1, jhep2} on quasi-integrability. Then, one considers an anomalous zero curvature representation of the deformed NLS model (\ref{nlsd}) with the connection given by  
\br
\label{con1}
A_{x} &=&- i\, T_{3}^{1} + \bar{\g} \bar{\psi} \,T_{+}^{0}+\g \psi \,T_{-}^{0},\\
\nonumber
A_{t} &=& i \,T^{2}_{3} + i \frac{\d V}{\d |\psi|^2} \,T_{3}^{0} -( \bar{\g} \bar{\psi} \,T_{+}^{1}+\g \psi \,T_{-}^{1}) -i (\bar{\g} \pa_x \bar{\psi} \,T_{+}^{0}-\g \pa_x \psi \,T_{-}^{0}),
\er  
where the above Lax potentials are based on a $sl(2)$ loop algebra with commutation relations
\br
[T^{m}_{3}\,, T_{\pm}^{n}] = \pm T^{m+n}_{\pm};\,\,\,\,\, [T^{m}_{+}\,, T_{-}^{n}] = 2\, T^{m+n}_{3}.
\er 
One can verify that the curvature of the connection (\ref{con1}) is given by
\br
\label{zero0}
F_{xt} & \equiv &  \pa_{t}A_x - \pa_{x} A_t + [A_x\,,A_t] \\
 &=&X T_{3}^{0}+ i \bar{\g} \Big[-i \pa_{t} \bar{\psi} + \pa_x^2 \bar{\psi} - \bar{\psi} \frac{\d V}{\d |\psi|^2} \Big] T_{+}^{0}-i \g \Big[i \pa_{t} \psi + \pa_x^2 \psi - \psi \frac{\d V}{\d |\psi|^2} \Big] T_{-}^{0}\label{zero1}
\er
with
\br
\label{an0}
X \equiv -i \pa_x \(\frac{\d V}{\d |\psi|^2} -2 \bar{\g} \g |\psi|^2 \).
\er

Note that when the equation of motion  (\ref{nlsd}) and its complex conjugate, i.e  $-i \pa_{t} \bar{\psi} + \pa_x^2 \bar{\psi} - \bar{\psi} \frac{\d V}{\d |\psi|^2} =0 $, are satisfied the terms proportional to the Lie algebra generators $T_{\pm}^0$ vanish.  In addition, the quantity $X$ vanishes for the usual non-linear Schr\"{o}dinger potential
\br
\label{pot0}
V(|\psi|^2) = \eta (|\psi|^2)^2,\,\,\,\,\eta \equiv   \bar{\g} \g.   
\er
Then, the curvature vanishes for the NLS model making this theory an integrable field theory. For the generalizations of this theory, i.e. deformations of the potential (\ref{pot0}), the above curvature does not vanish and the model is regarded as non-integrable \cite{jhep2}. Next, we use some algebraic techniques borrowed from the theory of integrable models in order to analyse the equation (\ref{zero1}) corresponding to a model with non-vanishing anomaly $X$ (\ref{an0}), or equivalently $F_{xt}$ in (\ref{zero0}). Let us denote
\br
\label{exp1}
\psi = \sqrt{R} e^{i \vp/2}
\er
and parametrize $\g = i \sqrt{\eta} e^{i \a},\,\,\g = -i \sqrt{\eta} e^{-i \a},\,\,\eta \equiv   \bar{\g} \g$. Substituting the parametrization  (\ref{exp1}) into (\ref{nlsd}) one gets the system of eqs. of motion
\br\label{eq111}
 \pa_t R + \pa_x \( R \pa_x \vp \) &=& 0\\
 \pa_t \vp + \frac{1}{2} (\pa_x \vp)^2 - \frac{\pa_x^2 R}{R} + \frac{1}{2} (\frac{\pa_x R}{R})^2+2 \, \frac{\d V}{\d R } &=&0\label{eq222} 
\er

We specialize the parametrizations for the case of the defocusing NLS, i.e. the case $\eta > 0$.  In addition, define a new basis of the $sl(2)$ loop algebra as
\br
b_n = T^{n}_{3},\,\,\,F_{1}^{n}= \frac{1}{2} ( T_{+}^{n}-T_{-}^{n}),\,\,\,F_{2}^{n}= \frac{1}{2} ( T_{+}^{n}+T_{-}^{n}),
\er
satisfying
\br
\label{sl2}
[b_{m}\,,b_{n}] =0,\,\,\,\,[b_{n}\, , F_{1}^{m}]= F_{2}^{n+m},\,\,\,\,[b_{n}\,, F_{2}^{m}]= F_{1}^{n+m},\,\,\,\,[F_{1}^{n}\,, F_{2}^{m}]= b_{n+m}.
\er

The connection in the new basis is obtained through the gauge transformation 
\br
\label{gauge1}
\widetilde{A}_{\mu} \equiv \widetilde{g}  A_{\mu} \widetilde{g}^{-1} + \pa_{\mu} \widetilde{g}\, \widetilde{g}^{-1},\,\,\,\,\mbox{with}\,\,\,\,      \widetilde{g} = \exp{\{i ( \frac{\vp}{2}+\a )\,b_0\}}.
\er
So, inserting the connection (\ref{con1}) into the above expression one gets 
\br
\label{con2}
\widetilde{A}_x &=&-i \,b_1 + \frac{i}{2} \pa_x \vp \,b_0 - 2 i \sqrt{\eta} \sqrt{R} \,F_1^{0}\\
\nonumber
\widetilde{A}_t &=&i \,b_2 + \frac{i}{2} \pa_t \vp \,b_0  + i \frac{\d V}{\d |\psi|^2}\, b_0 + 2 i \sqrt{\eta} \sqrt{R} \,F_1^{1} + \sqrt{\eta} \sqrt{R} \, [ - \frac{\pa_x R}{R} \, F_2^{0} + i \pa_x \vp \, F_1^{0}].
\er

So, considering the fields which satisfy the equation (\ref{nlsd}) and its complex conjugate, i.e. the terms in the directions of $T_{\pm}^{0}$ vanish in (\ref{zero1}), the curvature becomes    
\br
\label{curv2}
\widetilde{F}_{xt} =  \pa_{t}\widetilde{A}_x - \pa_{x} \widetilde{A}_t + [\widetilde{A}_x\,,\widetilde{A}_t] = X\, b_0; \,\,\,\mbox{with} \,\,\,X \equiv -i \pa_x \(\frac{\d V}{\d |\psi|^2} -2 \eta |\psi|^2 \).  
\er 

Next, in order to construct the quasi-conservation laws we perform a further gauge transformation 
\br
\label{gauge2}
\widetilde{A}_{\mu} \rightarrow a_{\mu} = g \widetilde{A}_{\mu} g^{-1} + \pa_{\mu} g\, g^{-1}
\er
with 
\br
\label{group}
g = e^{\sum_{n=1}^{\infty} {\cal F}^{(-n)}},\,\,\,\mbox{and} \,\,\, {\cal F}^{(-n)} \equiv \zeta_{1}^{(-n)} F_{1}^{-n} + \zeta_{2}^{(-n)} F_{2}^{-n}.
\er
The parameters $\zeta_{i}^{(-n)}$ of the transformation will be determined below so that the $a_x$ component of the new connection lies in the abelian sub-algebra generated by the elements $b_n$. Under the gauge transformation (\ref{gauge2}) the curvature (\ref{curv2}) transforms as
\br
\label{con3}
\widetilde{F}_{xt} \rightarrow g \widetilde{F}_{xt} g^{-1} = \pa_{t} a_x - \pa_{x} a_t + [a_t\,,a_x]= X g b_0 g^{-1}, 
\er

The $sl(2)$ loop algebra is furnished with an integer gradation associated to the grading operator $d$ defined by
\br
\label{grad}
d \equiv \l \frac{d}{d \l},\,\,\, [d, \, b_{n}] = n b_n, \,\,\, [d,\,F_i^{n}]= n F_i^{n}.
\er
Since the group element $g$ in (\ref{group}) is generated by exponentiating negative grade elements of the algebra and the $\widetilde{A}_x$ component of the connection (\ref{con2}) lies in the directions of generators of grades $0$ and $1$, the $a_x$ component of the new connection has generators with grades $n = 1, 0, -1, -2,....$. Then, expanding $a_x$ as a linear combination of the eigen-subspaces of the grading operator $d$, one has
\br
a_x &=& \sum_{n=-\infty}^{1} a_x^{(n)}\nonumber \\
 a_x^{(1)}   &=& -i b_1 \nonumber \\
 a_x^{(0)}  &=&  i [b_1,\,{\cal F}^{(-1)}] + \widetilde{A}_x^{(0)} \nonumber \\
 \label{newcon} a_x^{(-1)}   &=&  i [b_1,\,{\cal F}^{(-2)}] + [{\cal F}^{(-1)},\,\widetilde{A}_x^{(0)}]- \frac{i}{2\!} [{\cal F}^{(-1)},\,[{\cal F}^{(-1)} ,\,b_1]]+ \pa_x {\cal F}^{(-1)}\\
 a_x^{(-2)}   &=&  i [b_1,\,{\cal F}^{(-3)}] + [{\cal F}^{(-2)},\,\widetilde{A}_x^{(0)}]- \frac{i}{2\!} [{\cal F}^{(-2)},\,[{\cal F}^{(-1)} ,\,b_1]] \nonumber \\
 && -\frac{i}{2\!} [{\cal F}^{(-1)},\,[{\cal F}^{(-2)} ,\,b_1]] + \frac{i}{2\!} [{\cal F}^{(-1)},\,[{\cal F}^{(-1)} ,\,\widetilde{A}_x^{(0)}]]\nonumber \\
 &&- \frac{i}{3\!} [{\cal F}^{(-1)},\,[{\cal F}^{(-1)},\,[{\cal F}^{(-1)} ,\,b_1]]]+ \frac{i}{2\!} [{\cal F}^{(-1)} ,\,\pa_x {\cal F}^{(-1)}]] + \pa_x {\cal F}^{(-2)}\nonumber \\ 
    && \vdots \nonumber \\
  a_x^{(-(n-1))}  &=&  i [b_1,\,{\cal F}^{(-n)}]+ ...   \nonumber
\er  
where 
\br
\label{con0}
\widetilde{A}_x^{(0)} \equiv \frac{i}{2} \pa_x \vp b_0 - 2 i \sqrt{\eta} \sqrt{R}\, F_1^{0}.
\er
As usual we will use the generator $b_1$ as a semi simple element in the sense that its adjoint action decomposes the $sl(2)$ loop algebra $\cal G$ into its kernel and image as follows
\br
{\cal G} = \mbox{Ker} + \mbox{Im};\,\,\,\,\,[b_1,\,\mbox{Ker}]=0,\,\,\,\,{\mbox Im} = [b_1,\,{\cal G}],    
\er  
such that Ker  and  Im have no elements in common. According to the commutation relations in (\ref{sl2}) one has that the $b_n$'s form a basis of Ker, and $F_i^{n},\, i=1,2,$ a basis of Im. Next, one can make all the components of $a_x$ in the subspace Im to vanish by choosing conveniently the parameters $\zeta_i^{(-n)}$. This can be done recursively, for example $\zeta_{1}^{(-1)}=0,\,\zeta_{2}^{(-1)}= 2 \sqrt{\eta} \sqrt{R}$, and so on. So, the $a_x$ component get rotated onto the subspace Ker
\br
\label{ax}
a_x = -i b_1 + \sum_{n=0}^{+\infty} a_x^{(3, -n)} b_{-n}.
\er          
Notice that in this process it has not been used neither the  equation of motion (\ref{nlsd}) nor its complex conjugate. The components $a_x^{(3, -n)}$ will depend on the real fields $R$ and $\pa_x \vp$, and their $x$-derivatives,  but not on the potential $V$. In the Appendix B we provide the first components of $a_x$.

Next, the component $a_t$ of the connection does not get rotated onto the subspace Ker even when the equation of motion (\ref{nlsd}) and its complex conjugate are used. Then, one has
\br
a_t = i b_2 + \sum_{n=0}^{+ \infty} \Big[ a_{t}^{(3, -n)} b_{-n} + a_t^{(1, -n)} F_{1}^{-n} +a_t^{(2, -n)} F_{2}^{-n}\Big].
\er    
Since $a_x$ lies in Ker one has that the term $[a_t ,\, a_x]$ in (\ref{con3}) has components only in Im. Let us denote  
\br
\label{a1n}
g b_0 g^{-1} = \sum_{n=0}^{\infty} \Big[\a^{(3, -n)} b_{-n} + \a^{(1,-n)} F_{1}^{-n} + \a^{(2,-n)} F_{2}^{-n} \Big]
\er  
then, we find that

\br
\label{cons1}
\pa_{t} a_x^{(3, -n)} - \pa_{x} a_{t}^{(3, -n)} = X \a^{(3, -n)};\,\,\,\,n=0,1,2,... 
\er
In the appendix C we present the first quantities $\a^{(3, -n)}$.

Next, let us discuss the nvbc associated to dark soliton solutions. These solutions present a difficulty in the sense that the field $\psi$ does not vanish at $x \rightarrow  \pm \infty$. In general, the behaviour of the solutions are such that 
\br
\label{vac1}
\psi \rightarrow \psi_0 = \left\{ \begin{array}{ll}
  |\psi_0| e^{i(k x + w t + \d)} , &   x \rightarrow -\infty\\  
 |\psi_0| e^{i(k x + w t + \d)} \, e^{i \theta_0}, &   x \rightarrow +\infty
\end{array} \right.
\er 
where $\theta_0$ is a constant phase. The connection (\ref{con1}) is not suitable to construct the charges because
\br
A_t|_{x\rightarrow -\infty} \neq A_t|_{x\rightarrow +\infty} 
\er

However, making the gauge transformation (\ref{gauge1}) and the reparametrization (\ref{exp1}) the new connection $\widetilde{A}_{\mu}$ depends upon the modulus $R$ and its derivatives, and only on the derivatives of the phase $\vp$. Therefore, one has that for dark solitons 
\br
\widetilde{A}_t|_{x\rightarrow -\infty} =  \widetilde{A}_t|_{x\rightarrow +\infty} 
\er
In addition the abelianized potential $a_{\mu}$ in (\ref{gauge2}) also satisfy that condition, i.e.
 \br
a_t|_{x\rightarrow -\infty} =  a_t|_{x\rightarrow +\infty} 
\er
because the group element also satisfies
\br
g|_{x\rightarrow -\infty} = g|_{x\rightarrow +\infty}
\er
Indeed, look at the expressions of the first terms of $\zeta_{1}^{(-n)}$\,and $\zeta_{2}^{(-n)}$ in the appendix B and one notes that $g$ depends on $R$ and its derivatives and only on the derivatives of the phase $\vp$.  

Then, in the defocusing NLS case with dark soliton solutions one has that the $a_t$ component of the connection satisfies a non-vanishing boundary condition such that $a_{t}^{(3, -n)}(x=+\infty) = a_{t}^{(3, -n)}(x=-\infty)$, then from  (\ref{cons1}) we have the anomalous conservation laws
\br
\label{ano1}
\frac{d Q^{(n)}}{dt} = \beta^{(n)},\,\,\,\,\,\mbox{where}\,\,\,\,\,Q^{(n)}= -i \int_{-\infty}^{\infty} dx \, a_x^{(3, -n)}\,\,\,\,\,\,\mbox{and}\,\,\,\,\,\,\b^{(n)} = -i \int_{-\infty}^{\infty} dx \,X \, \a^{(3, -n)} 
\er
Thus, the non-vanishing of the quantity $X$ given in (\ref{curv2}) and the anomalies $\b_n$ above imply the non-conservation of the charges.  Therefore, the charges and anomalies in (\ref{ano1}) are valid for the deformed NLS model with {\sl nvbc} and dark soliton solutions.

\subsection{Space-time parity and asymptotically conserved charges}
\label{stp}

Next, let us discuss a special symmetry which plays an important role in the study of quasi-integrable theories. For certain solutions of the theory (\ref{lag}) the charges $Q^{(n)}$ satisfy the so-called {\sl mirror symmetry}. This symmetry is realized on a special subset of solutions  of the deformed NLS model (\ref{nlsd}). For every solution belonging to this subset one can find a  point  $(x_{\rho},t_{\rho})$ in space-time and define a parity transformation around this point as  
\br
\label{pari0}
P: (\widetilde{x},\,\widetilde{t}) \rightarrow  (-\widetilde{x},\,-\widetilde{t}),\,\,\,\,\widetilde{x}\equiv x-x_{\rho},\,\widetilde{t}\equiv t-t_{\rho},       
\er
such that the fields $R$ and $\vp$ transform as
\br
\label{pari1}
R \rightarrow R,\,\,\,\,\vp \rightarrow -\vp + const. 
\er

In order to realize this symmetry at the level of the anomalous zero-curvature equation let us combine the above parity transformation with the order two $\IZ_2$ automorphism of the $sl(2)$ loop algebra given by
\br
\Sigma(b_n) = - b_n,\,\,\,\,\Sigma(F_1^{n}) = -F_1^{n},\,\,\,\, \Sigma(F_2^{n}) = F_2^{n}. 
\er

So, consider the composition of the internal and space time transformations as
\br
\Omega \equiv \Sigma P.
\er
Then, the $x-$component of the connection (\ref{con2}) is odd, i.e.
\br
\Omega (\widetilde{A}_x) = -\widetilde{A}_x 
\er
This property can be used to show that the group element $g$ which enter in the gauge transformation (\ref{gauge2}) is even, i.e. 
\br
\Omega(g)=g  
\er
Then, one can use the last property to show that the factor $\a^{(3, -n)}$ in the integrand of the anomaly $\beta_n$ in (\ref{ano1}) is even under the space-time parity, i.e.
\br  
\label{factor1} 
P\(\a^{(3, -n)}\) = \a^{(3, -n)}.
\er
For more details of this demonstration we refer to \cite{jhep2}.  

Next, let us note that the $X$ in the integrand of  the anomaly $\b_n$ in (\ref{ano1}) is a $x-$derivative of a functional of $R$.  Since $R$ is assumed to satisfy (\ref{pari1}) one has that the time-integrated anomaly vanishes
\br
\int_{-\widetilde{t}_o}^{\widetilde{t}_o} dt \int_{-\widetilde{x}_o}^{\widetilde{x}_o} dx \, X\, \a^{(3, -n)} = 0,   
\er 
where the integration is meant to be performed on any rectangle with center in $(x_{\rho},\, t_{\rho})$, and  the parameters $\widetilde{t}_o, \widetilde{x}_o$ which define the integration intervals are any given fixed values of the shifted space-time coordinates $\widetilde{t}$ and $\widetilde{x}$, respectively, defined in (\ref{pari0}).

A remarkable consequence of the vanishing of the integrated anomaly is that the charges (\ref{ano1}) satisfy a mirror time-symmetry around the point $t_\rho$   
\br
\label{mirror}
Q^{(n)} (t=\widetilde{t}_o+t_{\rho}) = Q^{(n)} (t=-\widetilde{t}_o+t_{\rho}),\,\,\,\,n=0,1,2,...
\er

So, this property holds for the special subsets of solutions satisfying (\ref{pari1}) and belonging to the deformed NLS model (\ref{nlsd}) with an even potential $V$ under the parity (\ref{pari1}). This defines the quasi-integrable deformed NLS model. Then, by taking  $\widetilde{t}_0 \rightarrow \infty $ in (\ref{mirror}) one has
\br
\label{asymp1}
  Q^{(n)} (+\infty) =  Q^{(n)} (-\infty),\,\,\,\, n=0,1,2,3,...
\er
This relationship shows that the scattering of two-soliton solutions may present an infinite number of charges which are asymptotically conserved. For some very special two-bright soliton configurations in the context of focusing deformed NLS it has been verified that the charges are asymptotically conserved, i.e. even though they vary in time during collision their values
in the far past and the far future are the same \cite{jhep2}. As we will show below, for the case of defocusing NLS it is a remarkable fact that for any two-dark soliton configuration it is possible to find a point in space-time $(x_{\rho},\,t_{\rho})$, such that the fields $R$ and $\vp$ always satisfy the space-time parity symmetry (\ref{pari0})-(\ref{pari1}) without any additional restriction on the soliton parameters. 

Next we will summarize the main steps in the construction of solutions in perturbation theory,  for a detailed account see \cite{jhep2}. One can choose as a zero order solution a solution of the NLS equation, such that it satisfies (\ref{pari1}), i.e. it exhibits only the pair $\(R^{+}_0,\,\vp^{-}_0\)$ and nothing of the pair $\(R^{-}_0,\,\vp^{+}_0\)$.  So, let us expand the fields in powers of the deformation parameter $\epsilon$   
\br
R &=& R_{0}^{(+)} + \epsilon R_{1}+ \epsilon^2 R_{2}+...,\,\,\,\,\,\vp = \vp_{0}^{(-)} + \epsilon \vp_{1}+ \epsilon^2 \vp_{2}+...,\\
R_{n}^{(\pm)}& \equiv &\frac{1}{2} \(1 \pm P \) R_{n} ,\,\,\,\,\vp_{n}^{(\pm)}\equiv \frac{1}{2} \(1 \pm P \) \vp_{n}
\er       
Then the equations of motion for the components  $\(R_{1}^{+},\,\vp_{1}^{-}\)$ of the first order solution satisfy non-homogeneous equations and so they can never vanish.  However, the pair $\(R_{1}^{-},\,\vp_{1}^{+}\)$ satisfy homogeneous equations and so they can vanish. In addition, if the pair of components without definite parity $\(R_{1},\,\vp_{1}\)$ is a solution, so is the pair $\(R_{1}-R_{1}^{-} = R_{1}^{+},\, \vp_{1}-\vp_{1}^{+}=\vp_{1}^{-}\)$. Then, one can always choose the components of the first order solution whith definite parity. For this choice one can show that the second order solution has also similar properties, i.e. the components  $\(R_{2}^{+},\,\vp_{2}^{-}\)$ satisfy non-homogeneous equations and  the components $\(R_{1}^{-},\,\vp_{1}^{+}\)$ satisfy homogeneous equations. This enables one to choose the second order solution to be the pair  $\(R_{2}^{+},\,\vp_{2}^{-}\)$ with definite parity, and this process repeats in all orders.  

Summarizing, we may conclude that

1. If one has a two-dark-soliton-like solution of the equation of motion (\ref{nlsd}), transforming under the space-time parity (\ref{pari0}) as in (\ref{pari1}), i.e.
\br
\label{summ1}
P(R) = R,\,\,\,\,P(\vp)= - \vp + \mbox{constant,}
\er

and

2. If the potential $V(R)$ in (\ref{nlsd}) evaluated on such a solution is even under the parity $P$, i.e. 
\br
\label{summ2}
P(V) = V
\er
such that 
\br
\label{summ20}
P(X) = - X,
\er  

3. Then, one has an infinite set of asymptotically conserved charges, i.e.
\br
\label{summ3}
  Q^{(n)} (t=+\infty) =  Q^{(n)} (t=-\infty),\,\,\,\, n=0,1,2,3,...
\er

Therefore, the values of the charges in the infinite past, before the collision of the solitons,
are the same as in the infinite future, after the collision. Theories possessing such properties are dubbed as quasi-integrable theories. In particular,  the deformed  NLS models (\ref{cqnls2})  and (\ref{enns1}) can be shown to satisfy the requirements (\ref{summ1}) and (\ref{summ2}) for some field configurations, and then they belong to the class of quasi-integrable theories, as we will discuss in this work.

\subsection{Space-reflection symmetry and conserved charges}
\label{srs}

Next, we consider some special solutions which  exhibit  a space-reflection symmetry
\br
\label{spari11}
{\cal P}_x: (\widetilde{x},\,\widetilde{t}) \rightarrow  (-\widetilde{x},\,\widetilde{t}),\,\,\,\,\widetilde{x}\equiv x-x_{\rho},\,\widetilde{t}\equiv t-t_{\rho},       
\er
such that the fields $R$ and $\vp$ transform as
\br
\label{xsym} 
R(-\widetilde{x}, \widetilde{t}) \rightarrow R(\widetilde{x}, \widetilde{t}),\,\,\,\,\vp(-\widetilde{x}, \widetilde{t}) \rightarrow \vp(\widetilde{x}, \widetilde{t}). 
\er

As we will see below, in the special case of 2-dark solitons moving in opposite directions and  equal velocities, such that they undergo a head-on collision, we have a space-reflection symmetry. The implication of this additional symmetry of the fields under space-reflection (for any shifted time), i.e. $R$ and $\vp$ being even fields, on the behaviour of the quantities $\a^{(3,-n)}$ deserves a further analysis. We will show that some  $\a^{(3,-n)}$'s possess an even parity under ${\cal P}_x$ for any shifted time. Since $X$ is odd the anomaly $\b_n$ in (\ref{ano1}) will vanish identically providing an exact conservation equation for the charges $\frac{d Q^{(n)}}{dt} = 0,\, n = 0, 2, 4, ...$.  Therefore, these type of charges  will not vary during collision, implying their exact conservation in the whole process of scattering. 

In order to realize the symmetry (\ref{spari11})-(\ref{xsym}) at the level of the anomalous zero curvature equations  (\ref{curv2}) and (\ref{con3}) we introduce a transformation  $\widetilde{\Sigma}$ in the internal space along with the space-reflection symmetry ${\cal P}_x$. So,  let us consider another order two $\IZ_2$ automorphism $\widetilde{\Sigma}$ of the $sl(2)$ loop algebra given by
\br
\label{auto2}
\widetilde{\Sigma}(b_n) = e^{i \pi n} b_n,\,\,\,\,\widetilde{\Sigma}(F_1^{n}) = e^{i \pi (n+1)} F_1^{n},\,\,\,\, \widetilde{\Sigma}(F_2^{n}) =  e^{i \pi (n+1)} F_2^{n}. 
\er
Note that the connection component $\widetilde{A}_x$ in (\ref{con2}) does not have a definite space-reflection  ${\cal P}_x$ parity. In addition, it doest not lie in an eigenspace of the automorphism $\widetilde{\Sigma}$ of the loop algebra.    
So, consider the joint action of the space and internal transformations as
\br
\widetilde{\Omega} \equiv \widetilde{\Sigma} {\cal P}_x.
\er
Then, the $x-$component of the connection (\ref{con2}) transforms as
\br
 \widetilde{\Omega}  (\widetilde{A}_x) = -\widetilde{A}_x.
\er
Actually, each term of $\widetilde{A}_x$ in (\ref{con2}) possesses this property. Next, we analyse the transformation property of the new connection $a_x$ in (\ref{newcon}). Thus, one can have  
$ \widetilde{\Omega} \([b_1,\,{\cal F}^{(-n)}]\) = - [b_1,\,\widetilde{\Omega} \({\cal F}^{(-n)}\)]$, and so 
\br
\(1+ \widetilde{\Omega}\) \([b_1,\,{\cal F}^{(-n)}]\) =  [b_1,\,(1-\widetilde{\Omega}) \({\cal F}^{(-n)}\)]
\er
The term $\widetilde{A}_x^{(0)}$ in the third eq. of (\ref{newcon}) as defined in (\ref{con0}) is odd under $\widetilde{\Omega}$, then it follows from the third equation in (\ref{newcon}) that 
\br
\label{iden}
(1+ \widetilde{\Omega}) a_{x}^{(0)} = i [b_1,\,(1-\widetilde{\Omega}) {\cal F}^{(-1)}]
\er
The ${\cal F}^{(-n)}$ have been selected to rotate $a_x$ into the kernel of the adjoint action of $b_1$ and the r.h.s. of (\ref{iden}) is in the image of the adjoint action, so for consistency the both sides  of (\ref{iden}) must vanish  
\br
(1+ \widetilde{\Omega}) a_{x}^{(0)}=0,\,\,\,\,(1-\widetilde{\Omega}){\cal F}^{(-1)}=0.
\er
Therefore, ${\cal F}^{(-1)}$ is even under $\widetilde{\Omega}$. Taking into account this result we see from the fourth equation in (\ref{newcon}) that
\br
(1+ \widetilde{\Omega}) a_{x}^{(-1)} = i [b_1,\,(1-\widetilde{\Omega}) {\cal F}^{(-2)}].
\er
Then, following similar arguments we conclude that
\br
(1+ \widetilde{\Omega}) a_{x}^{(-1)}=0,\,\,\,\,(1-\widetilde{\Omega}){\cal F}^{(-2)}=0,
\er
and so that  ${\cal F}^{(-2)}$ is even under $\widetilde{\Omega}$ as well. Furthermore, from the fifth equation in (\ref{newcon}) one has that $(1+ \widetilde{\Omega}) a_{x}^{(-2)} = i [b_1,\,(1-\widetilde{\Omega}) {\cal F}^{(-3)}]$, and following similar arguments as above we find that 
\br
(1+ \widetilde{\Omega}) a_{x}^{(-2)}=0,\,\,\,\,(1-\widetilde{\Omega}){\cal F}^{(-3)}=0
\er
By repeating the same arguments as above one can show that all ${\cal F}^{(-n)}$ are even under $\widetilde{\Omega}$. This property can be used to show that the group element $g$ which enter in the gauge transformation (\ref{gauge2}) is even, i.e. 
\br
\label{gev}
\widetilde{\Omega}(g)=g  
\er
Since $\widetilde{A}_x$ and $\pa_x$ are odd under $\widetilde{\Omega}$, and since $g$ is even one can show that $a_x$ in (\ref{gauge2}) is odd under $\widetilde{\Omega}$. In addition, since the ${\cal F}^{(-n)}$ are even under $\widetilde{\Omega}$ and the generators satisfy (\ref{auto2}), it follows from (\ref{group}) that  $\widetilde{\Omega}(\zeta_a^{(-n)}) = e^{(-n+1)i \pi}\zeta_a^{(-n)},\,a=1,2 $. One can verify these results by inspecting the explicit expressions for the fields $\zeta_a^{-n}$ provided in the appendix B.
  
In order to find  the parity of the factor $\a^{(3, -n)}$ in the integrand of the anomaly $\beta_n$ in (\ref{ano1}) let us consider some properties of the $sl(2)$ loop algebra. The Killing form is given by  
\br
\mbox{Tr} \( b_n,\,b_m\) = \frac{1}{2} \d_{n+m,0};\,\,\,\,\mbox{Tr} \( b_n,\,F_{i}^m\) = 0,\,\,\,\,i=1,2.
\er
It is realized by  $\mbox{Tr} (\star) = \frac{1}{2\pi i} \oint \frac{d\l}{\l} \mbox{tr} (\star)$, where $\mbox{tr}$ is the usual finite matrix trace, and $T_{i}^{n}= \l T_i,\,i=3, \pm$. So, from (\ref{a1n}) one  has that 
\br
\a^{(3, -n)} = 2 \mbox{Tr}\( g b_0 g^{-1} b_n\) = 2 e^{i n\pi}\mbox{Tr}\( \widetilde{\Sigma}(g) b_0 \widetilde{\Sigma}(g^{-1}) b_n\),  
\er
where the invariance property of the Killing form under $\widetilde{\Sigma}$ has been used, and the fact that the $b_n$ transform as in (\ref{auto2}). Then, taking into account (\ref{gev}) one has that
\br
{\cal P}_x\(\a^{(3, -n)}\)= 2 e^{i n\pi}\mbox{Tr}\( \widetilde{\Omega}(g) b_0 \widetilde{\Omega}(g^{-1}) b_n\) =  2 e^{i n\pi}\mbox{Tr}\(g b_0 g^{-1} b_n\)=  e^{i n\pi}  \a^{(3, -n)}.
\er
and so one can see that the $\a^{(3, -n)}$'s with $n=0,2,4,...$ are even under ${\cal P}_x$. On the other hand, since we have assumed that $R$ is even under  ${\cal P}_x$ and that $X$ given in (\ref{curv2}) is a $x-$derivative of a functional of $R$, one has that $X$ is odd under ${\cal P}_x$, i.e. ${\cal P}_x(X) = - X$ and so that   
\br \nonumber
\b^{(n)} &=& -i \int_{-\infty}^{+\infty} dx \, X \a^{(3, -n)}\\
 &=& 0,\,\,\,\,\,\,\,\,\,\,\,\,\,\,\,\,\,\,\,\,\,\,\,\,\,\,n=0,2,4,...\label{qcon}
\er 
Therefore, the anomalous conservation laws (\ref{ano1}) imply 
\br
\label{charge11}
\frac{d Q^{(n)}}{dt} =0,\,\,\,\,\,\,\,\,\,\,n=0,2,4,...
\er 
Consequently, the even order charges are exactly conserved. Notice that it holds for any potential $V$ which depends only on the modulus $|\psi|$. The only requirement is that the field components satisfy (\ref{xsym}).
     
Next let us analyse the solutions which admit the system of equations (\ref{eq111})-(\ref{eq222})  such that the field components $R, \vp$ satisfy the space-reflection symmetry (\ref{spari11})-(\ref{xsym}). We perform this construction in perturbation theory around solutions of the NLS model, so we expand the equations of motion and the solutions into power series in $\epsilon$, as
\br
R &=& \sum_{n=0}^{\infty} \epsilon^n \( R^{(+)}_n + R^{(-)}_n\),\,\,\,\,\,\vp = \sum_{n=0}^{\infty} \epsilon^n \( \vp^{(+)}_n+ \vp^{(-)}_n\),\\
R^{(\pm)}_n &= &\frac{1}{2} \(1 \pm {\cal P}_x \) R_n ,\,\,\,\,\vp^{(\pm)}_n=\frac{1}{2} \(1 \pm {\cal P}_x \) \vp_n
\er

The deformed potential and its gradient can been expanded as
\br
V(R) = V|_{\epsilon=0} + \epsilon \Big[ \frac{\pa V}{\pa \epsilon}|_{\epsilon=0} +  \frac{\pa V}{\pa R}|_{\epsilon=0} R_1 \Big] + {\cal O}({\epsilon}^2)
\er
and
\br
\frac{\pa V}{\pa R} &=& \frac{\pa V}{\pa R}|_{\epsilon=0} + \epsilon \Big[ \frac{\pa^2 V}{\pa R \pa \epsilon}|_{\epsilon=0} +  \frac{\pa^2 V}{\pa R^2}|_{\epsilon=0} R_1 \Big]  + \frac{\epsilon^2}{2}  \Big[ \frac{\pa^3 V}{\pa R \pa \epsilon^2}|_{\epsilon=0} \\
&& + 2 \frac{\pa^3 V}{\pa R^3}|_{\epsilon=0} R_1^2 + 2 \frac{\pa^2 V}{\pa R^2}|_{\epsilon=0} R_2 \Big]+... 
\er

An interesting solution is the one satisfying $R=R^{+},\,\vp=\vp^{+}$, i.e. the odd components vanish   $R^{-}=0,\,\vp^{-}=0$. We will show that this property is maintained in all orders of perturbation theory. So, substituting the solution $R^{-}_0=0,\,\vp^{-}_0=0$ into the equations   
(\ref{eq111})-(\ref{eq222}) one gets the zeroth order ${\cal O}(\epsilon^0)$ system of equations for $ R_{0}^{+}$  and $\vp_{0}^{+}$
\br 
\label{p01}
\pa_{t} R_{0}^{+} &=& -\pa_{x} \( R_{0}^{+} \pa_{x} \vp_{0}^{+}\)\\
2 (R_{0}^{+} )^2  \pa_{t} \vp_{0}^{+} &=& - (R_{0}^{+})^2 ( \pa_{x} \vp_{0}^{+})^2 +2 R_{0}^{+} \pa^2_{x} R_{0}^{+}- (\pa_{x} R_{0}^{+})^2 - 4 \Big[\frac{\pa V}{\pa R}|_{_{\epsilon =0}}\Big]^{(+)} (R_{0}^{+})^2\label{p02},
\er 
Then, the solution of (\ref{p01})-(\ref{p02}) with vanishing odd components will satisfy
\br
\label{order0}
{\cal P}_x : R_{0} \rightarrow R_0,\,\,\,\,\vp_0 \rightarrow \vp_0    
\er
Now, assuming $\(R_0,\,\vp_0 \)= \(R_{0}^{+}, \vp^{+}_0 \)$ one has that the first order terms ${\cal O}(\epsilon)$ of the even and odd parts of the eqs. of motion (\ref{eq111})-(\ref{eq222}) satisfy    
\br 
\label{p11}
\pa_{t} R_{1}^{+} &=& -\pa_{x} \( R_{1}^{+} \pa_{x} \vp_{0}^{+} + R_{0}^{+} \pa_{x} \vp_{1}^{+}\)\\ 
\pa_{t} R_{1}^{-} &=& -\pa_{x} \( R_{1}^{-} \pa_{x} \vp_{0}^{+} + R_{0}^{+} \pa_{x} \vp_{1}^{-}\)\label{p12}\\
\nonumber
(R_{0}^{+})^2 \pa_{t} \vp_{1}^{+} &=& - 2 R_{0}^{+} R_{1}^{+} \pa_{t} \vp_{0}^{+}   - (R_{0}^{+})^2 \pa_{x} \vp_{0}^{+} \pa_{x} \vp_{1}^{+}-R_{0}^{+} R_{1}^{+} (\pa_{x} \vp_{0}^{+})^2+ R_{0}^{+} \pa_{x}^2 R_{1}^{+}\\ && +  R_{1}^{+} \pa_{x}^2 R_{0}^{+}   - \pa_{x}  R_{0}^{+} \pa_{x} R_{1}^{+}-2 (R_{0}^{+})^2  R_{1}^{+}  \Big[\frac{\pa^2 V}{\pa R^2}|_{_{\epsilon =0}}\Big]^{(+)}\label{p13} \\&&  - 4 R_{0}^{+} R_{1}^{+} \Big[\frac{\pa V}{\pa R}|_{_{\epsilon =0}}\Big]^{(+)} - 2 ( R_{0}^{+})^2 \Big[\frac{\pa^2 V}{\pa \epsilon \pa R}|_{_{\epsilon =0}}\Big]^{(+)} \nonumber\\
\nonumber
(R_{0}^{+})^2 \pa_{t} \vp_{1}^{-} &=& - 2 R_{0}^{+} R_{1}^{-} \pa_{t} \vp_{0}^{+}   -2 R_{0}^{+} R_{1}^{-} (\pa_{x} \vp_{0}^{+})^2- (R_{0}^{+})^2  \pa_{x} \vp_{1}^{-} \pa_{x} \vp_{0}^{+} + R_{1}^{-} \pa^2_{x} R^{+}_{0} - \\ && \pa_{x} R^{-}_{1}\pa_{x} R^{+}_{0}  - 2 (R_{0}^{+})^2 R_{1}^{-} \Big[\frac{\pa^2 V}{\pa R^2}|_{_{\epsilon =0}}\Big]^{(+)} + \pa^2_{x} R^{-}_{1} R^{+}_{0}-4 R^{-}_{1} R^{+}_{0} \Big[\frac{\pa V}{\pa R}|_{_{\epsilon =0}}\Big]^{(+)} \label{p14}
\er
Some comments are in order here. First, the pair of fields $R^{+}_1$ and  $\vp_{1}^{+}$ satisfies the linear system of equations (\ref{p11}) and (\ref{p13}) with variable coefficients, similarly the pair  $R^{-}_1$ and  $\vp_{1}^{-}$  satisfies the linear system of equations (\ref{p12}) and (\ref{p14}). The  couple of first order fields $(R^{+}_1,\, \vp_1^{+})$ and $(R^{-}_1,\, \vp_1^{-})$ in the eqs. above form two decoupled system of equations, respectively. Second, the equations of motion for the even components $\(R_{1}^{+},\,\vp_{1}^{+}\)$, (\ref{p11}) and (\ref{p13}),   satisfy a non-homogeneous coupled system of equations (the non-homogeneous term being the last term in (\ref{p13}), $- 2 ( R_{0}^{+})^2 \Big[\frac{\pa^2 V}{\pa \epsilon \pa R}|_{_{\epsilon =0}}\Big]^{(+)}$) and so they can never vanish.  However, the odd part components $\(R_{1}^{-},\,\vp_{1}^{-}\)$, (\ref{p12}) and (\ref{p14}), satisfy a homogeneous system of equations and so they can vanish. Third, if the pair of components without definite parity $\(R_{1},\,\vp_{1}\)$ is a solution, so is the pair $\(R_{1}-R_{1}^{-} = R_{1}^{+},\, \vp_{1}-\vp_{1}^{-}=\vp_{1}^{+}\)$. Then, one can always choose a first order solution which is even under the space-reflection parity, i.e.
\br
\label{order1}
{\cal P}_x : R_{1} \rightarrow R_1,\,\,\,\,\vp_1 \rightarrow \vp_1    
\er
Choosing the zeroth (\ref{order0}) and first order (\ref{order1}) solutions one has that the second order terms ${\cal O}(\epsilon^2)$ of  (\ref{eq111})-(\ref{eq222}) splits under ${\cal P}_x$ as
\br \label{p21}
\pa_{t} R_{2}^{+} &=& -\pa_{x} \( R_{2}^{+} \pa_{x} \vp_{0}^{+} + R_{0}^{+} \pa_{x} \vp_{2}^{+} + R_{1}^{+} \pa_{x} \vp_{1}^{+}\)\\ 
\pa_{t} R_{2}^{-} &=& -\pa_{x} \( R_{0}^{+} \pa_{x} \vp_{2}^{-} + R_{2}^{-} \pa_{x} \vp_{0}^{+}\) \label{p22}
\\
\nonumber
2 (R_{0}^{+})^2 \pa_{t} \vp_{2}^{+} &=& - 2 (R_{0}^{+})^2 \pa_{x} \vp_{0}^{+} \pa_{x} \vp_{2}^{+} -(R_{0}^{+})^2 (\pa_{x} \vp_{1}^{+})^2 -4   R_{0}^{+} R_{1}^{+} \pa_{x} \vp_{1}^{+}  \pa_{x} \vp_{0}^{+} \\\nonumber
&&-2  R_{0}^{+} R_{2}^{+} (\pa_{x} \vp_{0}^{+})^2 - (R_{1}^{+})^2 (\pa_{x} \vp_{0}^{+})^2 - 4  R_{0}^{+} R_{1}^{+} \pa_{t} \vp_{1}^{+}-4  R_{0}^{+} R_{2}^{+} \pa_{t} \vp_{0}^{+} \\\nonumber
&&- 2 ( R_{1}^{+})^2 \pa_{t} \vp_{0}^{+} + 2 R_{0}^{+} \pa^2_{x} R_{2}^{+} + 2 R_{1}^{+} \pa^2_{x} R_{1}^{+} + 2 R_{2}^{+} \pa^2_{x} R_{0}^{+} - 2  \pa_x R_{0}^{+} \pa_{x} R_{2}^{+} \\ \nonumber
&& -   (\pa_{x} R_{1}^{+})^2 - 4  (R_{0}^{+})^2  R_{2}^{+} \Big[\frac{\pa^2 V}{\pa R^2}|_{_{\epsilon =0}}\Big]^{(+)} - 8 R_{0}^{+} R_{2}^{+} \Big[\frac{\pa V}{\pa R}|_{_{\epsilon =0}}\Big]^{(+)} \\ \label{p23}
&&-2 (R_{0}^{+} R_{1}^{+})^2 \Big[\frac{\pa^3 V}{\pa R^3}|_{_{\epsilon =0}}\Big]^{(+)} - 8 R_{0}^{+} (R_{1}^{+})^2 \Big[\frac{\pa^2 V}{\pa R^2}|_{_{\epsilon =0}}\Big]^{(+)}\\ \nonumber
&&- 4 (R_{1}^{+})^2 \Big[\frac{\pa V}{\pa R}|_{_{\epsilon =0}}\Big]^{(+)} -4 (R_{0}^{+})^2 R_{1}^{+} \Big[\frac{\pa^3 V}{\pa \epsilon \pa R^2}|_{_{\epsilon =0}}\Big]^{(+)}\\ \nonumber
&& - 8 R_{0}^{+} R_{1}^{+} \Big[\frac{\pa^2 V}{\pa \epsilon \pa R}|_{_{\epsilon =0}}\Big]^{(+)}- 2 (R_{0}^{+})^2 \Big[\frac{\pa^3 V}{ \pa \epsilon^2 \pa R}|_{_{\epsilon =0}}\Big]^{(+)} \\ \nonumber
-(R_{0}^{+})^2 \pa_{t} \vp_{2}^{-} &=& R_{0}^{+} R_{2}^{-} (\pa_{x} \vp_{0}^{+})^2 +  ( R_{0}^{+})^2 \pa_{x} \vp_{2}^{-}  \pa_{x} \vp_{0}^{+} + 2   R_{2}^{-} R_{0}^{+} \pa_{t} \vp_{0}^{+}\\
&& - R_{2}^{-}  \pa^2_{x} R_{0}^{+} + \pa_{x} R_{2}^{-} \pa_{x} R_{0}^{+} + 2  R_{2}^{-} (R_{0}^{+})^2 \Big[\frac{\pa^2 V}{\pa R^2}|_{_{\epsilon =0}}\Big]^{(+)} \label{p24}\\
&& -\pa_{x}^2 R_{2}^{-}  R_{0}^{+} + 4 R_{2}^{-} R_{0}^{+} \Big[\frac{\pa V}{\pa R}|_{_{\epsilon =0}}\Big]^{(+)}  \nonumber
\er
For this choice one can show that the second order solution has similar properties, i.e. the components  $\(R_{2}^{+},\,\vp_{2}^{+}\)$   satisfy a non-homogeneous system of equations, (\ref{p21}) and (\ref{p23}), and  the odd components $\(R_{1}^{-},\,\vp_{1}^{-}\)$  satisfy homogeneous equations, (\ref{p22}) and (\ref{p24}). 
So, we can always choose the second order solutions such that $R_{2}^{-}=0,\,\vp_{2}^{-} =0$, i.e.
\br
\label{order2}
{\cal P}_x : R_{2} \rightarrow R_2,\,\,\,\,\vp_2\rightarrow \vp_2    
\er
Then one can choose again the second order solution to be even, and this process repeats in all orders ${\cal O}(\epsilon^n)$. By repeating this procedure, order by order, one can construct a perturbative solution which satisfies (\ref{spari11})-(\ref{xsym}), and so has
charges satisfying (\ref{charge11}). Therefore, the theory (\ref{nlsd}) possesses a subset of solutions such that the charges $Q^{(n)}$ of order $n=0,1,2,3,4,6,8,...,$ are exactly  conserved. In fact, the first four charges $n=0,1,2,3$ will be related below  to the phase jump, normalization, momentum and energy of the deformed NLS model. The set of even order conservation laws (\ref{charge11}) is extended to include the odd cases $n=1,3$, as we will show in the next section.  

Let us summarize our results. The deformed NLS model presents an infinite number of asymptotically  conserved charges as in (\ref{asymp1}) for solitons satisfying the space-time symmetry (\ref{pari0})-(\ref{pari1}). In addition, for solitons satisfying the same space-time symmetry, as well as the space-reflection symmetry (\ref{spari11})-(\ref{xsym}) one can say more. In this case,  the sequence of the even order charges become inded exactly conserved (\ref{charge11}). So, the model supports an infinite number of alternating conserved and asymptotically conserved charges associated to soliton solutions  satisfying the both space-time and space-reflection symmetries.  
   
\section{Renormalized charges and anomalies}  
\label{renor}

The difficulty with dark solitons is the fact that the charges and anomalies (\ref{ano1}) diverge when evaluated on the dark-solitons. However, the divergence comes from the vacuum behavior of the solution at $x \rightarrow \pm \infty$, i.e. the cwb solution given in (\ref{vac1}). So, one can renormalize them by substructing the vacuum contribution (which is infinite). Using  the notation  $\psi = \sqrt{R} e^{i \vp/2}$, we consider the following boundary condition
\br
\label{bc}
R(x =\pm \infty) =  |\psi_0|^2,\,\,\,\,\, \pa_{x} R|_{x \rightarrow \pm \infty} =  0, \,\,\,\,\,\pa_{x} \vp|_{x \rightarrow \pm \infty} = 2 k.   
\er   

The nvbc (\ref{bc}) is suitable for soliton solutions with the continuous wave background such that $R = |\psi_0|$  and   $\vp/2 = k x + w t + x_0$. 

The vacuum solution is given in (\ref{vac1}) and we have 
\br
\frac{d Q_{vac}^{(n)}}{dt} = \b^{(n)}_{vac}.
\er  
So, we define the renormalized charges and anomalies 
\br
Q_{r}^{(n)} =Q^{(n)}-Q_{vac}^{(n)};\,\,\,\,\,\,\,\b^{(n)}_{r} = \b^{(n)}-\b^{(n)}_{vac}
\er
and they satisfy
\br
\frac{d Q_{r}^{(n)}}{dt} = \b^{(n)}_{r}.
\er 

Let us evaluate the first charges and anomalies in (\ref{ano1}).
 
1. The $Q_{r}^{(0)}$ charge

One has 
\br
a_x^{(3,0)} &=& \frac{i}{2} \pa_x \vp,\,\,\,\,\a^{(3,0)} = 1\\
Q_{r}^{(0)} &=& \int  [\frac{1}{2} \pa_x \vp + 2i k]\\
&=& -i \int \(  \frac{\bar{\psi} \pa_x \psi - \psi \pa_x \bar{\psi}}{2 |\psi|^2} - 2 k \),
\er
The anomaly does not get renormalized and it vanishes  
\br
\b^{(0)} = -i \int_{-\infty}^{+\infty} X = 0
\er
The charge $Q_{r}^{(0)}$ is associated to the phase difference (or the phase jump) of the solutions.

2. The $Q_{r}^{(1)}$ charge 

We have
\br
a_{x}^{(3, -1)} = 2i \eta R =2i \eta |\psi|^2,\,\,\,\,\,\,\,\a_{x}^{(3, -1)}=0 
\er
Therefore
\br
Q_{r}^{(1)} = 2  \eta \int_{-\infty}^{+\infty} dx \,|\psi|^2 \(1- \frac{|\psi_0|^2}{|\psi|^2} \) 
\er 

This renormalized charge $Q_{r}^{(-1)}$ defines the normalization of the solution $\psi$ and it is related to the internal symmetry of the model: $\psi \rightarrow e^{i \a} \psi,\,\,\a=\mbox{const}.$ The anomaly $\b^{(1)}$ vanishes identically.

3. The $Q_{r}^{(2)}$ charge
\br
a_{x}^{(3, -2 )} = 2i \eta \pa_{x} \vp R,\,\,\,\,\,\,\,\a^{(3, -2)} = 2 \eta R
\er
So, it can be shown that the anomaly $\b_{2}$ vanishes 
\br
\b^{(2)} = 2 \eta \int_{-\infty}^{ +\infty} dx \, \pa_x \Big[ V - \eta R^2\Big] = 0
\er
and 
\br
Q_{r}^{(2)} &=& 2  \eta \int_{-\infty}^{ +\infty} dx\, \Big[ R \pa_x \vp - 2 k |\psi_0|^2 \Big].
\er
This charge is related to the space translation symmetry of the model and using the expression (\ref{exp1}) it can be written as
\br
Q_{r}^{(2)}  = 2  \eta \int_{-\infty}^{+\infty} dx\, \Big[\bar{\psi} \pa_x \psi- \psi \pa_x \bar{\psi}  - 2 i k |\psi_0|^2 \Big], 
\er

Since the solution $\psi$ incorporates the {\sl cwb} solutions along with the dark soliton itself, it is a composite object. The momentum associated to the dark soliton itself must be renormalized even for $k=0$, and it is provided by \cite{kivshar}

\br
\label{momentum}
P = i \eta  \int_{-\infty}^{+\infty} dx\, \Big[\bar{\psi} \pa_x \psi- \psi \pa_x \bar{\psi}\Big] \Big[1- \frac{|\psi_0|^2}{|\psi|^2} \Big].
\er

4. The $Q_{r}^{(3)}$ charge
\br
\label{a33}
a_x^{(3, -3)} &=& \frac{i}{2} \Big[ 4 \eta^2 R^2 + \eta R (\pa_x \vp)^2 - 2 \eta \pa_{x}^2 R + \eta \frac{(\pa_x R)^2}{R}\Big]\\
 \a^{(3, -3)}&=& 2 \eta R \pa_x\vp
\er
So
\br
\b^{(3)} &=& - 2  \eta \int_{-\infty}^{ +\infty} dx\, \pa_x \Big[ \frac{\d V}{\d R} - 2 \eta R\Big] R \pa_x \vp\\
&=& - 2  \eta \int_{-\infty}^{ +\infty} dx\, \pa_t \( V- \eta R^2\)
\er
where we have used the eq. of motion for $R$ given in the appendix. Therefore (\ref{ano1}) implies 
\br
\frac{d}{dt} \int_{-\infty}^{ +\infty} dx\,\Big[ -i a_{x}^{(3, -3)} + 2 i \eta \( V- \eta R^2\) \Big]=0
\er

We then define the fake charge as 
\br
\widetilde{Q}^{(3)} = \int_{-\infty}^{ +\infty} dx\,\Big[-i a_{x}^{(3, -3)} + 2  \eta \( V- \eta R^2\) \Big]
\er
Discarding the total derivative term $\pa_x (\pa_x R) $ of $a_x^{(3, -3)}$ in (\ref{a33}) and using the relationship $|\pa_x \psi|^2 =\frac{1}{4}\( R (\pa_x \vp)^2 + (\pa_xR)^2/R\)$ one can define the charge as 
\br
Q^{(3)} = \int_{-\infty}^{ +\infty} dx\,\Big[ |\pa_x \psi|^2 + V\Big]
\er
and so the renormalized charge becomes
\br
Q^{(3)}_{r} = \int_{-\infty}^{ +\infty} dx\,\Big[ |\pa_x \psi|^2 + V - |\psi_0|^2 k^2 - V_{vac}\Big],
\er
such that  $\frac{d}{dt} Q^{(-3)}_{r} =0$. However, the renormalized Hamiltonian associated to the dark soliton itself for $k=0$ becomes \cite{kivshar}
\br
\label{hamiltonian}
H = \int_{-\infty}^{ +\infty} dx\,\Big\{ |\pa_x \psi|^2 + \int_{|\psi_0|^2}^{|\psi|^2} dI \Big[ V'[I] - V'[|\psi_0|^2] \Big]\Big\}.
\er

So, the renormalized Hamiltonian of the usual NLS model ($V = \eta |\psi|^4 $) can be written as
\br
\label{hamiltonian1}
H_{NLS} = \int_{-\infty}^{ +\infty} dx\,\Big\{ |\pa_x \psi|^2 + \eta (|\psi|^2-|\psi_0|^2)^2\Big\}.
\er

Therefore, the NLS Hamiltonian can be written as a linear combination $ H_{NLS}= Q^{(3)}_{r} + |\psi_0|^2 \, Q^{(1)}_{r}$. Notice that the expressions (\ref{momentum}) and (\ref{hamiltonian1}) evaluated for a NLS  1-dark soliton solution will provide a standard relationship $\frac{\pa H}{\pa P} = v$ as in classical mechanics, where $v$ is the velocity of the soliton. This indicates that the dark soliton can be considered as an effective particle \cite{kivshar}.

5. The $Q_{r}^{(4)}$ charge: the first asymptotically-conserved charge
\br
a_{x}^{(3, -4)} &=& \frac{i}{4 R} \Big[ 12 \eta^2 R^3 \pa_x \vp - 6 \eta  R \pa_x (\pa_x \vp \pa_x R)+ 3 \eta \pa_x \vp (\pa_x R)^2+ \eta R^2 \( (\pa_x \vp)^3 - 4 \pa_x^3 \vp\) \Big] \\
 \a^{(3, -4)}&=& 6 \eta^2R^2 + \frac{3}{2} \eta (\pa_x \vp)^2 R - 2 \eta \pa^2_x R+ \frac{3}{2} \eta \frac{(\pa_x R)^2}{R}
\er
Therefore we have that the renormalized charge and anomaly are
\br
Q_{r}^{(4)} &=& \frac{\eta}{4}   \int_{-\infty}^{ +\infty} dx\, \Big[ 12 \eta R^2 \pa_x \vp + 3  \pa_x \vp \frac{(\pa_x R)^2}{R}+  R \( (\pa_x \vp)^3 - 4 \pa_x^3 \vp\) - 24 \eta k |\psi_0|^4 - 8|\psi_0|^2 k^3 \Big]\\
\b_{r}^{(4)}&=&- \eta  \int_{-\infty}^{ +\infty} dx\,\pa_x \( \frac{\d V}{\d R} - 2 \eta R\) \Big[ 6 \eta R^2 + \frac{3}{2} R (\pa_x \vp)^2 - 2 \pa^2_x R+ \frac{3}{2} \frac{(\pa_x R)^2}{R} \Big]
\er 
Note that $\b_{r}^{(4)}= \b^{(4)}$ since $\pa_x \Big[\( \frac{\d V}{\d R} - 2 \eta R\) |_{R=|\psi_0|^2}\Big]=0$, i.e. the anomaly $\b^{(4)}$ does not get renormalized. So, consider the renormalized quasi-conservation law
\br
\label{qcon11}
\frac{dQ^{(4)}_{r}}{dt} = \b_{r}^{(4)}
\er
with 
\br
\label{charge}
Q^{(4)}_{r} \equiv \frac{\eta}{4}  \int_{-\infty}^{+\infty} dx \Big[ 12 \eta R^2 \pa_x \vp + 3 \pa_x \vp \frac{(\pa_x R)^2}{R} + R \( (\pa_x \vp )^3-4 \pa_x^3 \vp \) - 24 \eta k |\psi_0|^4 - 8 k^3 |\psi_0|^2 \Big]
\er
and
\br
\b^{(4)}_{r} &\equiv & -\eta \int_{-\infty}^{+\infty} dx \(V''[R] -2\eta \) \Big\{ 2 \eta \pa_x R^3 + \frac{3}{4} \pa_x R^2 (\pa_x \vp)^2 -\pa_x (\pa_x R)^2 + \frac{3}{2} \frac{(\pa_x R)^3}{R}\Big\}
\label{bet0}
\er

This is the first non-trivial non-vanishing anomaly associated to an asymptotically conserved charge for solutions satisfying the  symmetry (\ref{pari1}). Moreover, one can have vanishing anomaly $\b^{(4)}_{r}$ associated to the exactly conserved charge $Q_{r}^{(4)} $ for solutions satisfying the symmetry  (\ref{xsym}). In addition,  we will show below that this anomaly  vanishes for a general solitary wave solution of the deformed NLS.
 
Discarding a total derivative associated to the first term in (\ref{bet0}) (i.e. the term $\(V'' -2\eta \) 2 \eta \pa_x R^3$  must vanish once integrated in the whole line taking into account the b.c. (\ref{bc})) the anomaly  $\b^{(-4)}_r $  can be rewritten in a form amenable to numerical calculations. Then, discarding the  `surface' term, one is left with the expression
\br
\b^{(4)}_r (t)
 & = &- \eta  \int_{-\infty}^{+\infty} dx \(V''[R] -2\eta \) \Big\{   \frac{3}{4} \pa_x R^2 (\pa_x \vp)^2 -\pa_x (\pa_x R)^2 + \frac{3}{2} \frac{(\pa_x R)^3}{R}\Big\}
\label{beta11},\\
&\equiv& \int_{-\infty}^{+\infty} \, dx\,  \g(x,t) \label{func11}
\er
For later purposes we defined the integrand function $\g(x,t)$.  The time integrated anomaly  becomes  
\br 
\label{tint} 
\int_{-\widetilde{t}_0}^{+\widetilde{t}_0}\b^{(4)}_r(t') dt' = \int_{-\widetilde{t}_0}^{+\widetilde{t}_0} \int_{-\infty}^{+\infty} \, dx \, dt'\, \g(x,t).
\er    

One notices that in the limit $\epsilon \rightarrow 0$ the anomaly $\b^{(4)}_r$ vanishes identically since $V''[R] \rightarrow 2 \eta$ in this limit. In addition, this anomaly vanishes when $|\psi| = |\psi_0|$, i.e. when evaluated on the {\sl cwb}. 
We will compute numerically the anomaly $\b^{(4)}_{r}$ for certain dark soliton configurations in the both type of deformed NLS described above.

\section{Space-time symmetries of defocusing NLS and dark solitons}

\label{dark}
Next we discuss the both space-time and space-reflection symmetries in the general one and two dark soliton solutions of the integrable defocusing NLS model. So,  consider the defocusing NLS equation 
\br
\label{nls1}
i \pa_{t} \psi +  \pa_{xx} \psi - 2 \eta |\psi|^{2} \psi=0,\,\,\,\, \eta > 0
\er
Here we consider the Hirota method and construct the one and two dark solitons. Let us introduce the Hirota tau functions as \cite{jpa1}
\br
\label{tau1}
\psi = h(x,t)\, \frac{G(x,t)}{F(x,t)},
\er
where $h(x,t) = |\psi_0| \exp{[i(k x + w t + x_0)]}$ \,and  $|\psi_0| , k, w,x_0$ are real constants. The tau function $G$ is a complex function, whereas $F$ is real. Notice that for $F=G=1$ in (\ref{tau1}) one has the solution   
\br
\label{cw}
\psi = |\psi_0|  \exp{[i(k x + w t + x_0)]},
\er
which is  a continuous wave background (cwb) solution of (\ref{nls1}) provided that $w= -k^2 - 2 \eta |\psi_0| ^2$. 

Substituting the expression (\ref{tau1}) into (\ref{nls1}) one gets 
\br
i[\pa_{t}h\,  F^2 G + h F (\pa_{t}G \,F - G \,\pa_{t}F)]+ \pa^2_{x}h\, F^2 G + 2 \pa_{x}h\, F ( F \pa_{x}G \,  - G \,\pa_{x} F)&+& \\
h [ F^2\, \pa^2_{x}G - 2 F \pa_{x} F \pa_{x}G - F\, G\, \pa^2_{x}F+ 2 G\, (\pa_{x}F)^2]+ 2 \eta |h|^2 h\, \, G\, |G|^2 &=& 0.
\er

The {\bf $1$-dark soliton} solution is given by
\br
G(x, t) &=&  1 + y \exp{[\theta(x, t) +\bar{\theta}(x, t) ]}; \\
F(x, t) &=&  1 +  \exp{[\theta(x, t) +\bar{\theta}(x, t) ]}
\er 
where   $y$ is a complex variable. The functions $\theta, \bar{\theta}$ are defined as
\br
\theta(x, t) = p\, x - i p^2\, t + \theta_0,\,\,\,\,\bar{\theta}(x, t) = \bar{p}\, x + i \bar{p}^2 \,t + \bar{\theta}_0,\,\,
\er 
where the bar stands for complex conjugation. The parameters satisfy the relationships
\br
\label{param1}
w = -k^2 - 2 |\psi_0|^2 \,\eta ;\,\,\,y = \frac{k + i \bar{p}}{k - i p},\,\,|\psi_0|  = \sqrt{\frac{1}{|\eta|}} \, |k- i \, p|
\er
Putting these tau functions into the NLS function $\psi$ in (\ref{tau1}) and making use of the identity $\frac{1+ y \exp{\phi}}{1+ \exp{\phi}} = \frac{1}{2} [1+y + (y-1) \tanh{(\phi/2)}]$ one gets 
\br
\label{1dark}
\psi(x,t) = \frac{|\psi_0| }{2} \exp{[i(k x + w t + x_0)]} \,\, \Big\{1+ y + (y-1) \tanh{[\frac{\theta(x, t)+\bar{\theta}(x, t)}{2}]}\Big\}
\er
This solution possesses {\sl three} arbitrary real parameters, say $|\psi_0| , k$ and the phase associated to $y$. In fact, from the relationships (\ref{param1}) one can get the remaining ones $w, p$ in terms of the set $\{|\psi_0| , k, y\}$. From the third relationship in (\ref{param1}) one gets the condition (assume  $k>0$,\,\,$p = p_R+ i \,p_I$)
\br
\label{const1}
|p_{I} +k| \leq \sqrt{\eta}\,\,|\psi_0| .
\er

The intensity function $|\psi|$ moves at the velocity $v = -2 p_I$, which is the velocity of the dark soliton. The dark soliton approaches constant amplitude $|\psi_0| $ as $|x| \rightarrow \pm \infty$. As $x$ varies from $-\infty$ to $+\infty$ the soliton acquires a $2 \d $ phase, where this quantity is the phase of the parameter $y$
\br
y = e^{2 i \d}.
\er   
We restrict $-\pi/2< \d <\pi/2$. Moreover, at the center of the soliton, i.e. for $\theta(x, t)+\bar{\theta}(x, t)+\rho \equiv 0$, one has that the intensity becomes 
\br
|\psi|_{center} = |\psi_0| \cos \d.
\er
This center intensity is lower than the asymptotic amplitude $|\psi_0| $ and this property characterizes a dark soliton. Notice that this center intensity is controlled by the parameter $\d$; i.e. this parameter defines the ``darkness'' of the soliton. 

From (\ref{const1}) one can see that the velocity $v =-2 p_I$ of the dark soliton satisfy the relationship
\br
\label{limit1}
v \leq v_s - 2 k,\,\,\,\,\,\,\,\, v_{s} \equiv \, \sqrt{4 \eta} \, |\psi_0| ,
 \er  
where $v_s$ is the  sound speed. Therefore the speed of the dark soliton must satisfy 
\br
\label{limit2}
|v|  < v_s
\er

The {\bf $2$-dark soliton} solution is given by
\br
\nonumber
G(x, t) &=&  1 + y_1 \exp{[\theta_1 (x, t) +\bar{\theta}_1 (x, t)]  } +y_2 \exp{[\theta_2 (x, t) +\bar{\theta}_2 (x, t)  ]}+ \\
&& r\, y_1\, y_2 \exp{[\theta_1 (x, t) +\bar{\theta}_1 (x, t)  ]} \exp{[\theta_2 (x, t) +\bar{\theta}_2 (x, t)  ]}; l\label{tau21}\\
\nonumber
F(x, t) &=&  1 +  \exp{[\theta_1 (x, t) +\bar{\theta}_1 (x, t) ]}+ \exp{[\theta_2 (x, t) +\bar{\theta}_2 (x, t)  ]} + \\ \label{tau22}
&& r \exp{[\theta_1 (x, t) +\bar{\theta}_1 (x, t)}  ] \exp{[\theta_2 (x, t) +\bar{\theta}_2 (x, t)]  },
\er 
where   $r$ is a  real constant, and $y_j\,(j=1,2)$ are constant phases. The functions $\theta_j, \bar{\theta}_j$ are defined as
\br
\theta_j (x, t) = p_j\, x - i p_j^2\, t + \theta_0^{(j)},\,\,\,\,\bar{\theta}_j(x, t) = \bar{p}_j\, x + i \bar{p_j}^2 \,t + \bar{\theta}_0^{(j)},\,\,
\er 
where the bar stands for complex conjugation. The parameters satisfy the following relationships
\br
\label{param11}
w &=& -k^2 - 2 |\psi_0|^2 \,\eta ,\,\,\,\,y_j = \frac{k + i \bar{p_j}}{k - i p_j},\,\,\,\,r= \frac{|p_1 - p_2|^2 }{ |\bar{p_1} + p_2|^2}\\ k &=& i \, \frac{|p_2|^2- |p_1|^2}{(p_1 - \bar{p_1}) - (p_2 - \bar{p_2})},\,\,\,\, |\psi_0|
 =  \sqrt{\frac{2}{\eta}} \,\,\, |k- i \, p_j|,\,\,\,j=1,2
\er
 
For later purposes let us record the next relationship between the soliton parameters in the above 2-dark soliton solution
\br
\label{param2}
4 k = v_1 + v_2 + \frac{a_1^2-a_2^2}{v_1-v_2},
\er
where $v_j = - 2 p_{j I},\,\,a_j = 2 p_{j R},\,p_j \equiv p_{jR}+ i p_{jI},\,j=1,2$. The $v_{j}$ are the soliton velocities and the $a_{j}$ are the parameters associated to the width ($\sim \frac{1}{|a_{j}|}$) of each soliton.

\subsection{Space-time parity transformation}
\label{spacetime}

Let us verify the parity properties of the above solutions under the transformation 
\br
\label{parity1}
P: \,\,
\widetilde{x} \rightarrow -\widetilde{x},\,\,\,\,\widetilde{t} \rightarrow -\widetilde{t},\,\,\,\,\mbox{with}\,\,\,\,\widetilde{x}\equiv x- x_{\theta},\,\,\,\,\widetilde{t} \equiv t-t_{\theta}.
\er

{\bf 1-dark soliton}

Let us consider the 1-dark soliton in (\ref{1dark}) rewritten as 
\br
\label{dark111}
\psi(x,t) = |\psi_0|  \exp{[i(k x + w t + x_0)]} \,\,e^{i \d}  \Big\{\cos{\d} + i \sin{\d}  \tanh{ \( a \widetilde{x}+ b \widetilde{t}\, \) } \Big\}
\er
where $a=p+\bar{p},\,\,\,b = i a (\bar{p}-p),\,\,\,\,a \,x_{\theta} + b \, t_{\theta} = -(\theta_0+\bar{\theta}_0  )$.

Writing the function as $\psi = \sqrt{R} e^{i \frac{\vp}{2}}$ such that 
\br
\vp & = & 2 (k x + w t + x_0) + 2 \phi + 2  \d,\,\,\,\phi \equiv \arctan{\Big[ \tan{\d}\, \tanh{ \(a \widetilde{x}+ b \widetilde{t} \)\Big]}},\\
R &=&  |\psi_0|^2 [1- \sin^2(\delta) \, \mbox{sech}^2\(a \widetilde{x}+ b \widetilde{t} \)]
\er
we can show  that 
\br
P: R \rightarrow R,\,\,\,\,\,\vp \rightarrow -\vp + \mbox{const.}
\er

{\bf 2-dark soliton}

Using the relationship (\ref{tau1}) and the tau functions (\ref{tau21})-(\ref{tau22}), the 2-dark soliton can be written as 

\br
\label{22dark}
\psi_2 &=& |\psi_0|  e^{[i(k x + w t + x_0)]}\,\,
\frac{1+ y_1   e^{\Gamma_1} + y_2  e^{\Gamma_2} + r y_1 y_2     e^{\Gamma_1} e^{\Gamma_2}}{1+    e^{\Gamma_1} +   e^{\Gamma_2} + r   e^{\Gamma_1} e^{\Gamma_2}},\,\,\,\,y_{j}=e^{2 i \d_j},\,j=1,2,
\er
where we have defined $\Gamma_j \(x, t \)  \equiv  \theta_j (x, t)+\bar{\theta}_j (x, t),\,\,j=1,2$.

The last expression can be rewritten as 
\br
\label{ND}
\psi_2 &=& |\psi_0|  e^{[i(k x + w t + x_0)]}\, \(\frac{e^{[i(\d_1+\d_2 )]}}{2}\)\,\times \frac{\cal{N}}{\cal{D}} \er 
where we have defined 
\br
\label{numer1}
\cal{N} &\equiv & e^{[i(\d_1+\d_2 ) ]} e^{z_{+}} +e^{[-i(\d_1+\d_2 ) ]} e^{- z_{+}} + e^{-\D/2} \(  e^{[i(\d_1-\d_2 ) ] } e^{z_{-}} + e^{[-i(\d_1-\d_2 ) ] } e^{- z_{-} } \)
\\
\cal{D} &  \equiv &
\cosh{z_{+}} + e^{-\D/2} \,\cosh{z_{-}}\\
 z_{+} &\equiv& \( \G_1 + \G_2   + \D\)/2,\,\,\,\,\,\,\,\,\,z_{-} \equiv  \( \G_1 - \G_2  \)/2,\,\,\,r \equiv e^{\D}
\er
\br
z_{\pm} &\equiv & a_{\pm} \, \widetilde{x} + b_{\pm} \, \widetilde{t},\,\,\,\,\,\,\,\,\,\,\,\widetilde{x}=x-x_{\rho},\,\,\,\,\,\widetilde{t}=t-t_{\rho}\\
a_{\pm}&=&\frac{1}{2} [(p_1+\bar{p}_1)\pm(p_2+\bar{p}_2)],\,\,\,\,\,b_{\pm}=\frac{-i}{2} [(p_1^2-\bar{p}_1^2)\pm(p_2^2-\bar{p}_2^2)]
\label{abp}
\er
and
\br
\nonumber
x_{\rho}&=&\frac{(\L_2-\L_1)(p_1^2-\bar{p}_1^2)+(\L_2+\L_1) (p_2^2-\bar{p}_2^2)}{(p_1 + \bar{p}_1) (p_1 - \bar{p}_1 - p_2 + \bar{p}_2) (p_2 + \bar{p}_2)},\,\,\,\,t_{\rho} = i\, \frac{(p_1 + \bar{p}_1 - p_2 - \bar{p}_2) \L_1 - (p_1 + \bar{p}_1 + p_2 + \bar{p}_2) \L_2}{(p_1 + \bar{p}_1) (p_1 - \bar{p}_1 - p_2 + \bar{p}_2) (p_2 + \bar{p}_2)}\\
\L_1 &\equiv &\frac{1}{2} \( \D +\theta_0^{(1)} + \bar{\theta}_0^{(1)} + \theta_0^{(2)} + \bar{\theta}_0^{(2)}\),\,\,\,\, \L_2 \equiv \frac{1}{2} \(  \theta_0^{(1)} + \bar{\theta}_0^{(1)} - \theta_0^{(2)} - \bar{\theta}_0^{(2)}\)\nonumber
\er
Notice that the parameters $x_{\rho},\,t_{\rho}$ and  $\L_1, \L_2$ are real. 

The parity transformation  (\ref{parity1}) can be written in terms of $z_{\pm} $ as
\br
P: \( z_{+}\,,\, z_{-} \) \rightarrow   \(- z_{+} \,,\, -z_{-}\).
\er 

Then, one has
\br
 P ({\cal D}) = {\cal D},\,\,\,\,\,P ({\cal N}) = {\cal N}^{\star}.
\er
The phase $\phi$  of $\cal{N} $ in (\ref{numer1}) can be written as 
\br
\label{phase1}
\phi = \arctan{\(\frac{\sinh{z_{+}} \sin{\d_{+}} + e^{-\D/2} \sinh{z_{-}} \sin{\d_{-}} }{\cosh{z_{+}} \cos{\d_{+}} + e^{-\D/2} \cosh{z_{-}} \cos{\d_{-}} }\)},\,\,\,\, \,\,\d_{\pm} \equiv  \d_1\pm \d_2, \er
which satisfies 
\br
P: \phi \rightarrow -\phi 
\er  

Therefore it can directly be verified that the squared absolute value $R = |\psi|^2$\, and phase $\vp = 2(k x + w t + x_0) + 2 \phi $ of  $\psi = \sqrt{R} e^{i \frac{\vp}{2}}$ satisfy 
\br
P: R \rightarrow R,\,\,\,\,\,\vp \rightarrow -\vp + \mbox{const.}
\er

It must be emphasized that the above parity symmetry is satisfied for any set of values $\{ \d_1, \d_2\}$ characterizing the `darkness' of each one-dark soliton component. In fact, there is no further restriction on the parameter space of this two-dark soliton solution in order to satisfy the parity property. Therefore, for this type of solutions it is expected that the integrated anomalies to vanish and the associated charges to be asymptotically conserved.  This behaviour is in contradistinction from the one described in the two-bright analytical soliton solution of the focusing NLS model, in which only a subset of the general  solution satisfies that symmetry (i.e. the solutions such that the 2-soliton parameter $c$ satisfies $c=n \frac{\pi}{2},\,n\in \IZ$ in the eq. (4.23) of \cite{jhep1}). In \cite{jhep1} the initial conditions for the  simulations have used $c=0$ in order to show  the vanishing of the integrated anomalies associated to asymptotically conserved charges for the collision of bright solitons in the deformed focusing NLS model. The $c\neq 0$ cases exhibit non-vanishing integrated anomalies for the soliton collisions.   

\begin{figure}
\centering
\label{fig0}
\includegraphics[width=12cm,scale=3, angle=0,height=4.5cm]{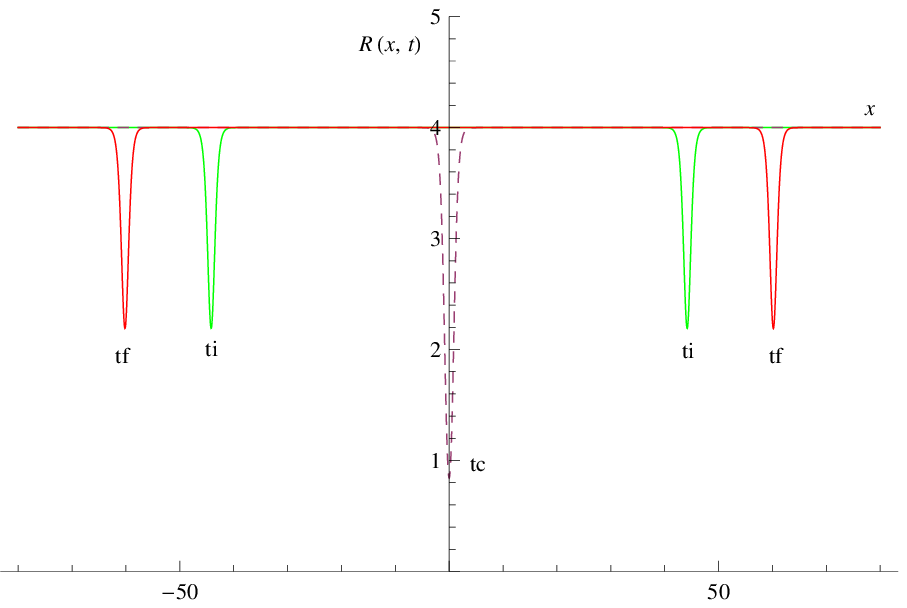}
\includegraphics[width=12cm,scale=3, angle=0,height=4.5cm]{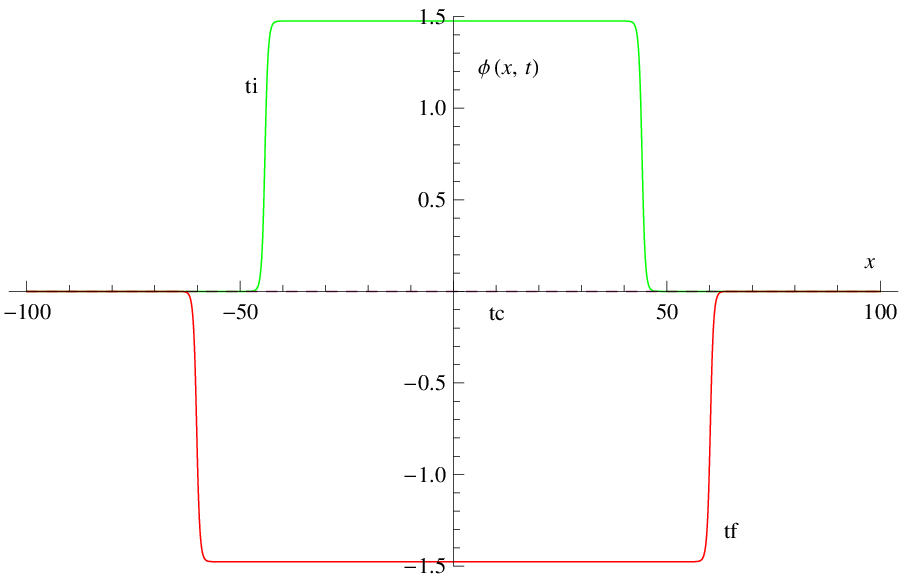}
\parbox{5in}{\caption{(color online) The amplitude R (a) and   phase $\phi$ (b) with  space-reflection symmetries for two-solitons sent at $v_1=-v_2=2$,\,$v_s = 4,\,|\psi_0| = 2, \,\eta =1$,\,$p_{1R}=1.1$,\,$x_{\rho} = 0,\,t_{\rho}=0$, for initial $t_i$ (green), collision  $t_c$ (dashed) and final $t_f$ (red) times, respectively. At $t_c$ the solitons have merged and their phases cancel each other. After collision, $t_f$, each soliton again reveals its phase rise or fall.}}
\end{figure}

The general form of the one and two dark soliton solutions as presented in (\ref{dark111}) and (\ref{22dark}), respectively, have been derived in \cite{jpa1, ohta}. In the first reference a hybrid of the dressing transformation and tau function methods has been used, whereas in the second one they have been derived by the KP-hierarchy reduction method. The ansatz (\ref{tau1}) for the general N-soliton solution in terms of the tau functions $G$ and $F$ follows the construction presented in \cite{jpa1}.  

\subsection{Space-reflection parity transformation}
\label{space1}

It is interesting to see the behaviour of the fields $R$ and $\phi$ for the two-soliton solutions possessing a space-reflection  symmetric  amplitude $R(-\widetilde{x}, \widetilde{t}) = R(\widetilde{x}, \widetilde{t}) $ and phase $\phi(-\widetilde{x}, \widetilde{t})=\phi(\widetilde{x}, \widetilde{t})$ for any given shifted time $\widetilde{t}$. In fact,  when the continuous wave background is at rest, i.e. $k=0$, one has the parameter relationships   $\d_2=\d_1,\, p_{2R}=-p_{1R},\,p_{2I}=-p_{1I}$,\, which satisfy the eq. (\ref{param2}) for two-soliton solutions. These parameters correspond to two dark solitons moving in opposite directions  with equal velocities such that they undergo a head-on collision (see Fig. 1). So, for this choice of parameters, from eqs. (\ref{ND})-(\ref{abp}) and (\ref{phase1}) we have  
\br \nonumber
R(\widetilde{x},\widetilde{t}) &=& \frac{1}{4} |\psi_0|^2 \frac{{\cal N}(x_1,t_1) [{\cal N}(x_1,t_1)]^{\star}}{{\cal D}^2};\,\,\,\,\,x_1\equiv 2 p_{1R}\widetilde{x},\,\,t_1 \equiv 4 p_{1R} p_{1I} \widetilde{t}\\ 
\nonumber
&=& \frac{1}{2} |\psi_0|^2\times\\
&&  \frac{e^{-\D} + \cos{4 \d_1} +e^{-\D} \cosh{2 x_1} +4 e^{-\D/2} \cos{2 \d_1}\, \cosh{t_1}\, \cosh{x_1} + \cosh{2 t_1} }{[e^{-\D/2} \cosh{x_1} + \cosh{t_1} \,]^2 }
\\
\phi(\widetilde{x},\widetilde{t}) &=& \arctan{\Big[\frac{\sin{2 \d_1} \sinh{t_1}}{e^{-\D/2} \cosh{x_1} +\cos{2 \d_1} \cosh{t_1}} \Big]}. \\
\nonumber
x_0 &\equiv & \frac{-\D -2 Re(\theta_{o}^{1}) - 2 Re(\theta_{o}^{2})}{4 p_{1R}},\,\,t_o \equiv \frac{Re(\theta_{o}^{2})-Re(\theta_{o}^{1})}{8 p_{1R} p_{1I}}
\er 
In Fig. 1 we plot the functions $R$ and $\phi$ for three successive times. Note that they are symmetric   under the reflection $\widetilde{x} \rightarrow -\widetilde{x}$ for  any value of the variable $\widetilde{t}=t$.  There is another solution with similar behaviour for the set of parameters $\{k=0,\,\d_2=-\d_1,\, p_{2R} = p_{1R},\,p_{2I}=-p_{1I}\}$.

Therefore, it is clear that the above solution satisfies
\br
\label{parxx}
{\cal P}_x : \,\,R(-\widetilde{x}, \widetilde{t}) \rightarrow  R(\widetilde{x}, \widetilde{t}),\,\,\,\,\, \phi(-\widetilde{x}, \widetilde{t}) \rightarrow \phi(\widetilde{x}, \widetilde{t}). 
\er
This type of solution will be associated to the even order exactly conserved charges as discussed in section \ref{srs}. In our numerical simulations we will verify the properties for this kind of solutions of deformed NLS models, as well as for configurations without this type of symmetry.    

\section{Solitary waves and anomalies} 
\label{solitary1}

Next we discuss the solitary wave solutions of the model (\ref{nlsd}) and their associated symmetries and vanishing anomalies. Below we discuss the cubic-quintic NLS solitary wave as a particular example of solution possessing the space-time parity symmetry $P$. In section \ref{svanish1} we present a general computation for any anomaly $\b_r^{(n)}$ of the deformed NLS (\ref{nlsd}). We will show that in the co-moving frame of the solitary wave the anomalies $\b_r^{(n)}$ vanish. Once  the anomaly is computed in the co-moving frame one performs a Galilean boost in order to compute for travelling waves. So, the computation is performed for a general deformed NLS required that it possesses a solitary wave solution. The cubic-quintic explicit solution is merely used to motivate the discussion since it possesses the symmetry $P$. In section \ref{svanish2} it is computed the particular anomaly $\b_r^{(4)}$, through another method, for solitary waves of deformed NLS (\ref{nlsd}). In this case the computation process considers the relationship between the solitary wave components  $R$ and $\vp$ arising from the equations of motion in appendix \ref{solitary}.      

We will consider translationally invariant solutions of (\ref{nlsd}) of the form 
\br
\label{dark00}
\psi(x,t) = \Phi(z) \mbox{exp}[i\Theta(z)+ i w t],\,\,\,\,z= x- v t,
\er
satisfying the following non-vanishing boundary condition ({\sl nvbc})
\br
\label{bc1}
|\Phi(z \rightarrow \pm \infty)| = |\psi_0|,\,\,\,\,\, \pa_{z} \Phi_{(z \rightarrow \pm \infty)} =  0, \,\,\,\,\,\pa_{z} \Theta_{(z \rightarrow \pm \infty)} = 0.   
\er   

The {\sl nvbc} (\ref{bc1}) is suitable for  wave solutions in the form of localized `dark' pulses created on the continuous wave background ({\sl cwb}) at rest  
\br
\label{cwb}
\psi_0(x,t) = |\psi_0| e^{i \( w t + \d\)},
\er
where $ w = - V'[|\psi_0|^2]$  is the phase shift experienced by the {\sl cwb} with intensity $|\psi_0|^2$, in the non-linear medium with intensity $|\psi|^2$ and varying according to the law $-  \frac{\pa V[|\psi|^2]}{\pa |\psi|^2}$.  Moreover, we will consider a family of equations such that small oscillations around $|\psi|=|\psi_0|$ of (\ref{nlsd}) defines a hyperbolic system with sound speed
\br
\label{sound1}
v_s = |\psi_0| \sqrt{2 V''[|\psi_0|^2]},\,\,\,\,\,V''[|\psi_0|^2]> 0.
\er           
Then the dark solitons of the model (\ref{nlsd}) must propagate with the velocity $v$ such that $|v| < v_s$.  
 
As discussed above, some deformed NLS models support analytical dark solitary waves. In particular, the cubic-quintic non-integrable NLS system (\ref{cqnls2}) possesses solitary waves and they are obtained considering the situations with  $\epsilon >0$ and $\epsilon < 0 $ separately. So, the system (\ref{cqnls2}) possesses dark soliton solutions of the form (\ref{dark00}) provided by \cite{crosta, kamchatnov}
\br
\label{dark22}
\Phi^2_{\pm}(z) &=& \frac{\xi_{1} + r \xi_{2} \tanh^2{[k^{\pm} (z-z_0)]}}{ 1 + r \tanh^2{[k^{\pm} (z-z_0)]}}\\
\label{dark221}
\Theta_{\pm}(z) &=& \mp \arctan{\Big[ \sqrt{r \frac{\xi_{2}}{\xi_{1}}}  \, \,\tanh{[k^{\pm} (z-z_0)]}\Big]}
\er
where
\br
r & \equiv & \frac{|\psi_0|^2-\xi_{1}}{\xi_{2}-|\psi_0|^2},\,\,\,\,\,\,k^{\pm} \equiv \sqrt{\frac{|\epsilon|}{6}} \sqrt{(\pm ) (\xi_{2} - |\psi_0|^2) (|\psi_0|^2-\xi_{1})}\label{s22},\\
\xi_{1} &=& \frac{B - \sqrt{B^2- 6 v^2 \epsilon}}{2 \epsilon},\\
\xi_{2} &=& \frac{B + \sqrt{B^2- 6 v^2 \epsilon}}{2 \epsilon},\,\,\,\,\,\,
B \equiv 6 \eta -2 \epsilon |\psi_0|^2.
\er
The signs $\pm $ in $\Phi_{\pm}$ and $\Theta_{\pm}$ correspond to the case $\epsilon>0$ and $\epsilon<0$, respectively. The parameter $\xi_1$ is always positive and $\xi_2$ becomes positive (negative) for $\epsilon > 0$ ($\epsilon <0$) . For $ \epsilon > 0$ they satisfy $\xi_{1} < |\psi_0|^2 < \xi_{2}$, whereas for $\epsilon < 0$ it holds $\xi_{2} < \xi_{1}< |\psi_0|^2 $. Notice that $k^{\pm}$ characterizes  the inverse soliton width and $\sqrt{\xi_{1}}$ is the minimum intensity (dip) of the dark soliton. The maximum  intensity of the dark soliton approaches $\Phi_2(\pm \infty)=|\psi_0|$. Moreover, for $v=0$ the parameter $\xi_{1}$ vanishes and the dark soliton becomes a black soliton.  
 
The space-time parity transformation (\ref{parity1}) can be performed defining 
\br
z-z_0 = \widetilde{x} - v \widetilde{t},\,\,\,\,z_0 \equiv x_{\theta}- v t_{\theta}.
\er

In fact, under (\ref{parity1}) the solution (\ref{dark22})-(\ref{dark221}) transforms as

\br
\Phi_{\pm}(z) \rightarrow \Phi_{\pm}(z),\,\,\,\,\,\Theta_{\pm}(z) \rightarrow - \Theta_{\pm}(z)
\er 

Therefore, it can be verified that the squared absolute value $R = |\psi|^2=\Phi_{\pm}^2(z)$\, and phase $\vp = 2( w t + \d) + 2 \Theta_{\pm} $\, of\,  $\psi = \sqrt{R} e^{i \frac{\vp}{2}}$ satisfy 
\br
P: R \rightarrow R,\,\,\,\,\,\vp \rightarrow -\vp + \mbox{const.}
\er

\subsection{Vanishing of  anomaly $\b_r^{(n)}$  for solitary waves of deformed NLS}
\label{svanish1}

Note that the factor $X$ of the anomaly expression in (\ref{ano1}) when evaluated on the cwb solution (\ref{cwb}) satisfies $X=\pa_x \Big[\( \frac{\d V}{\d R} - 2 \eta R\)|_{R=|\psi_0|^2}\Big]=0$ then one has  that $\b_{r}^{(n)}= \b^{(n)}$, i.e. the anomaly $\b^{(n)}$ does not get renormalized.  
In order to compute the general renormalized anomaly $\b_r^{(n)}$ associated to a solitary wave of the deformed NLS model it is convenient to consider the co-moving coordinate of the soliton. In this frame, the soliton is
stationary. In this frame consider the cwb solution
of the deformed NLS model as a stationary oscillatory phase as in (\ref{cwb}). In addition, a stationary soliton can be transformed into a
non-stationary form through the Galilean symmetry. In fact, assume that $\psi(x,t)$ is a solution of the deformed NLS equation (\ref{nlsd}) and $v$ a constant, then 
\br
\label{news}
\psi(x, t ) \rightarrow  \psi'(x, t ) \equiv  e^{i[v x - \frac{v^2}{2} t ]} \psi(x - v t, t ) \er
 is also a solution. So, in the co-moving coordinate of the solitary wave one can write the solitary wave solutions  $\Phi^2({\widetilde{x}})$ and $\Theta(\widetilde{x})$ as dependents only on the variable $\widetilde{x}$. As an example, we can rewrite the solitary wave in (\ref{dark22})-(\ref{dark221}) as
\br 
\label{dark223}
\Phi^2_{\pm}(\widetilde{x}) &=& \frac{\xi_{1} + r \xi_{2} \tanh^2{[k^{\pm} \, \widetilde{x}]}}{ 1 + r \tanh^2{[k^{\pm} \, \widetilde{x}]}}\\
\label{dark2231}
\Theta_{\pm}(\widetilde{x}) &=& \mp \arctan{\Big[ \sqrt{r \frac{\xi_{2}}{\xi_{1}}}  \, \,\tanh{[k^{\pm} \, \widetilde{x}]}\Big]},\,\,\,\,\widetilde{x}= x- x_{\rho}.
\er

Then, the space-time transformation (\ref{pari0}) applied to this form of solution furnishes the next  symmetry transformation for 
$R(\widetilde{x}) = |\psi|^2=\Phi_{\pm}^2(\widetilde{x})$\, and phase $\vp(\widetilde{x},\widetilde{t}) = 2( w \, \widetilde{t} + t_0) + 2 \Theta_{\pm}(\widetilde{x}) $\, of\,  $\psi = \sqrt{R} e^{i \frac{\vp}{2}}$ 
\br
\label{t11}
P: \,\,\,\,\,\,\,\,\,\,R \,\,\,\, & \rightarrow & \,\,\,\,\,R ;\\
\vp \,\,\,\,&\rightarrow &\,\,\,\,\, -\vp  + \mbox{const.} \label{t22}
\er  
So, for solutions satisfying this type of space-time symmetry one has that the factor $\a^{(3, -n)}$ in the anomaly $\beta_{n}$ in (\ref{ano1}) is even according to (\ref{factor1}). The first expressions of $\a^{(3, -n)}$ provided in the appendix \ref{apen2} exhibit this property. Therefore, the associated charges $Q^{(n)}$ satisfy a mirror type symmetry (\ref{mirror}) and an asymptotically conservation law (\ref{asymp1}).

Moreover, in this case we can also explore the space-reflection symmetry of the anomalies which can be useful in order to estimate their values, by using general symmetry considerations. So, note that the time dependence of the solution $\psi(x,t)$ in (\ref{dark00}) for functions of the form  $\Phi^2(\widetilde{x})$ and $\Theta(\widetilde{x})$ appears only in the cwb phase (\ref{cwb}). In addition, the components $\a^{(3, -n)}$ are polynomials in $R(\widetilde{x})$  and $\pa_x \vp(\widetilde{x})$, and their $x-$derivatives (see the first five expressions in the appendix \ref{apen2}).  So, the space-time parity symmetry can be used in order to see the properties of these expressions. In fact, the parities of  $R$ and $\pa_{x} \vp$ of this special soliton solution can be obtained by  taking  $\widetilde{t}=0$ in this frame. Therefore, using the result (\ref{factor1}) one has that the $\a^{(3, -n)}$ are even, whereas $X$ is odd under the above space-reflection, then one has that the anomalies $\b^{(n)}$ vanish for this type of solitary wave. Therefore, in the co-moving frame one has
\br
\frac{d Q^{(n)}}{dt} = 0,\,\,\,\,\,\,\,\,n=0,1,2,3,...
\er

Let us discuss this result for travelling waves. For travelling waves the anomaly associated to a solitary wave can be written as
\br
\label{betatrav}
\b_n = -i \int_{-\infty}^{\infty} dx \,X[R(z)] \, \a^{(3, -n)}[R(z), \pa_x R(z),...,\pa_x \vp(z),\pa^2_x \vp(z),...];\,\,\,\,\,\,z=x-v t. 
\er
In fact, the integrand of these expressions are some polynomials containing the functions  $R(z)$ and  $\pa_{x}\vp(z)$, and their $x-$derivatives. Next, we can rewrite the integration  (\ref{betatrav}) by making a coordinate transformation $x\rightarrow z,\,\,t\rightarrow t$. Furthermore, one can relabel $z \rightarrow x$, and the outcome would be an equivalent integral expression to the one in the co-moving frame we have already analized above. Therefore, also for the travelling waves the anomalies vanish and  the associated charges will be conserved.  

\subsection{Vanishing of  anomaly $\b_r^{(4)}$  for solitary waves of deformed NLS}
\label{svanish2}

Since the first deformation, the cubic-quintic NLS model (\ref{cqnls2}), possesses analytic solitary wave solutions one can compute the anomalies for this type of solutions by substituting directly the solution (\ref{dark22})-(\ref{dark221}) into the expression of the anomaly $\b_r^{(4)}$ in (\ref{beta11}). However, we perform this computation using the relationships regarding this type of solutions presented in appendix \ref{solitary} for general deformations of the NLS model of type (\ref{nlsd}). One proceeds by discarding some `surface terms' and then  rewriting the improper anomaly integral $\lim_{L \to \infty} [\int_{-L}^{L} (...) dx ]$ as a definite integral in the field space (with $R$ as an integration variable) of type $\pm \int_{R_{min}}^{|\psi_0|^2} (...)dR$, where the limits of integration $R_{min}$ and $|\psi_0|^2$ characterize the minimum intensity dip of the dark soliton and the cwb amplitude, respectively. Let us make a coordinate transformation $x\rightarrow z,\,\,t\rightarrow t$ and rewrite (\ref{beta11}) as 
\br
\b^{(4)}_r
  =   -\eta \int_{-\infty}^{+\infty} dz \(V''[R] -2\eta \) \Big\{   \frac{3}{2} R(z) R'(z) [\vp'(z)]^2 - 2\(R'(z)R''(z) - \frac{[R'(z)]^3}{2 R}\) +     \frac{1}{2} \frac{[R'(z)]^3}{R} \Big\}.
\label{beta22}
\er 
Due to (\ref{darkap}) one can write $\vp'(z)= 2 \Theta'(z)$ and considering (\ref{theta11}) one notices that  $\Theta'(z)$ depends only on $R$. Moreover, using (\ref{sec1}) one has that the term inside parenthesis can be written as $\(R'(z)R''(z) - \frac{[R'(z)]^3}{2 R}\)\equiv H[R] R'(z)$. Therefore, all the terms in the integrand above, except the last one, can be rewritten in the form $K[R] R'$, which can be written as a total derivative in $z$, then this 'surface term' does not contribute to the anomaly for the boundary condition (\ref{bc}) with $k=0$. Regarding the last term, we can rewrite the integration in field space as follows
\br\nonumber
\b^{(4)}_r
  &=&  - \eta \int_{-\infty}^{+\infty} dz \(V''[R] -2\eta \) \Big\{\frac{1}{2} \frac{[R'(z)]^3}{R} \Big\}
\label{beta23}\\
\nonumber
  &=& -\frac{\eta}{2}  \int_{-\infty}^{+\infty} \( R' dz\) \frac{1}{R}\( V''[R] -2\eta \) [R'(z)]^2 \\
\nonumber
  &=& -2 \eta  \int^{R_{min}}_{|\psi_0|^2} dR \Big\{ \( V''[R] -2\eta \) \(J[R]\)^2\Big\} - 2 \eta  \int_{R_{min}}^{|\psi_0|^2} dR \Big\{ \( V''[R] -2\eta \) \(J[R]\)^2\Big\}\\
  &=& 0,\label{vanish1}
 \er  
in the third line we have used the relationships (\ref{j11})-(\ref{j12}) and converted the integration space variable $z$ to the field space $R$ variable integration. In this way the general solitary wave configurations of the deformed NLS model (\ref{nlsd}) provide vanishing $\b^{(4)}_r$ anomaly.

\section{Simulations}

\label{simul}

Firstly, we have solved numerically the ordinary NLS model (\ref{nls1}) without deformation ($\epsilon =0$). We develop this procedure for a general initial condition with a cwb, for the one-dark and two-dark soliton solutions, respectively. Then, it is convenient to rewrite the dark soliton (\ref{1dark}) as
\br
\label{dark0}
\psi(x,t) = \frac{1}{\sqrt{4 |\eta|}} e^{i\(k x + w t + \rho_0\)} \Big\{ i v + \a \tanh{\Big[\a(x-v t - x_0)\Big]}\Big\},
\er 
where $\a \equiv 2 p_{R},\,v \equiv - 2 p_I$. Notice that the dark soliton (\ref{dark0}) is characterized by the width $1/\a$ and the velocity $v$. The background density becomes $|\psi_0| = (\a^2 + v^2)^{1/2}/\sqrt{4 |\eta|}$. 

In order to study the interaction of 2-dark solitons we will choose the initial data ($t=0$) as (consider $\eta=1/4 $)
\br
\label{dark02}
\psi(x) &=& e^{i k x} \left\{ \begin{array}{ll}
 i (v_1 - k)+ \a \tanh{[\a( x+ x_0) ]}, &   -L < x < 0\\
-\Big[ i (v_2 - k) + \a \tanh{[\a(x - x_0)]} \Big], &   0 < x < L
\end{array} \right. 
\er  
The two solitons are initially centered at $\pm x_0$ ($x_0>0$), the soliton centered initially at $-x_0$ ($t=0$) moves to the right with velocity $v_1$ ($v_1>0$), whereas $-v_2$ ($v_2>0$) is the velocity of the soliton initially ($t=0$) centered  at $x_0$ and travels to the left,  $\a>0$ is a constant characterizing the width of the solitons. In the numerical simulation we will consider initially well-separated solitons, i.e. the parameter $2x_0$ is chosen to be  several times the width of the solitons. The equation (\ref{nls1}) is solved numerically by considering  the initial condition $\psi(x)$ defined in (\ref{dark02}) for $v_1= - v_2=v$. Notice that  the initial condition (\ref{dark02}) satisfy the boundary condition (\ref{bc}) at $|x| \rightarrow +\infty$ for  $|\psi_0| = \sqrt{\a^2 + (v-k)^2}$. The domain considered is ${\cal D} =[-L,L] $ with $L=25$  and sometimes $L=15$, mesh size $h = 0.02$ and time step $\tau = 0.00011$ (sometimes $\tau = 0.000085$). The domain ${\cal D} $ is chosen such that the effect of the extreme regions near the points $x = \pm L$ do not interfere the dynamics of the solitons, i.e. the boundary condition (\ref{bc}) is satisfied for each time step. In our numerical simulations we will  use the so-called time-splitting cosine pseudo-spectral finite difference (TSCP) and the time-splitting finite difference with transformation (TSFD-T) methods  \cite{bao} for $k=0$  and $k\neq 0$, respectively.

Our numerical simulations reproduce the main properties already known for dark soliton interactions in the integrable defocusing NLS model. Next, we summarize the main properties of the interaction of two dark solitons:

a). The interaction is repulsive, and the initial solitons recover completely their forms and velocities after the collision.

b). For a given asymptotic value $|\psi_{0}|$ there exists  a critical velocity $v_c=v_s/2,$ where $v_s$ is the sound speed (\ref{limit1}) which defines two types of solitons characterized by their behaviour under the collision: when $0<v<v_c$ one has the {\sl low speed} solitons which are reflected by each other, and when  $v_c<v<v_s$  they are the   {\sl high speed} solitons which are transmitted through each other \cite{frant}.   

c). After collision the two solitons recover completely their properties, i.e.  the NLS model is integrable, then all anomalies vanish (all charges are conserved).  

The numerical method TSFD is specialized in order to compute the soliton type solutions of the deformed  NLS model (\ref{nlsd}). The first deformed model (\ref{cqnls2}) possesses a solitary wave presented in (\ref{dark22})-(\ref{dark221}). Then,  we can take two one dark solitons some distance apart as the initial condition for our numerical simulations. For the second deformed model (\ref{enns1}) $\epsilon \neq 0,\, q\in \IZ_{+}$  we do not know any analytical expression for its solitary wave; however one can generate numerically a solitary wave, as we will explain below. We will study the dark soliton collisions numerically for the both deformed NLS models and compare the outcomes with the properties summarized above for the integrable NLS model. 

\subsection{First deformation: $V = \eta |\psi|^4 -\frac{\epsilon}{6} |\psi|^6$}
\label{first}
The collision of dark solitons in the cubic-quintic NLS equation (\ref{cqnls2}) can been simulated  numerically by considering the initial condition $\psi(x)$ defined as
\br
\label{cqsol2}
\psi(x) &=&\left\{ \begin{array}{ll}
  \Phi(x+x_0)  e^{i \Theta (x+ x_0)} , &   -L < x < 0\\
 \Phi(x-x_0)  e^{-i \Theta(x- x_0)}, &   0 < x < L
\end{array} \right. \\
\Phi(x)  &\equiv & \sqrt{\frac{\xi_{1} + r \xi_{2} \tanh^2{\(k^{\pm} \, x  \)}}{ 1 + r \tanh^2{\(k^{\pm} \, x\)}}}\\
 \Theta(x) &=& \arctan{\Big[ \sqrt{r \frac{\xi_{2}}{\xi_{1}}}  \, \,\tanh{\(k^{\pm} \,x\)}\Big]},
\er  
where two solitary wave solutions (\ref{dark22})-(\ref{dark221}) of the cubic-quintic NLS  model have been located some distance apart.  The function $\psi$ is continuous at $x=0$, i.e. it has continuous absolute value $|\psi|= \Phi(x_0)$ and phase equal to $\Theta(x_0)$. The parameters $\{ \xi_1, \xi_2, r, k^{\pm}\}$ are defined in (\ref{dark22})-(\ref{dark221}), so the $\pm$ signs correspond to the relevant sign of the deformation parameter $\epsilon$. The two solitons are initially centered at $\pm x_0$ ($x_0>0$), the soliton centered initially at $-x_0$ ($t=0$) moves to the right with velocity $v>0$, whereas the soliton initially ($t=0$) centered  at $x_0$ travels to the left with velocity ($-v$). Notice that the direction of motion of the soliton is related to the sign of the phase slope in (\ref{cqsol2}). In addition, we will consider initially well-separated solitons, i.e. the parameter $2x_0$ is chosen to be  several times the width of the solitons ($\sim \frac{1}{k^{\pm}}$)  and $2x_0 <  2 L$. So, the initial condition uses two of its static analytic solitary waves which are stitched together at the middle point, and then we allow  the scattering of them, absorbing the radiation at the
edges of the grid. It amounts to maintain the nvbc at the edges of the grid for each time step of the numerical simulation.   For solitons colliding with different velocities (different amplitudes) we must modify conveniently this initial condition.

In the Figs. 2-10 the results for the simulations of soliton collisions in the cubic-quintic NLS model (\ref{cqsol2}) are presented. The relevant anomaly $\beta^{(4)}_r(t)$  (\ref{beta11}) and the integrated anomaly  $\int^t dt' \beta^{(4)}_r(t')$ (\ref{tint}) as functions of time, are plotted for several parameters. The two-soliton collisions with equal velocities (equal amplitudes) and the relevant  anomalies have been considered in Figs. 2-6 with parameters  $\epsilon = \pm 0.01;  0.06$ for fast and slow solitons. In the Figs. 7, 8, 9 and  10 we have simulated the cubic-quintic NLS model with $\epsilon = \pm 0.05,\, -0.06,\,0.04$ for the collision process of two-solitons with different velocities (different amplitudes). In the Figs. 2, 3, 7 and 8 (top right figures) we also plot the anomaly integrand $\g(x,t)$ (\ref{func11}) as function of the space variable $x$ for three successive times and different parameters in order to see qualitatively the behaviour of this function that would imply the vanishing of the anomalies.

\begin{figure}
\centering
\label{fig1}
\includegraphics[width=10cm,scale=4, angle=0, height=6cm]{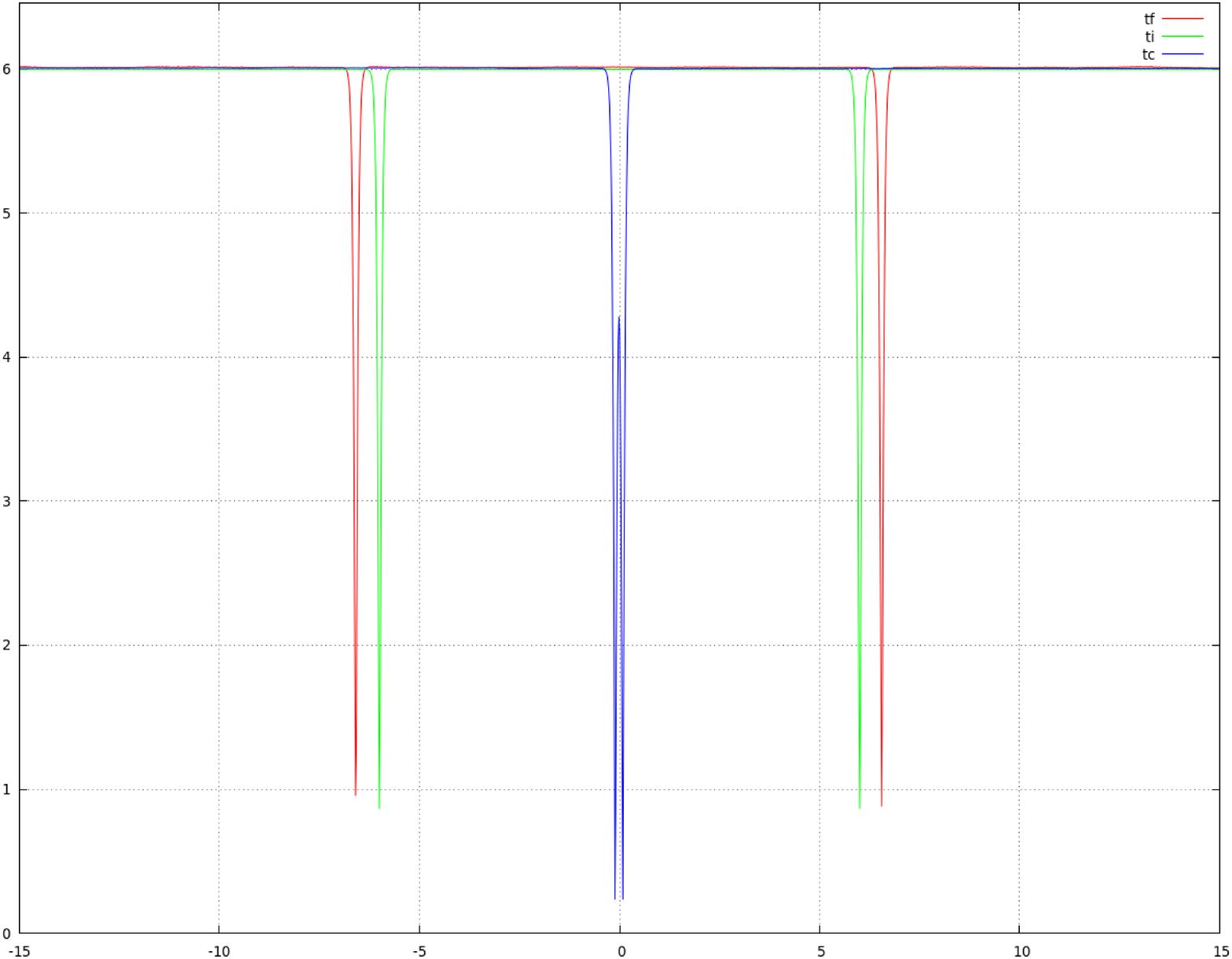} 
\includegraphics[width=10cm,scale=4, angle=0, height=6cm]{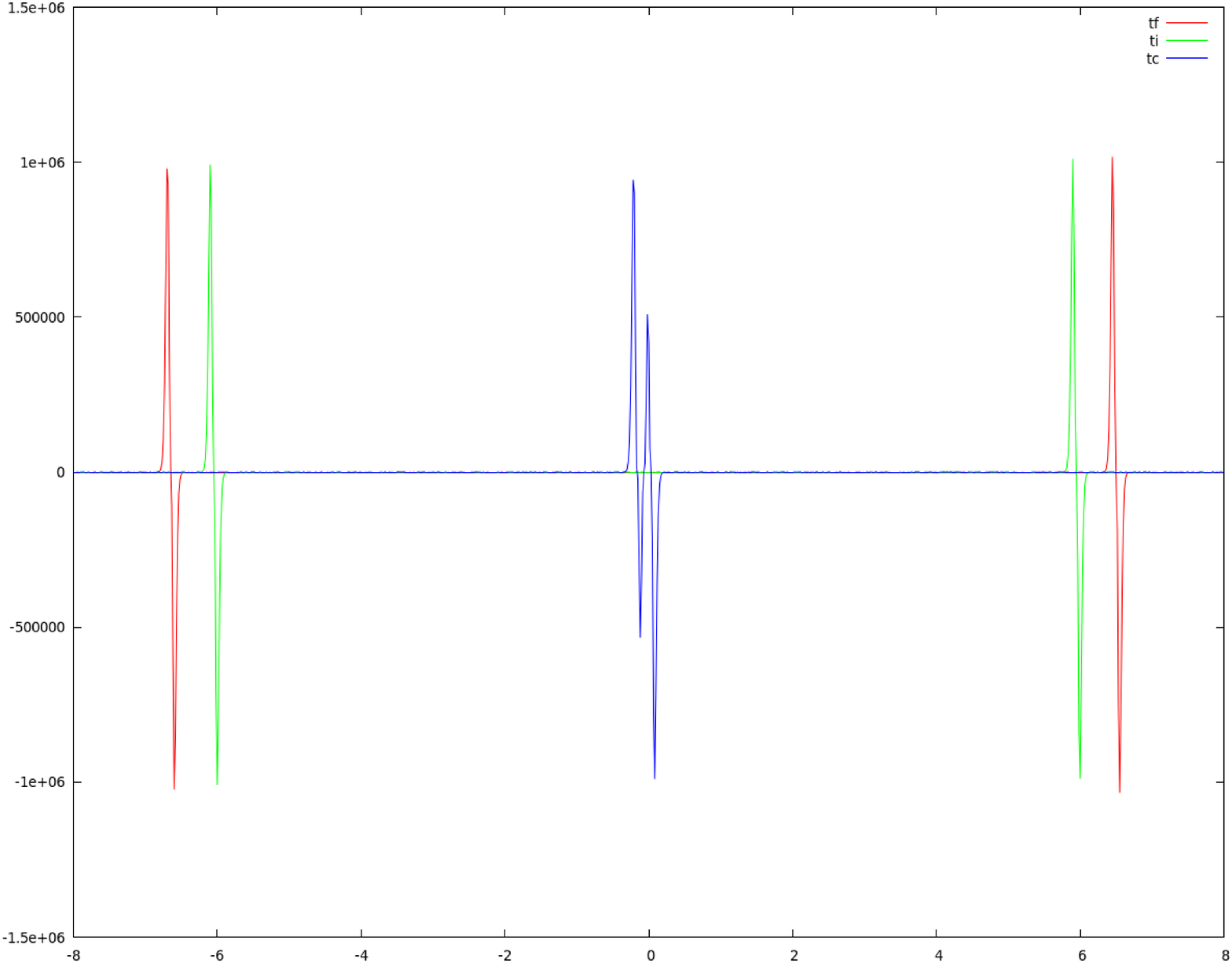}
\includegraphics[width=10cm,scale=4, angle=0, height=6cm]{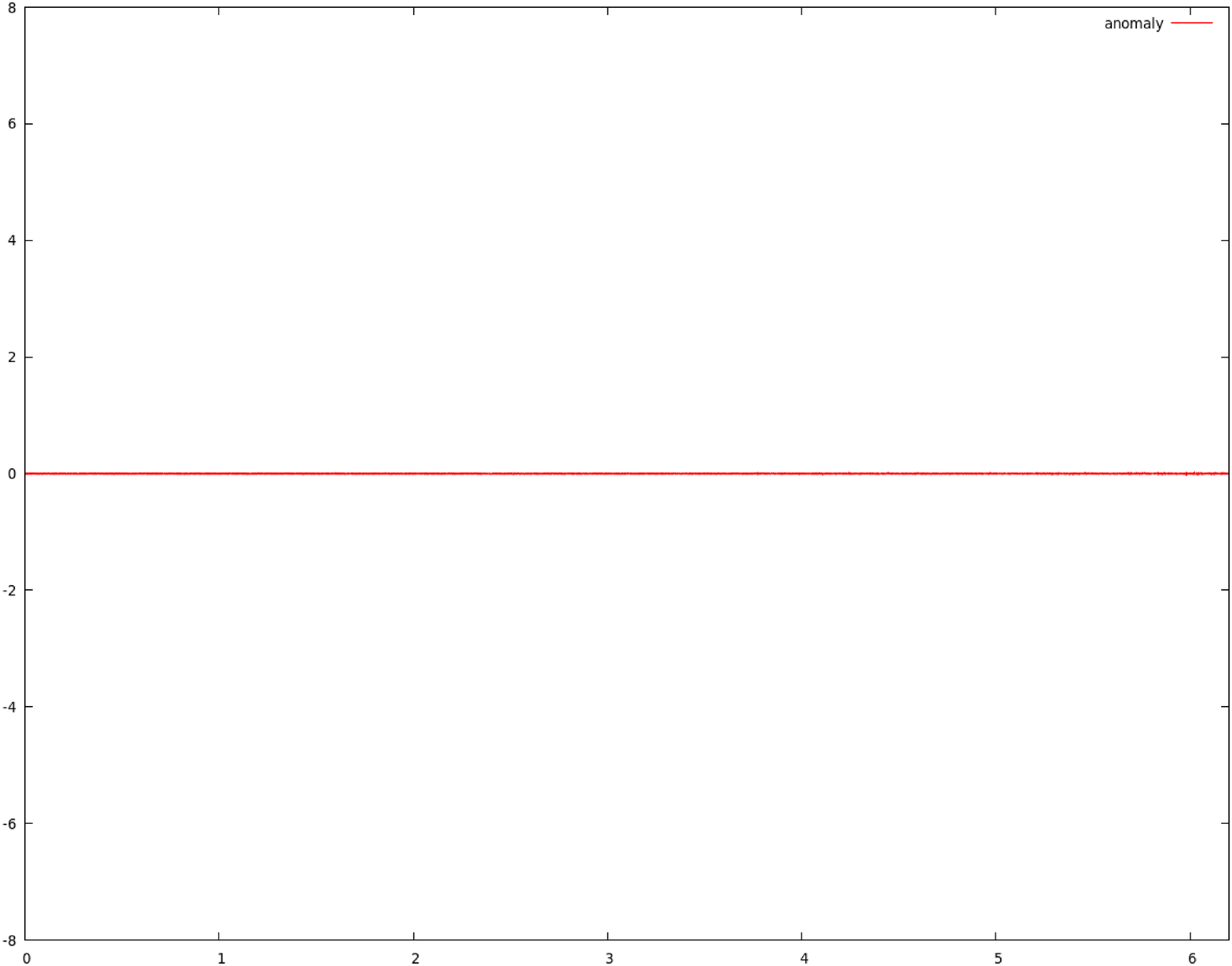}
\includegraphics[width=10cm,scale=4, angle=0, height=6cm]{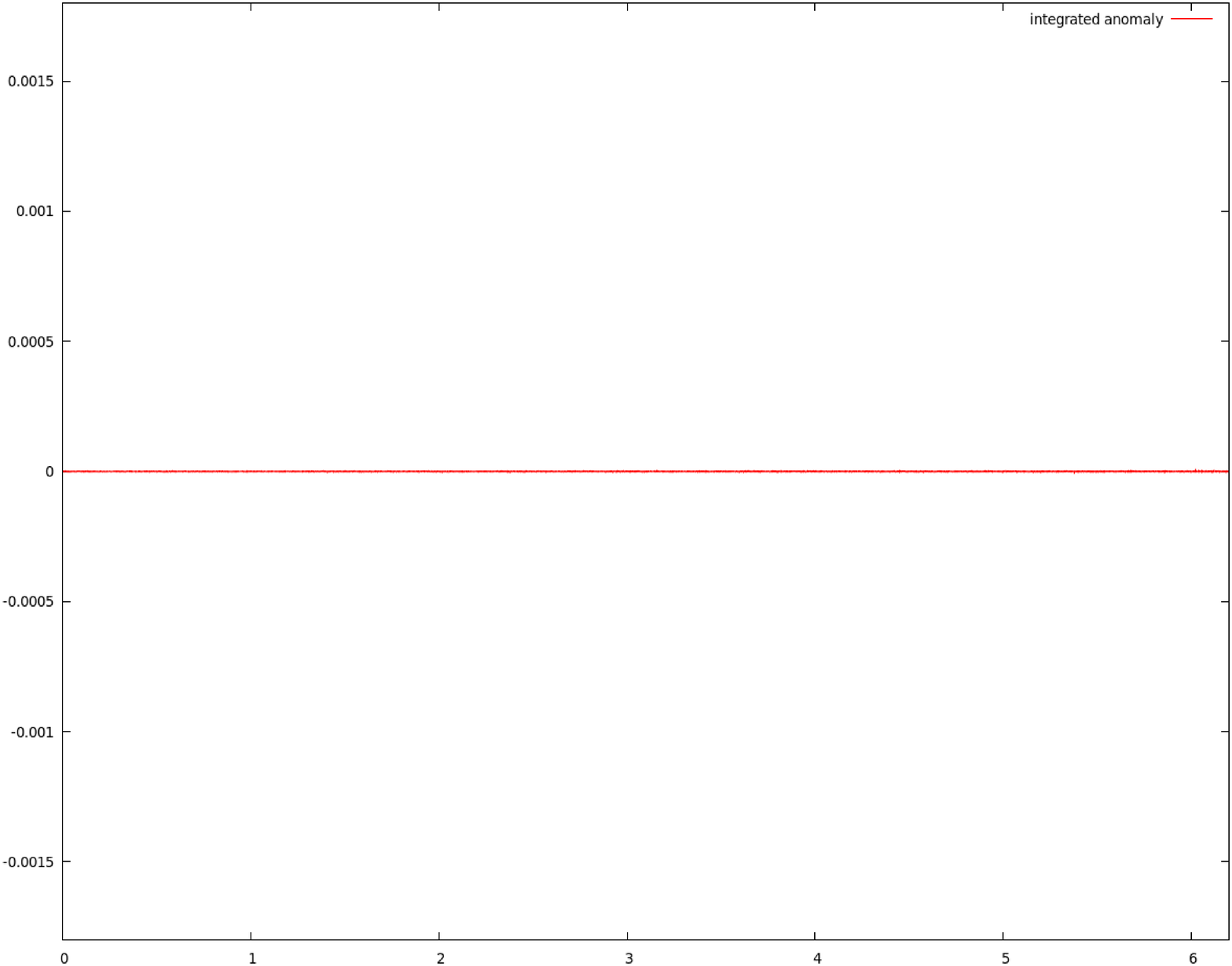}   
\parbox{5in}{\caption {(color online) Top left: The reflection of two dark solitons  of the cubic-quintic NLS model (\ref{cqsol2}) is plotted for $\epsilon=-0.01,\,|\psi_0|=6,\,\eta = 2.5$. The initial gray solitons ($t_i$ =green line) travel in opposite direction with velocity $v \approx 1.97 \sqrt{2}$. They  partially overlap ($t_c$= blue line) in their closest approximation and then reflect to each other. The gray solitons after collision are plotted as a red line ($t_f$). Note that  $v < v_s/2$ ($v_s \approx 13.89 \sqrt{2}$). Top right: the integrand $\g(x,t)$ of the anomaly plotted for three successive times ($t_i$, $t_c$ and $t_f$). Bottom: the anomaly $\beta^{(4)}_r(t)$ and time integrated anomaly $\int^t dt' \beta^{(4)}_r(t')$, respectively.}}
\end{figure}

\begin{figure}
\centering
\label{fig2}
\includegraphics[width=9cm,scale=5, angle=0, height=5cm]{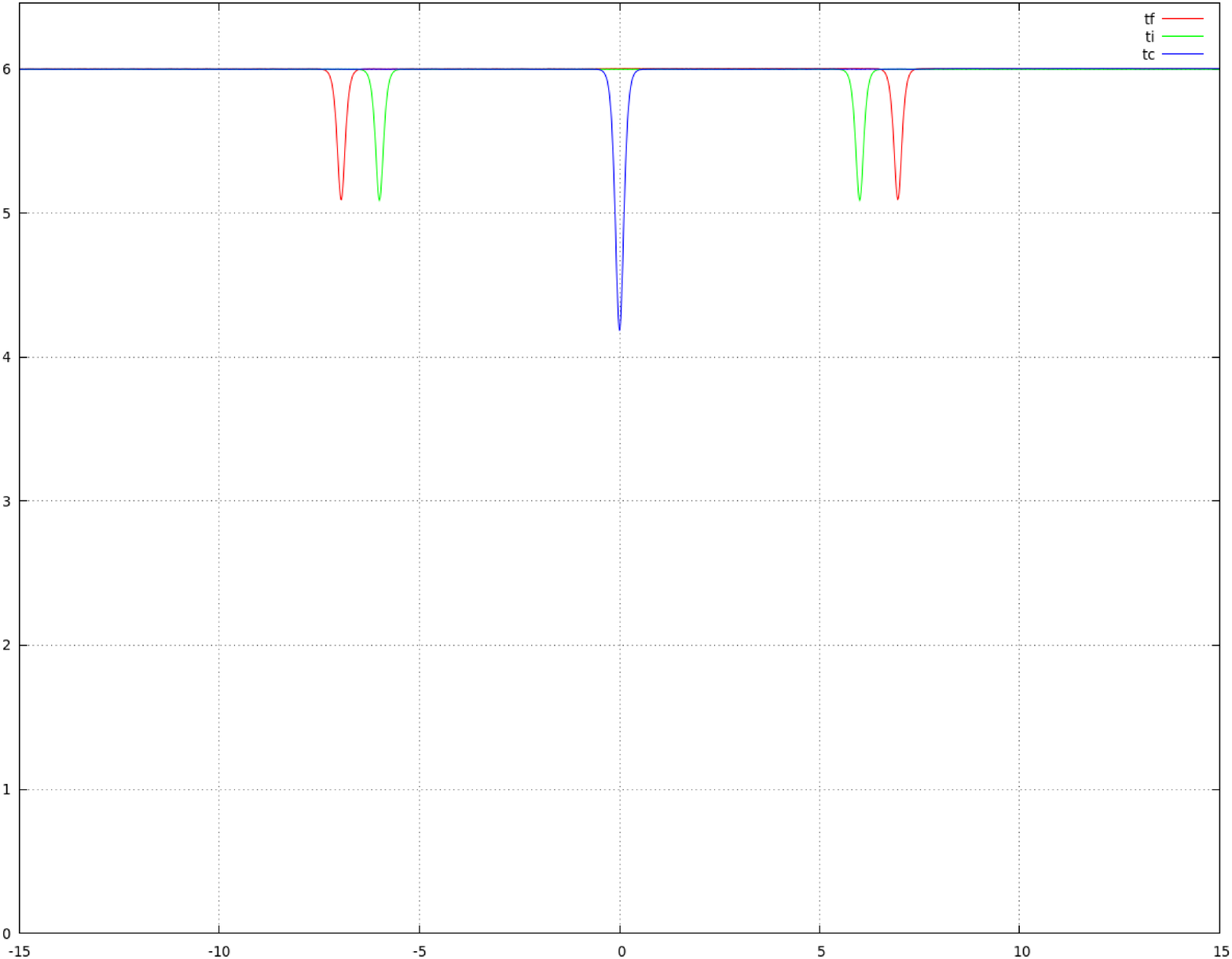}
\includegraphics[width=9cm,scale=5, angle=0, height=5cm]{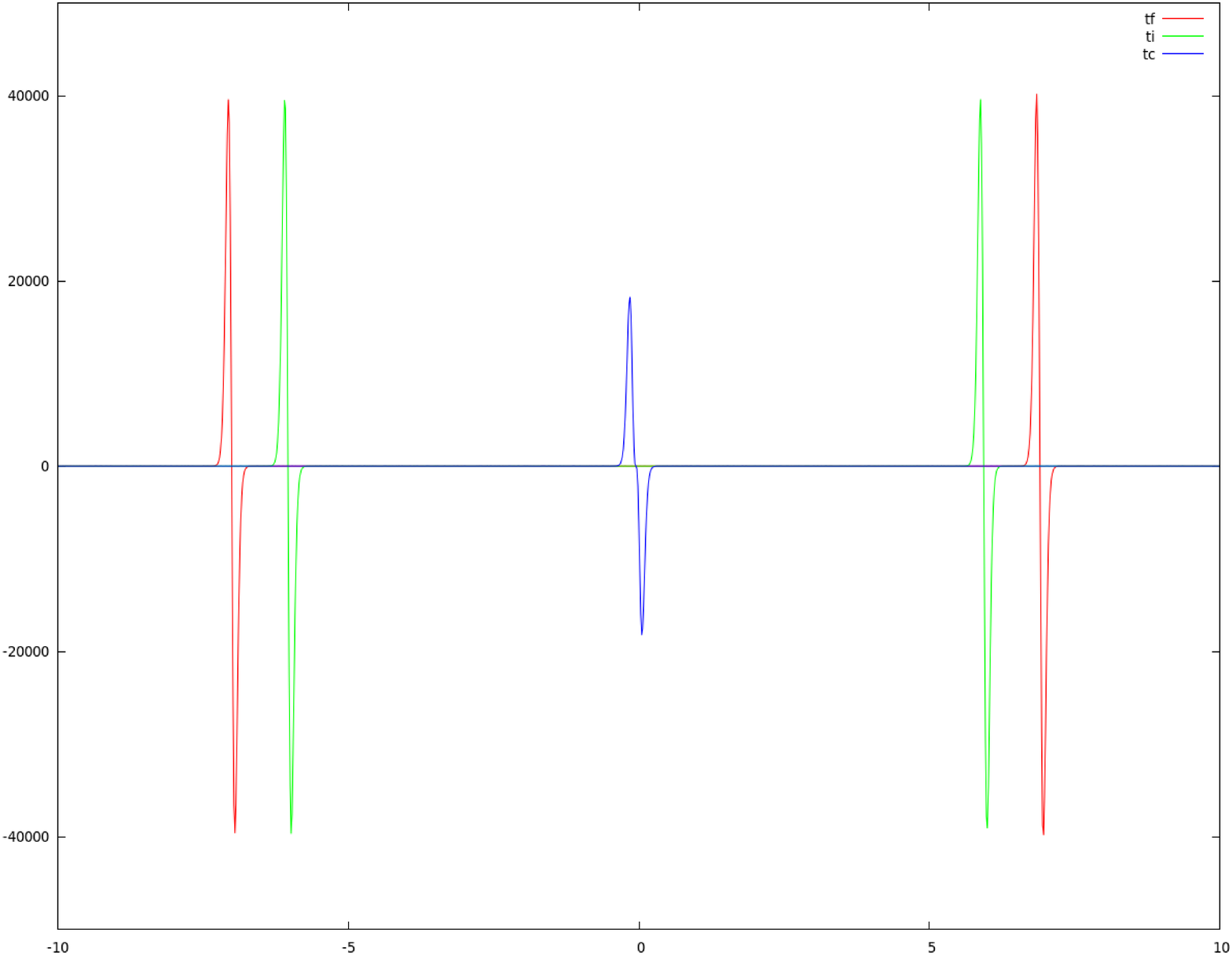}
\includegraphics[width=8cm,scale=5, angle=0, height=4cm]{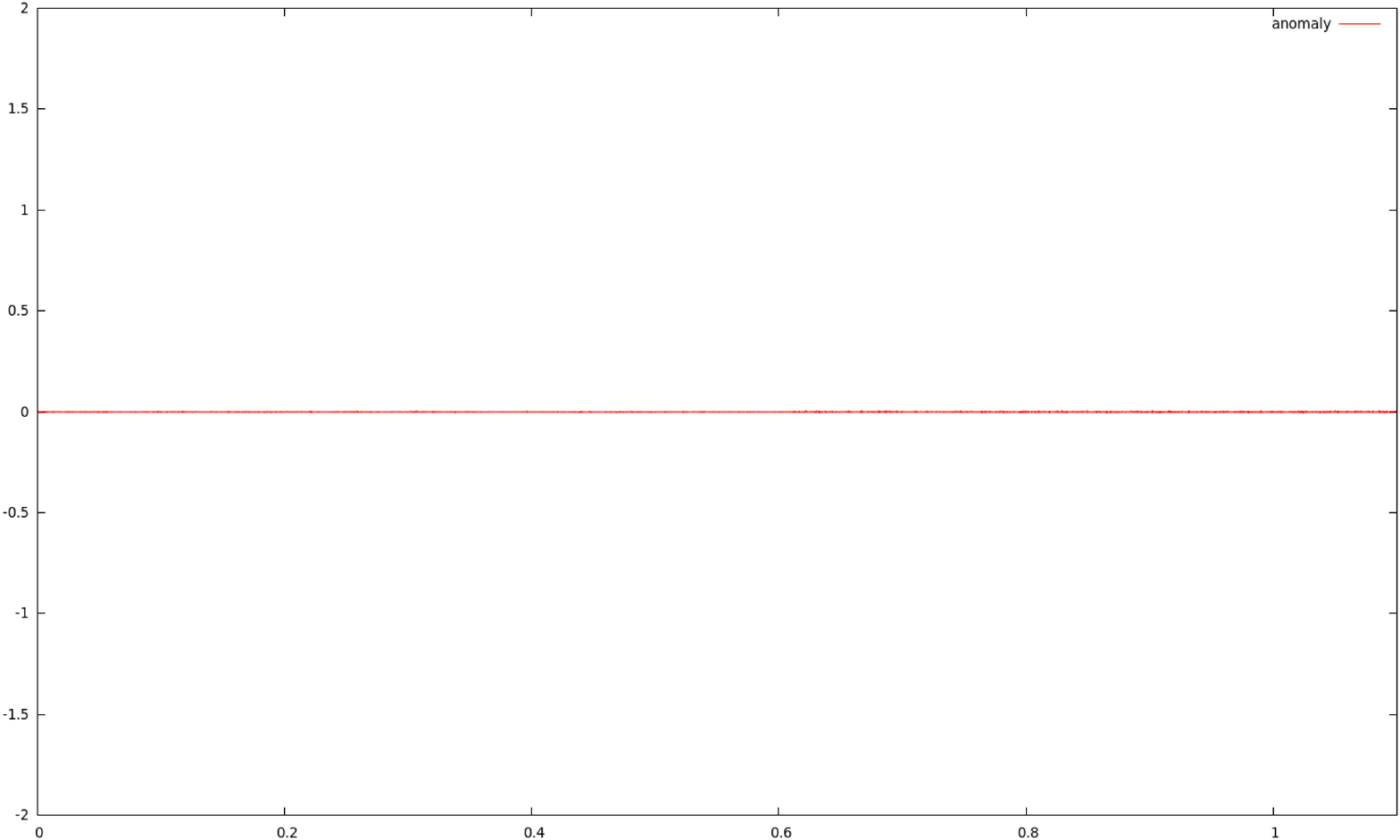}
\includegraphics[width=8cm,scale=5, angle=0, height=4cm]{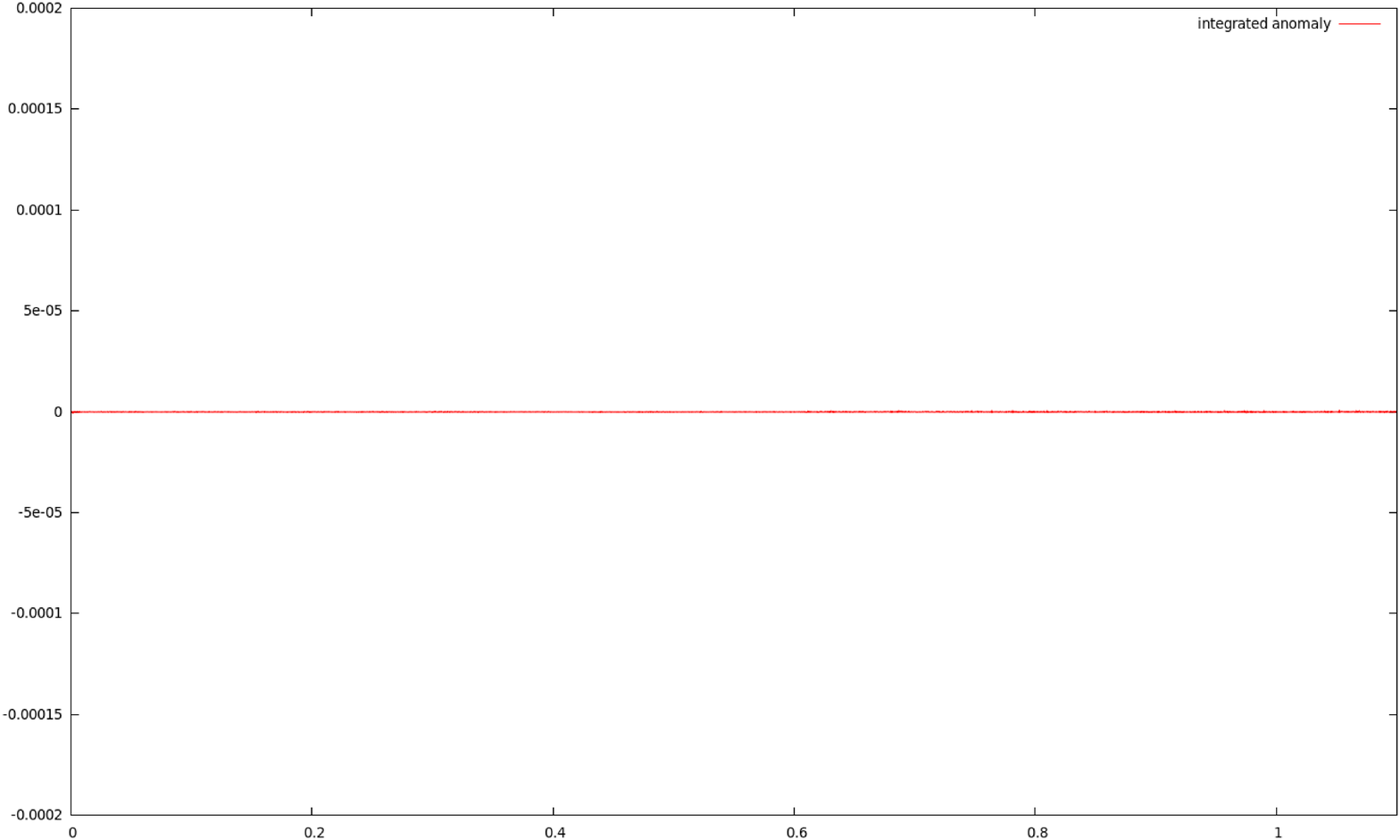} 
\parbox{5in}{\caption {(color online) Top left: the transmission of two dark solitons of the cubic-quintic NLS model (\ref{cqsol2}) is plotted for $\epsilon=-0.01,\,|\psi_0|=6,\,\eta = 2.5$. The gray solitons are initially at time $t_i$ (green line)  and travel in opposite direction with velocity $v = 11.75 \sqrt{2}$. They overlap completely ($t_c$ =blue  line) and then transmit to each other. The red line shows the solitons after collision ($t_f$). Note that  $v > v_s/2$ ($v_s = 13.89\sqrt{2}$). Top right: the integrand $\g(x,t)$ of the anomaly plotted for three successive times ($t_i$, $t_c$ and $t_f$). Bottom: the anomaly $\beta^{(4)}_r(t)$ and time integrated anomaly $\int^t dt' \beta^{(4)}_r(t')$, respectively.}}
\end{figure}

\begin{figure}
\centering

\label{fig3}
\includegraphics[width=12cm,scale=3, angle=0,height=4.5cm]{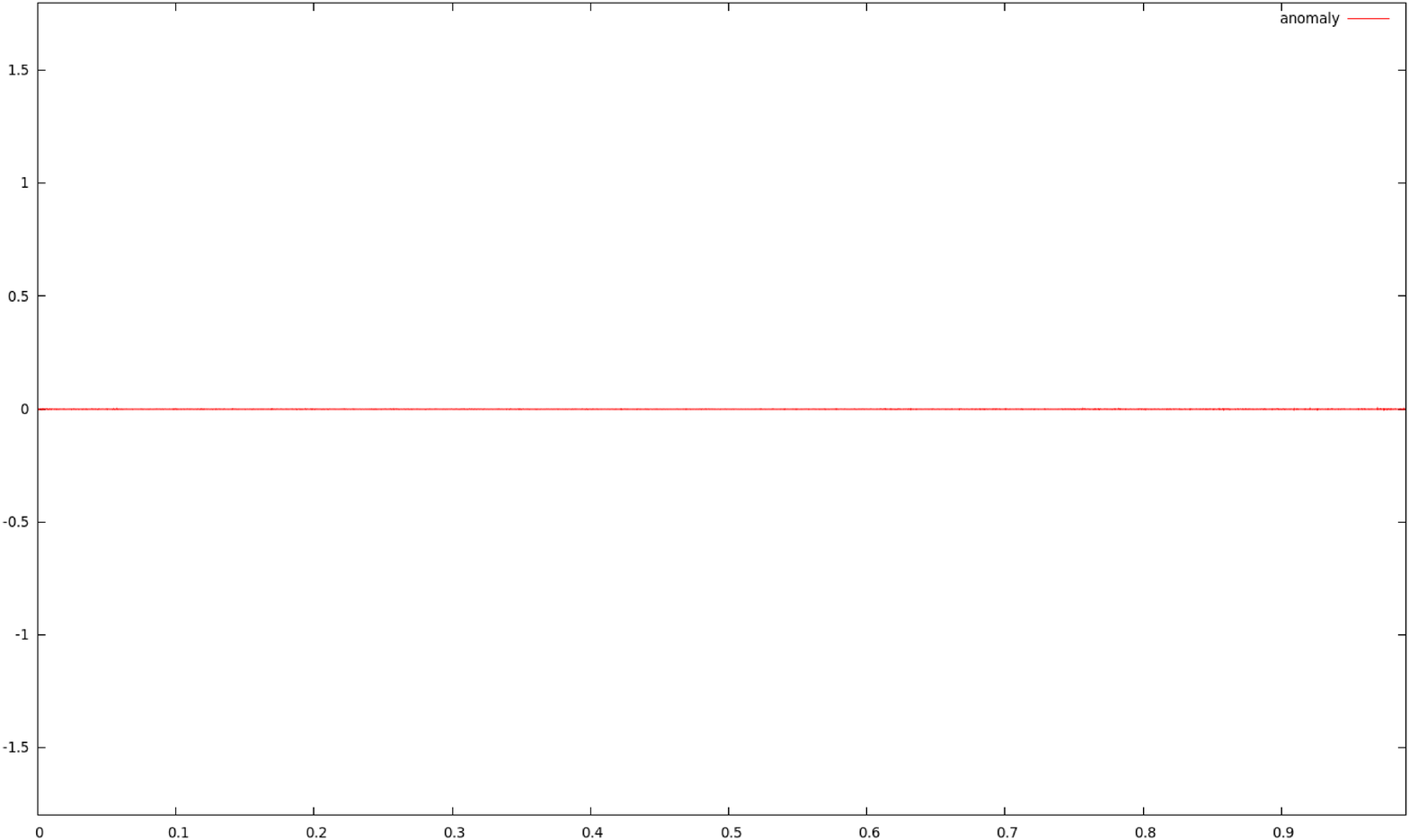}
\includegraphics[width=12cm,scale=3, angle=0,height=4.5cm]{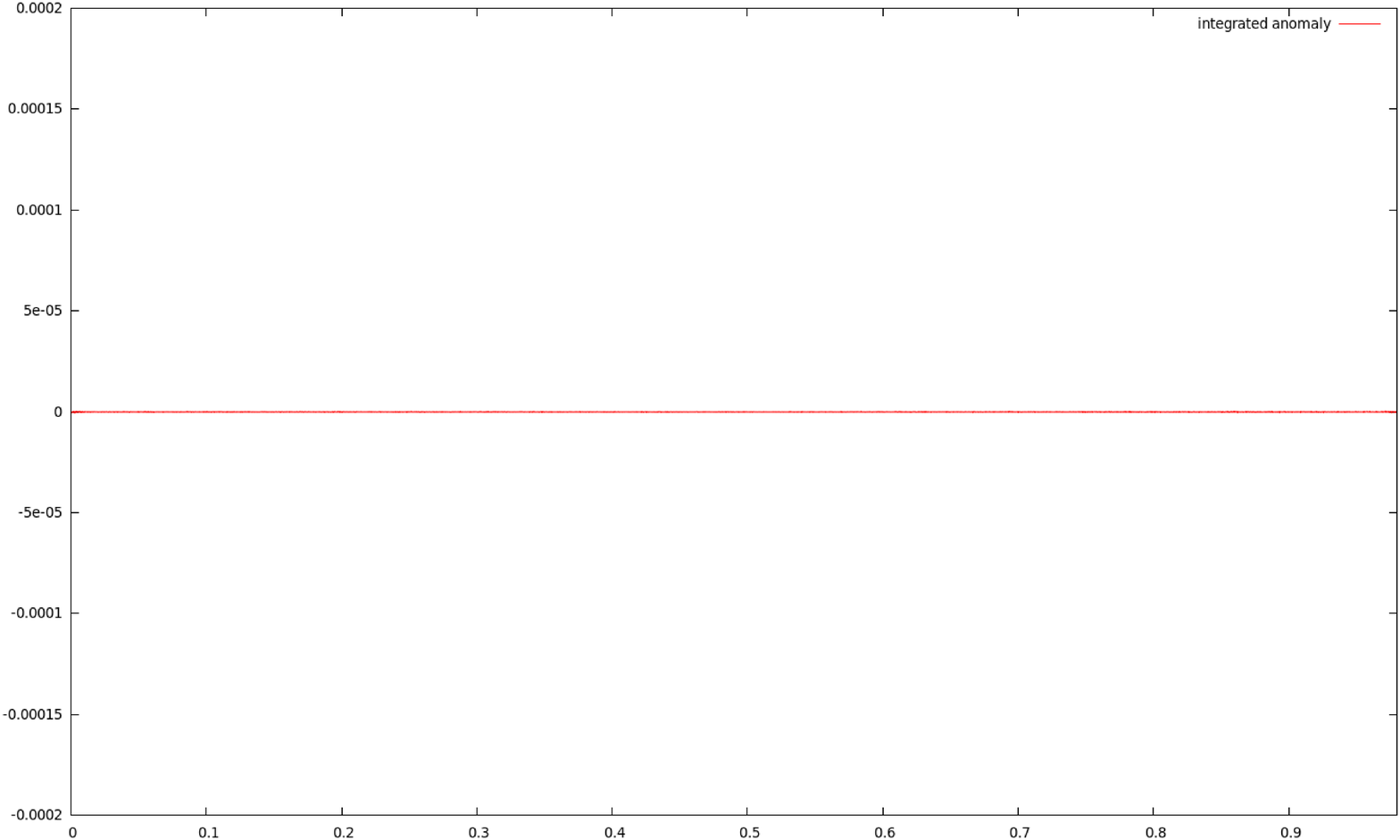}
\parbox{5in}{\caption{Anomaly $\beta^{(4)}_r(t)$ (a) and time integrated anomaly $\int^t dt' \beta^{(4)}_r(t')$ (b) for fast two-solitons sent at $v_1=-v_2=13.62\sqrt{2}$,\,$v_s = 13.89\sqrt{2},\,\epsilon = -0.01,\,\,\,|\psi_0| = 6, \eta = 2.5. $}}
\end{figure}

\begin{figure}
\centering

\label{fig4}
\includegraphics[width=12cm,scale=3, angle=0,height=4.5cm]{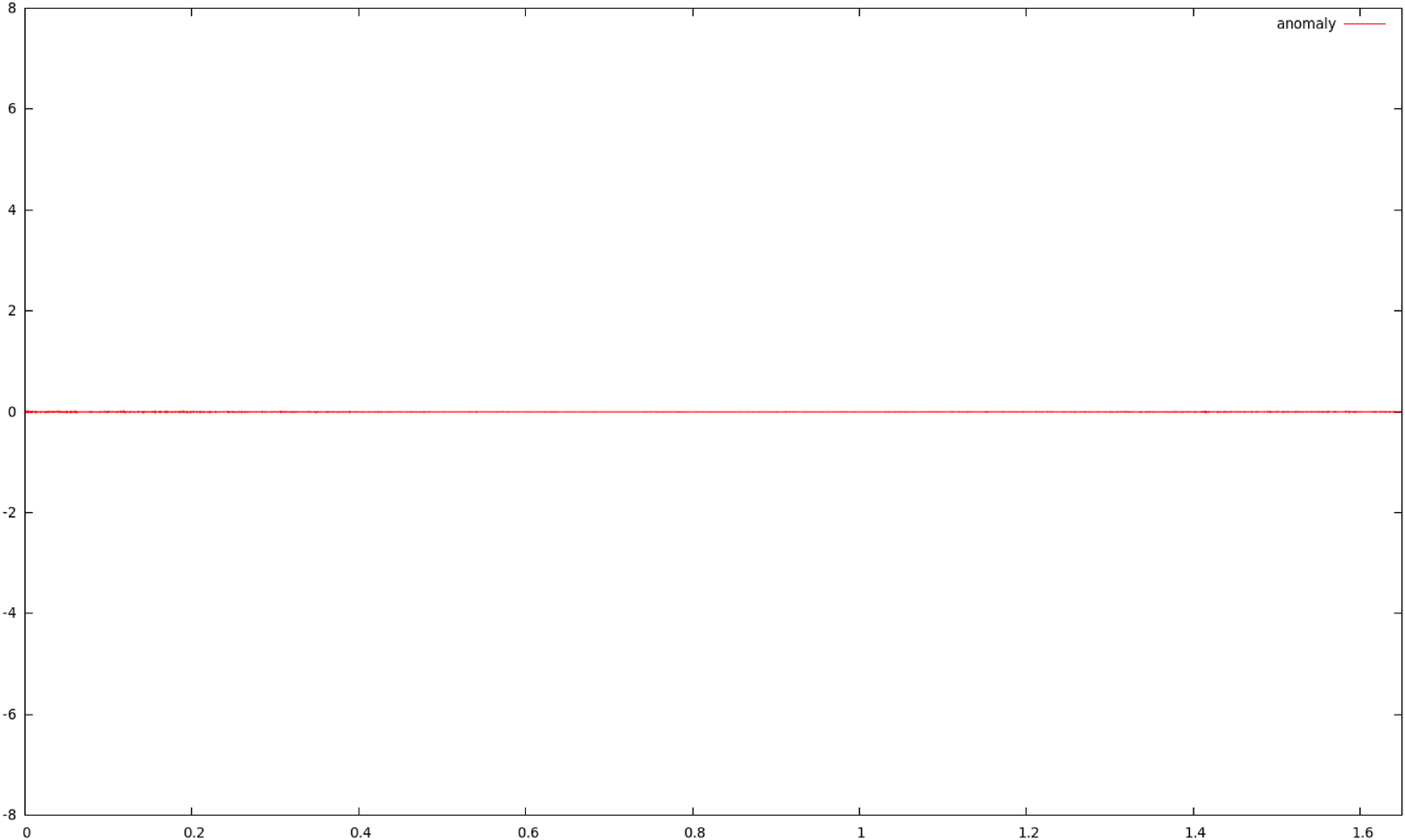}
\includegraphics[width=12cm,scale=3, angle=0,height=4.5cm]{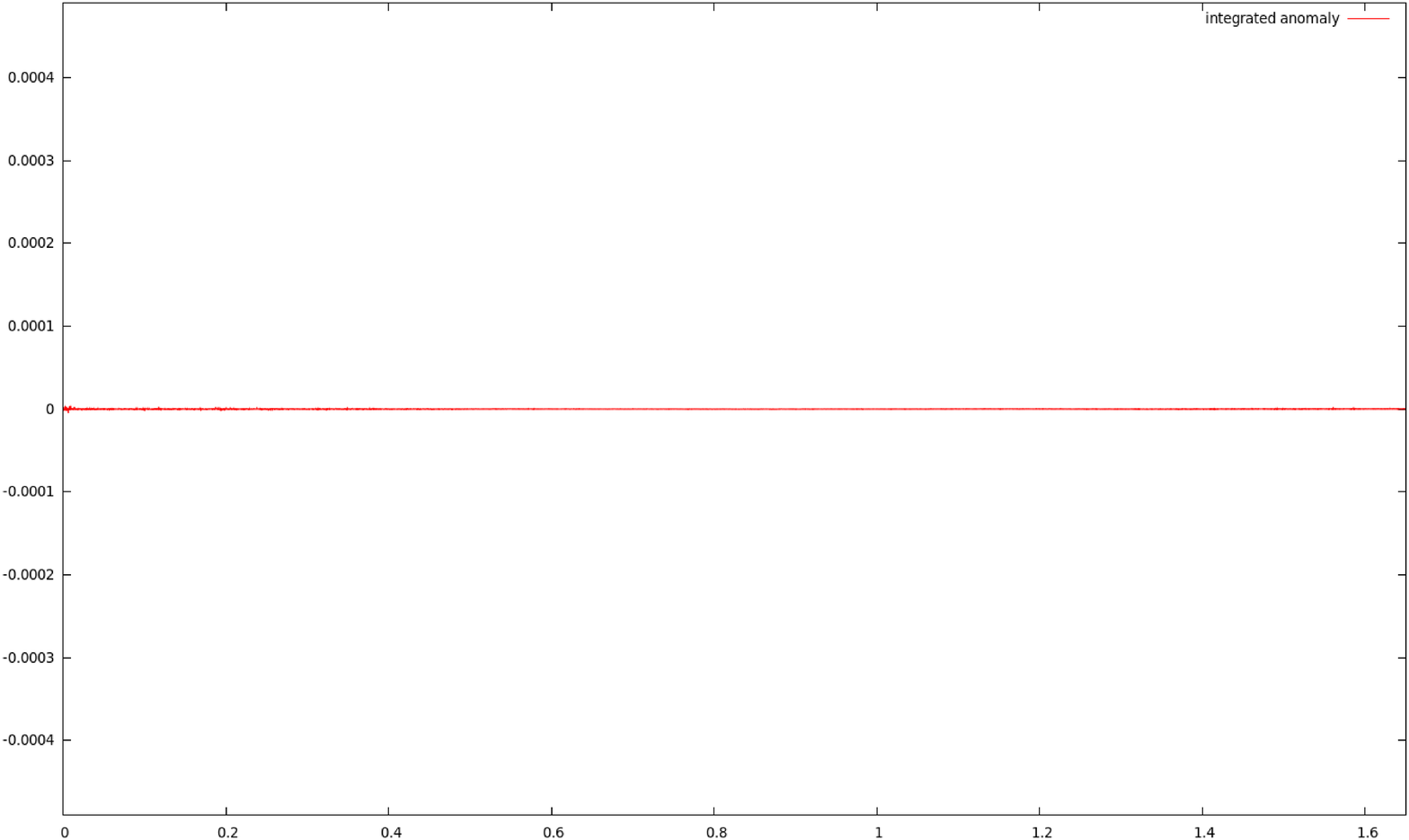}
\parbox{5in}{\caption{Anomaly $\beta^{(4)}_r(t)$ (a) and time integrated anomaly $\int^t dt' \beta^{(4)}_r(t')$ (b) for fast two-solitons sent at $v_1=-v_2=9.8 \sqrt{2}$,\,$v_s = 10.11 \sqrt{2},\,\epsilon = +0.06,\,\,\,|\psi_0| = 6, \eta = 2.5. $}}
\end{figure}
 
\begin{figure}
\centering

\label{fig5}
\includegraphics[width=12cm,scale=3, angle=0,height=4.5cm]{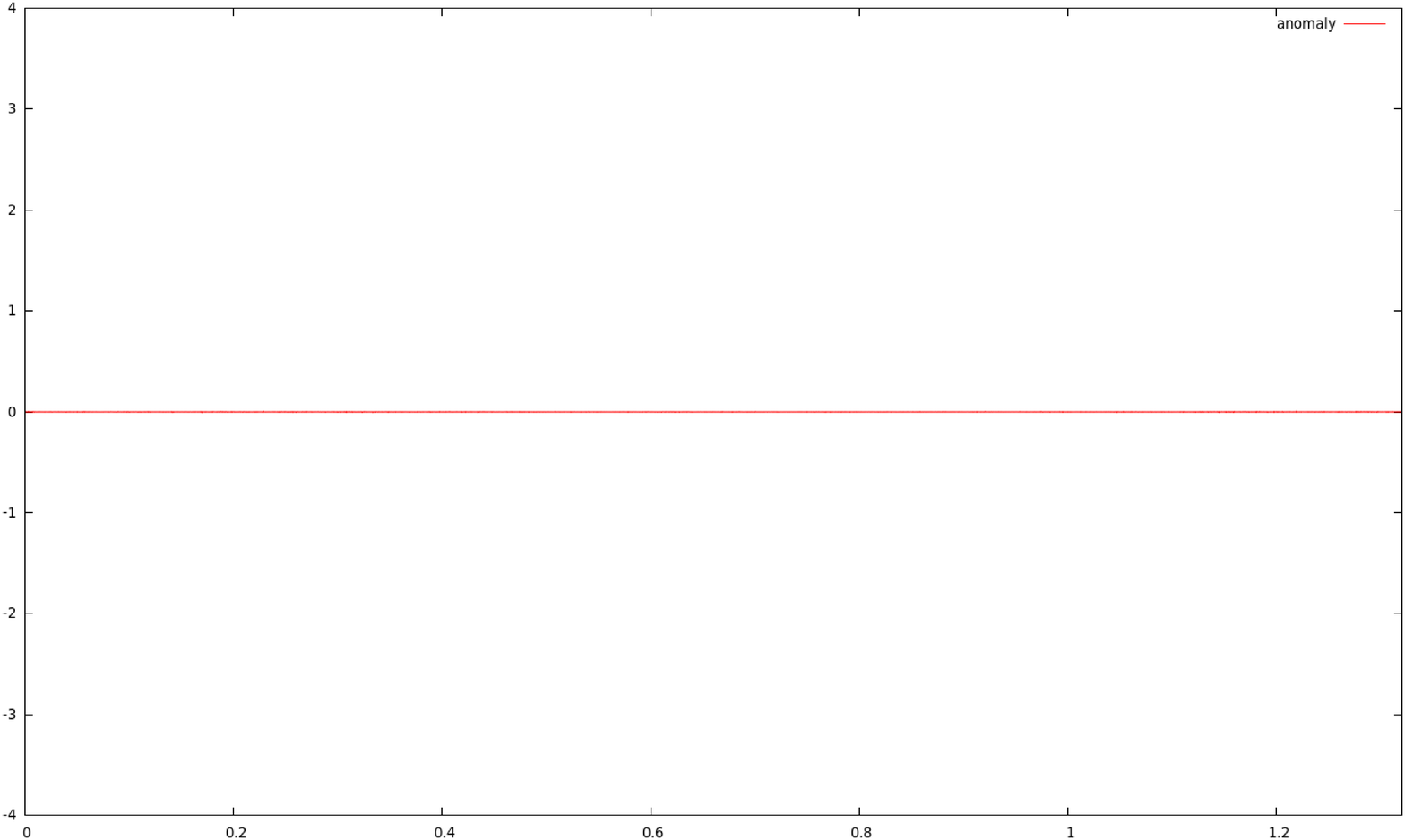}
\includegraphics[width=12cm,scale=3, angle=0,height=4.5cm]{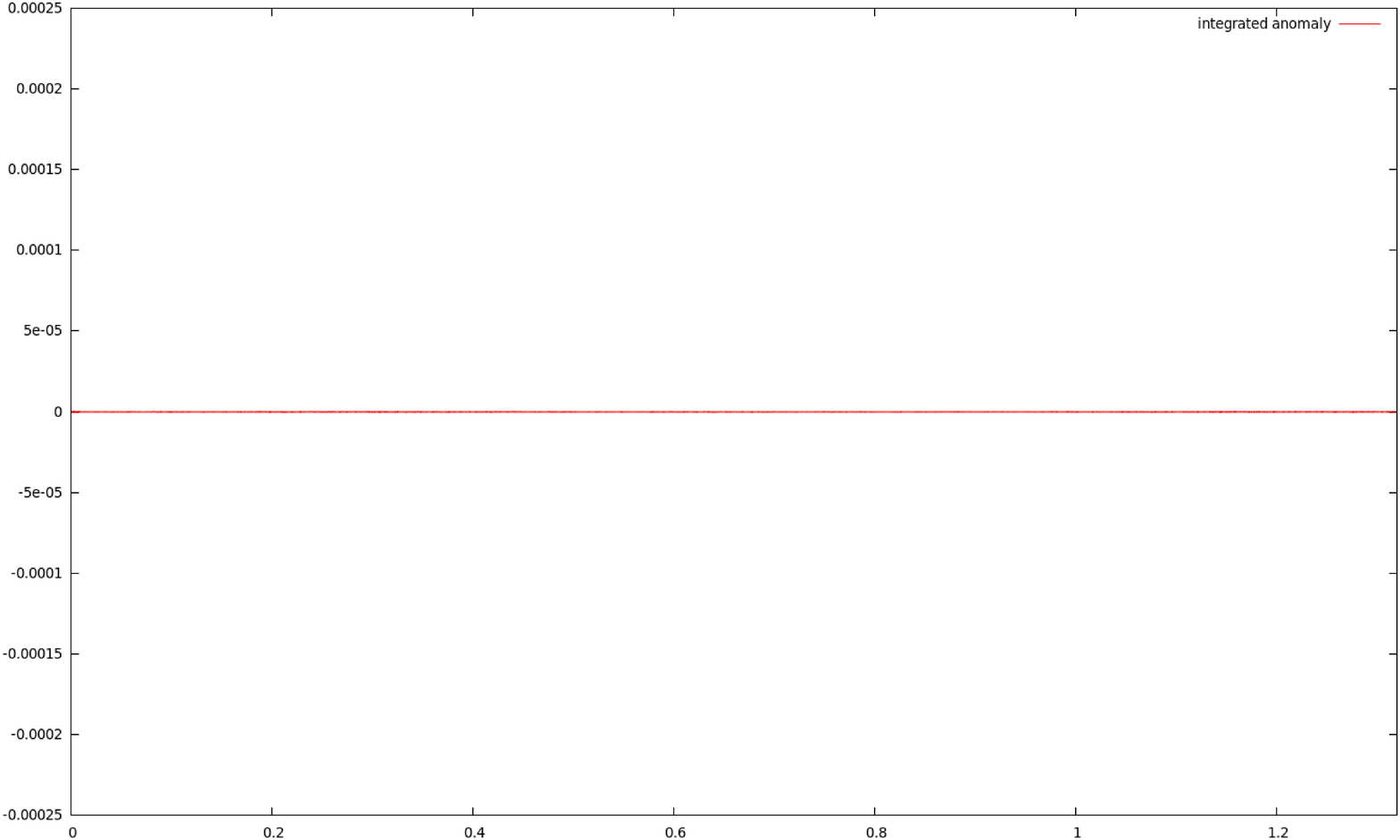}
\parbox{5in}{\caption{Anomaly $\beta^{(4)}_r(t)$ (a) and time integrated anomaly $\int^t dt' \beta^{(4)}_r(t')$ (b) for fast two-solitons sent at $v_1=-v_2=12.30 \sqrt{2}$,\,$v_s = 12.92 \sqrt{2},\,\epsilon = +0.01,\,\,\,|\psi_0| = 6, \eta = 2.5. $}}
\end{figure}

 \begin{figure}
\centering
\label{fig6}
\includegraphics[width=8cm,scale=4, angle=0, height=4cm]{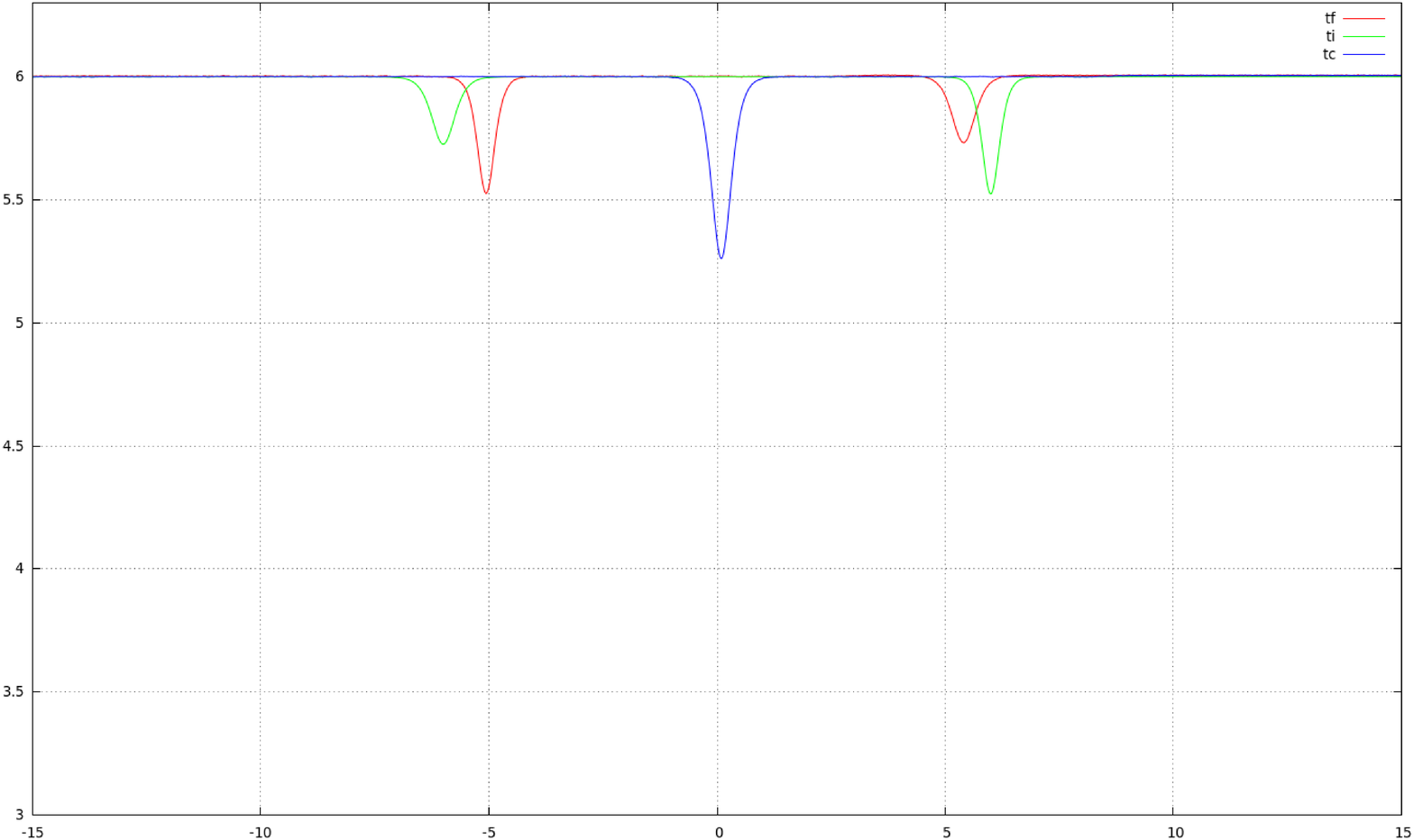} 
\includegraphics[width=8cm,scale=4, angle=0, height=4cm]{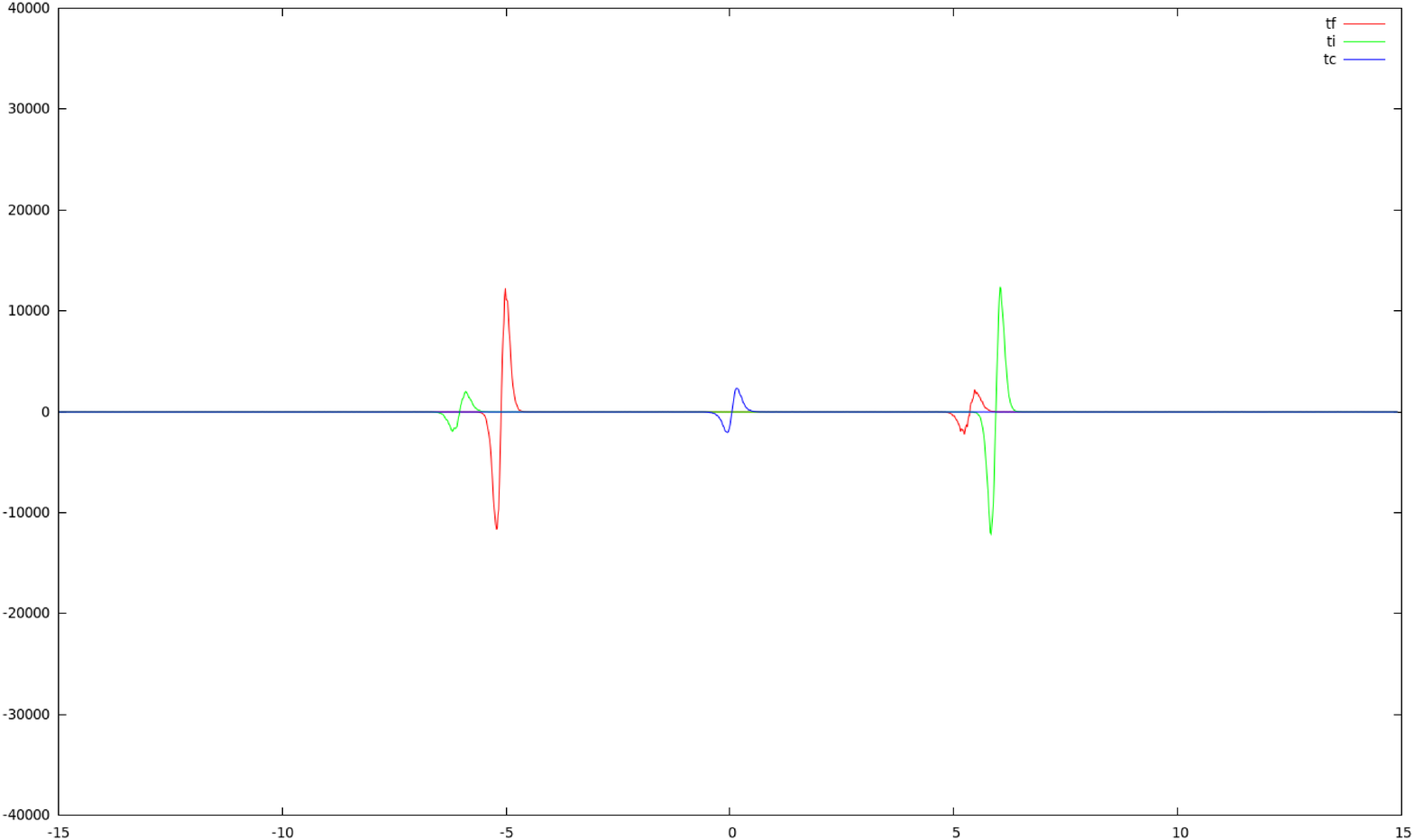}
\includegraphics[width=8cm,scale=4, angle=0, height=4cm]{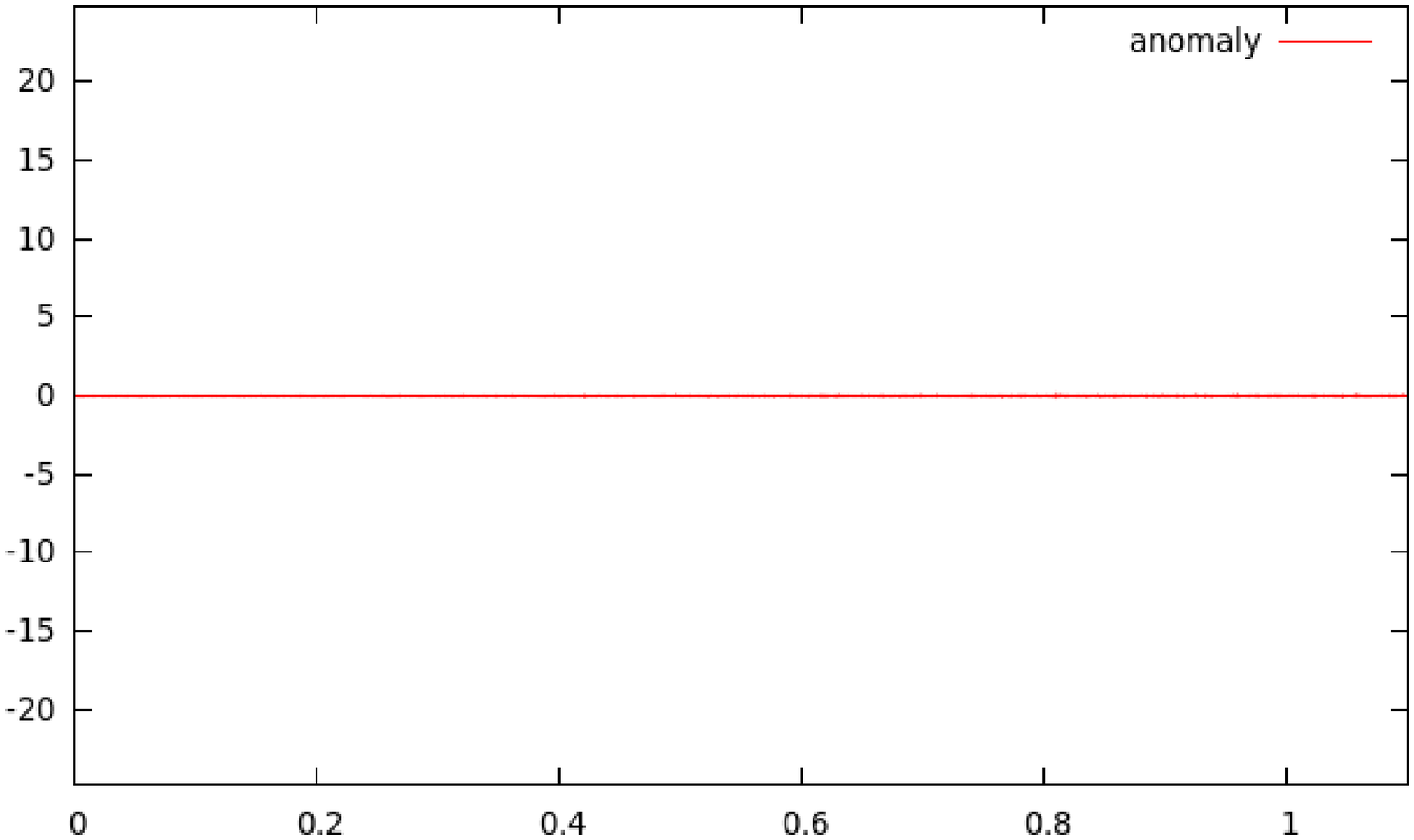}
\includegraphics[width=8cm,scale=4, angle=0, height=4cm]{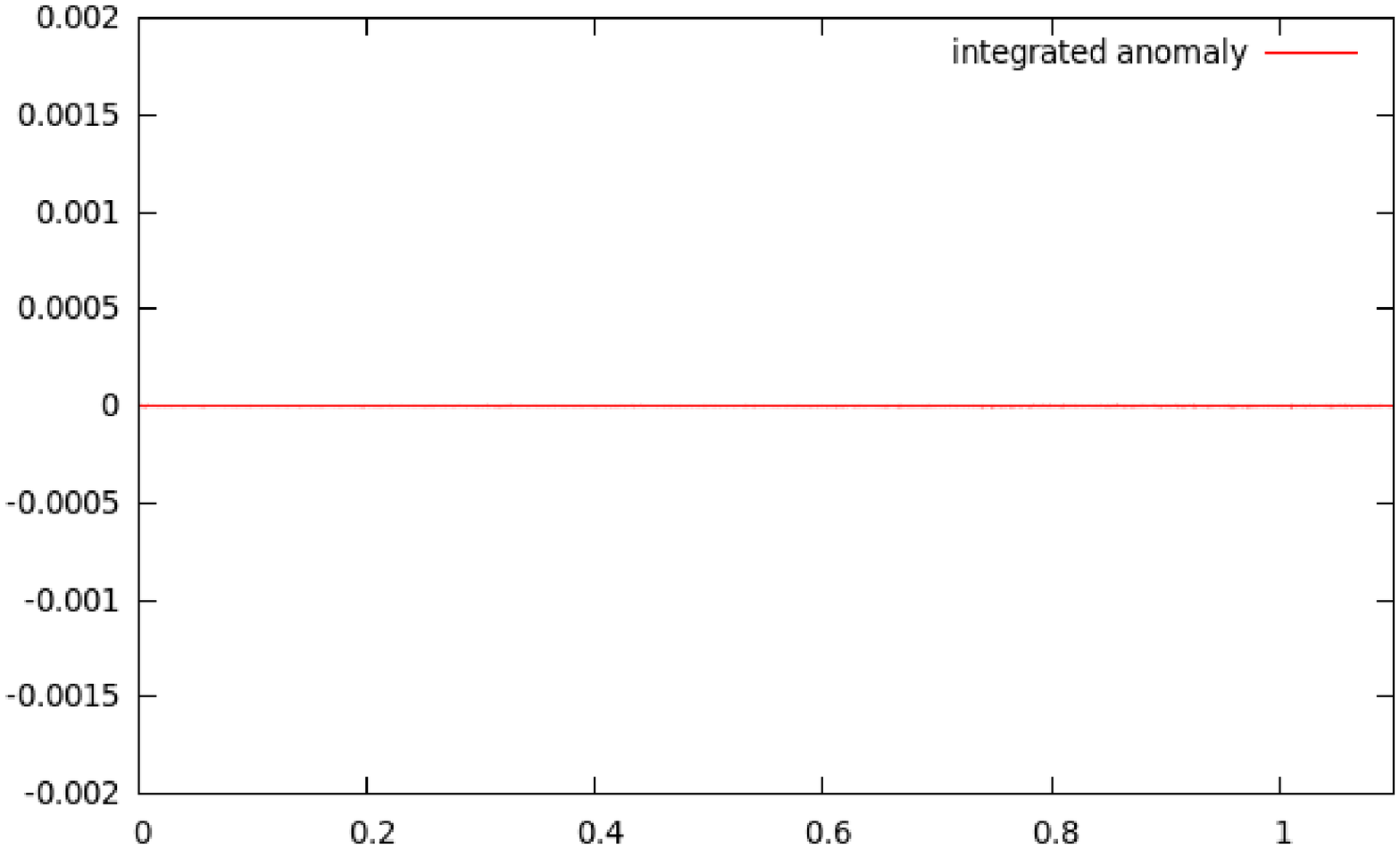}   
\parbox{5in}{\caption {(color online) Top left: transmission  of two dark solitons of the cubic-quintic NLS model (\ref{cqsol2}) with different velocities (different amplitudes) is plotted for $\epsilon=0.05,\,|\psi_0|=6,\,\eta = 2.5$. The initial gray solitons ($t_i$=green line) travel in opposite direction with velocities $v_1 = -10.02 \sqrt{2}$ (right soliton),\,$v_2 = 10.33 \sqrt{2}$ (left soliton). They completely  overlap ($t_c$=blue line) and then transmit to each other. The gray solitons after collision are plotted as a red line ($t_f$). Note that  $|v_1| + v_2  > v_s$ ($v_s = 10.73 \sqrt{2}$). Top right: the integrand $\g(x,t)$ of the anomaly plotted for three successive times ($t_i$, $t_c$ and $t_f$). Bottom: the anomaly $\beta^{(4)}_r(t)$ and time integrated anomaly $\int^t dt' \beta^{(4)}_r(t')$, respectively.}}
\end{figure}

\begin{figure}
\centering
\label{fig7}
\includegraphics[width=8cm,scale=4, angle=0, height=5cm]{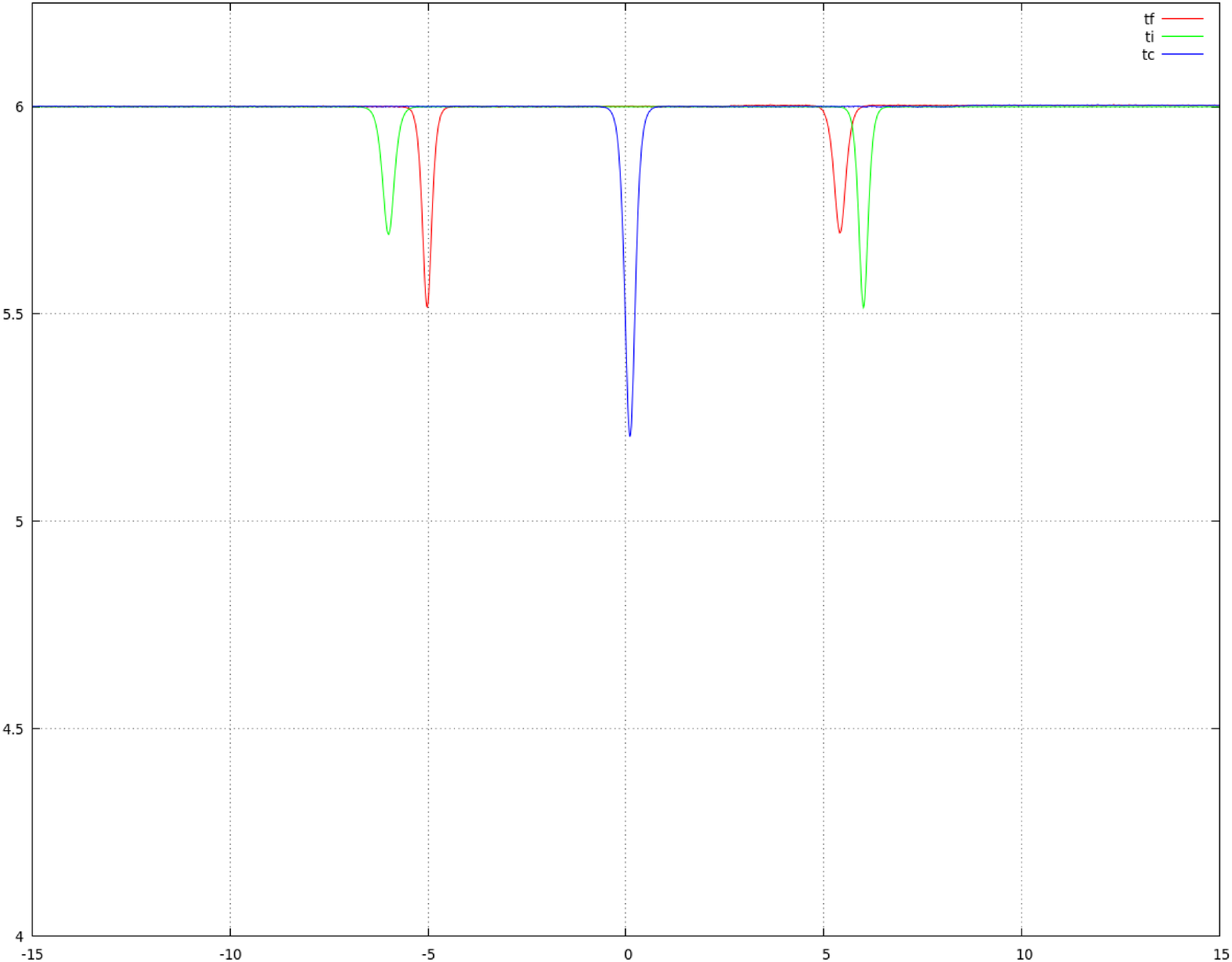} 
\includegraphics[width=8cm,scale=4, angle=0, height=5cm]{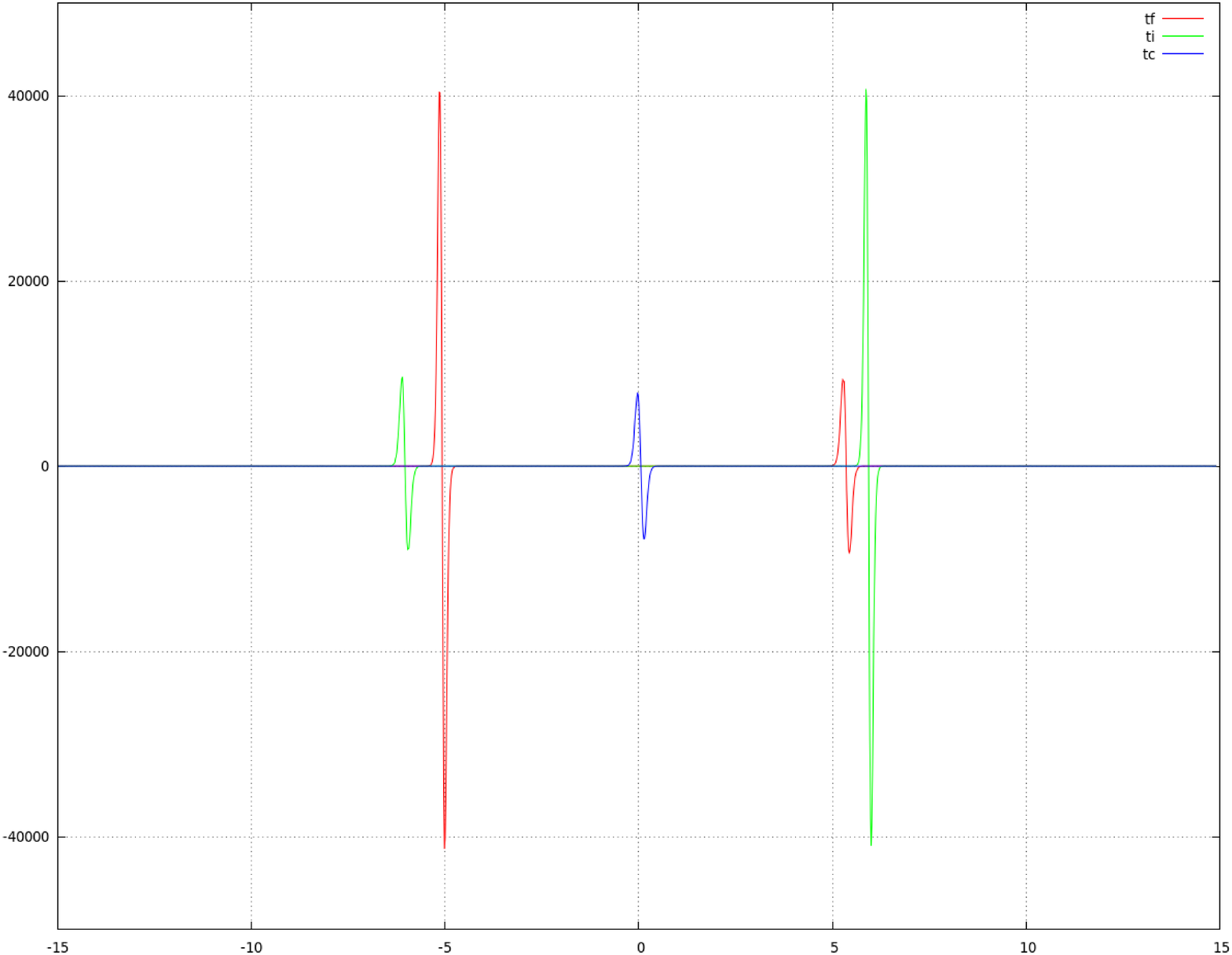}
\includegraphics[width=8cm,scale=5, angle=0, height=5cm]{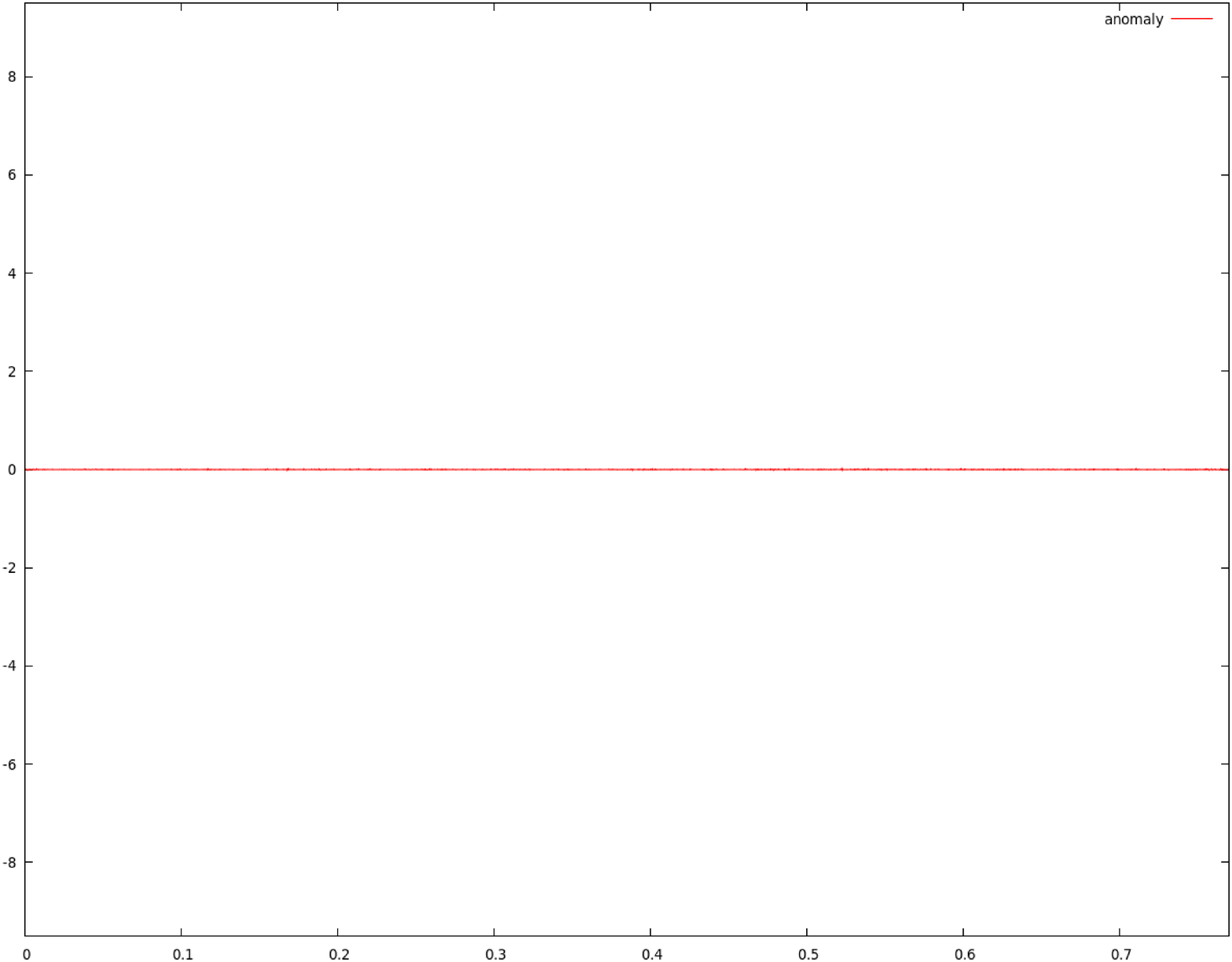}
\includegraphics[width=8cm,scale=5, angle=0, height=5cm]{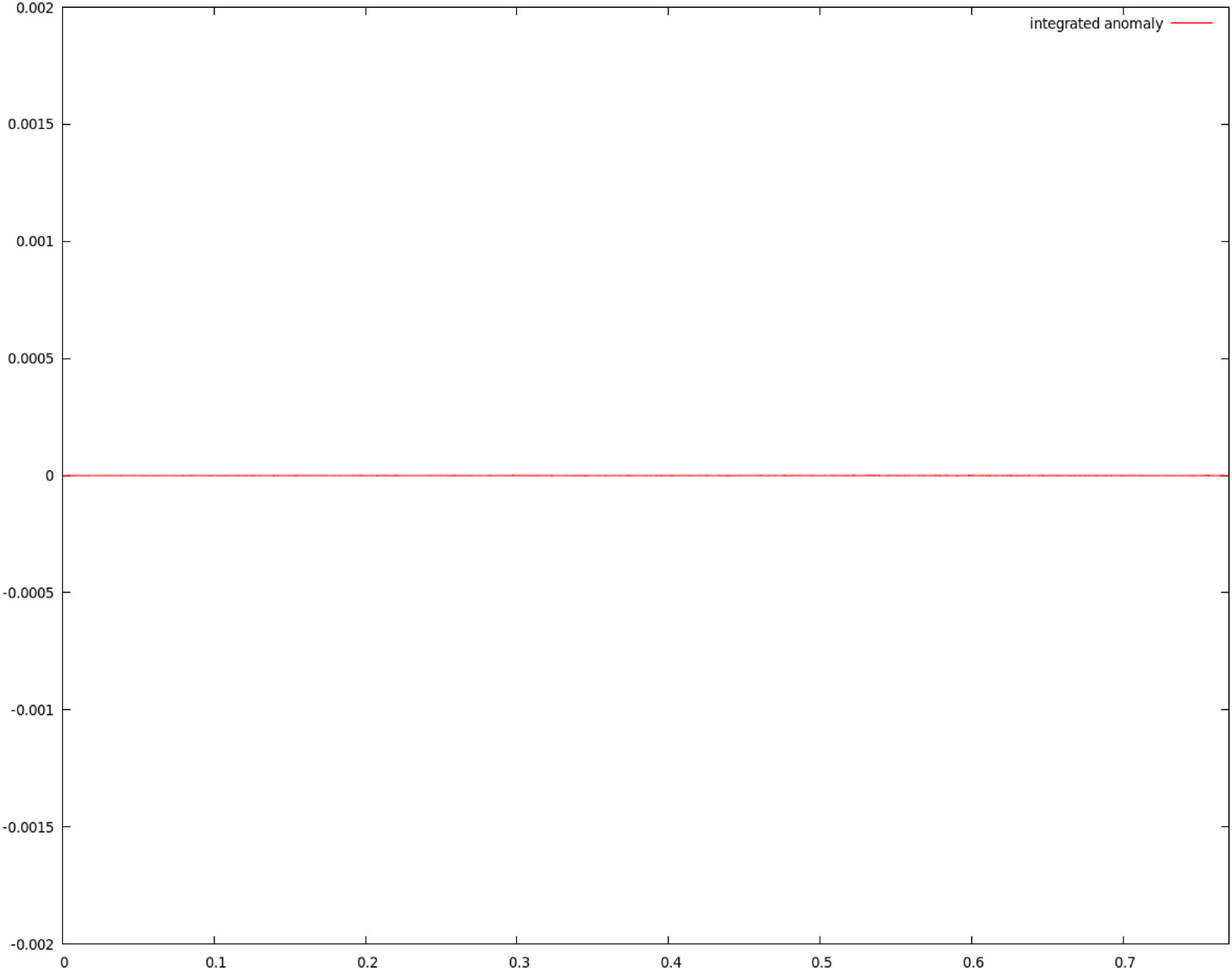}   
\parbox{5in}{\caption {(color online) Top left: transmission  of two dark solitons of the cubic-quintic NLS model (\ref{cqsol2}) with different velocities (different amplitudes) is plotted for $\epsilon=-0.05,\,|\psi_0|=6,\,\eta = 2.5$. The initial gray solitons ($t_i$=green line) travel in opposite direction with velocities $v_1 = -14.28 \sqrt{2}$ (right soliton),\,$v_2 = 14.78 \sqrt{2}$ (left soliton). They completely  overlap ($t_c$=blue line) and then transmit to each other. The gray solitons after collision are plotted as a red line ($t_f$). Note that  $|v_1| + v_2 >  v_s$ ($v_s \approx 15.65\sqrt{2}$). Top right: the integrand $\g(x,t)$ of the anomaly plotted for three successive times ($t_i$, $t_c$ and $t_f$). Bottom: the anomaly $\beta^{(4)}_r(t)$ and time integrated anomaly $\int^t dt' \beta^{(4)}_r(t')$, respectively.}}
\end{figure}

\begin{figure}
\centering
\label{fig63}
\includegraphics[width=8cm,scale=5, angle=0, height=5cm]{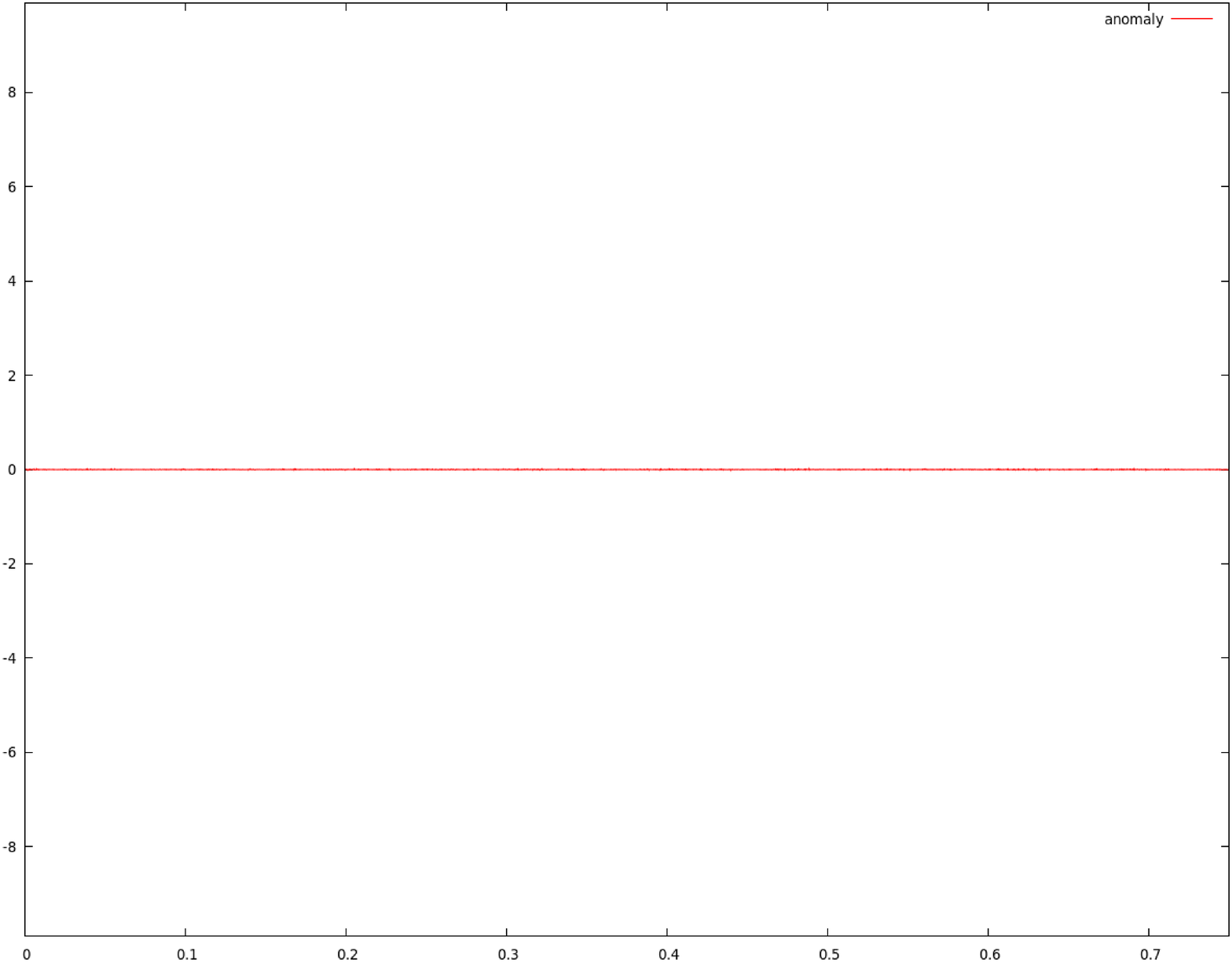}
\includegraphics[width=8cm,scale=5, angle=0, height=5cm]{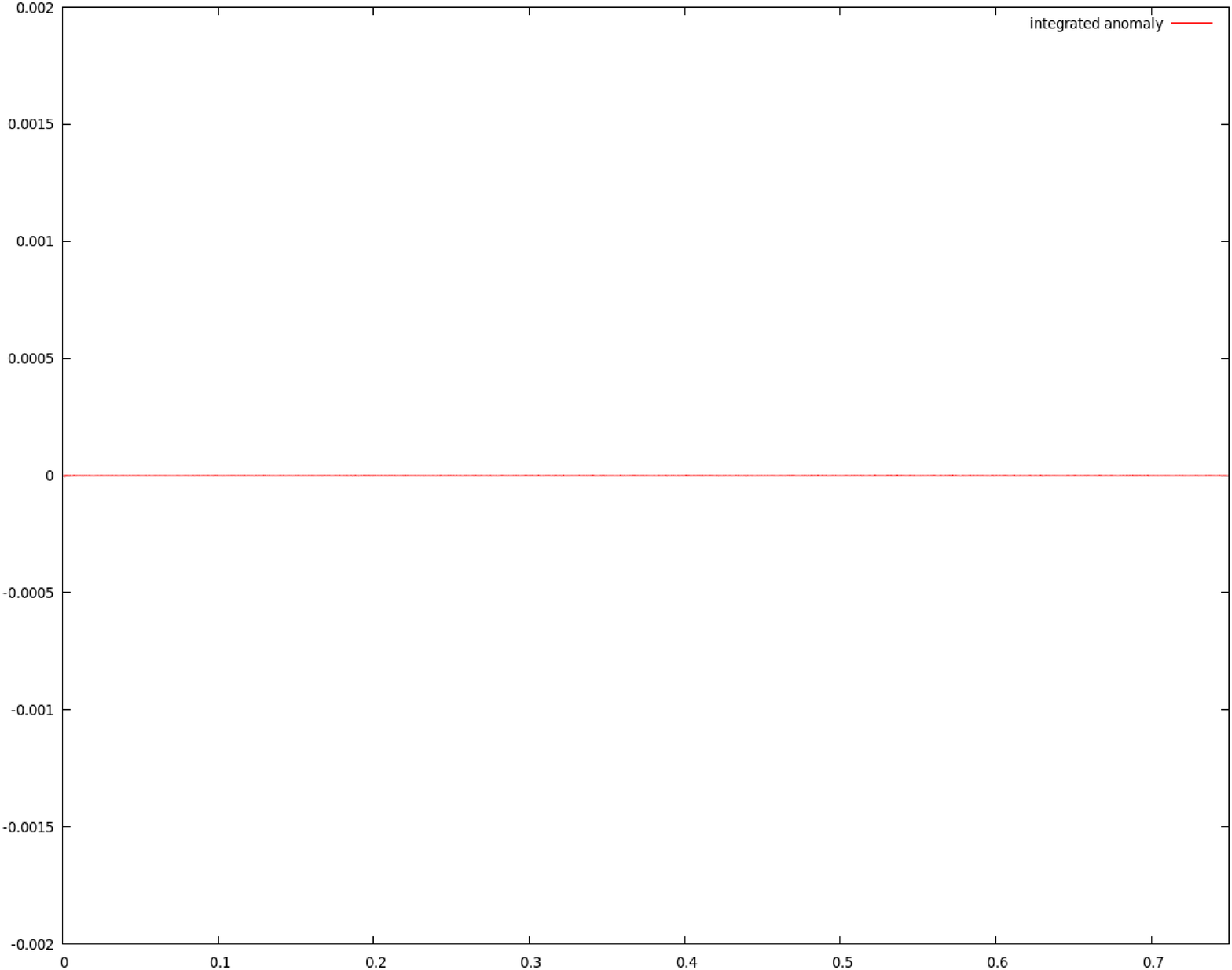}   
\parbox{5in}{\caption {(color online)  
Anomaly $\beta^{(4)}_r(t)$ (a) and time integrated anomaly $\int^t dt' \beta^{(4)}_r(t')$ (b) for the transmission of two dark solitons of the cubic-quintic NLS model (\ref{cqsol2}) with different velocities (different amplitudes) are plotted for $\epsilon=-0.06,\,|\psi_0|=6,\,\eta = 2.5$, sent at   $v_1 = -15.12 \sqrt{2}$ (right soliton),\,$v_2 = 15.59 \sqrt{2}$ (left soliton). $v_s = 16.05 \sqrt{2}$}}
\end{figure}

\begin{figure}
\centering
\label{fig64}
\includegraphics[width=8cm,scale=5, angle=0, height=5cm]{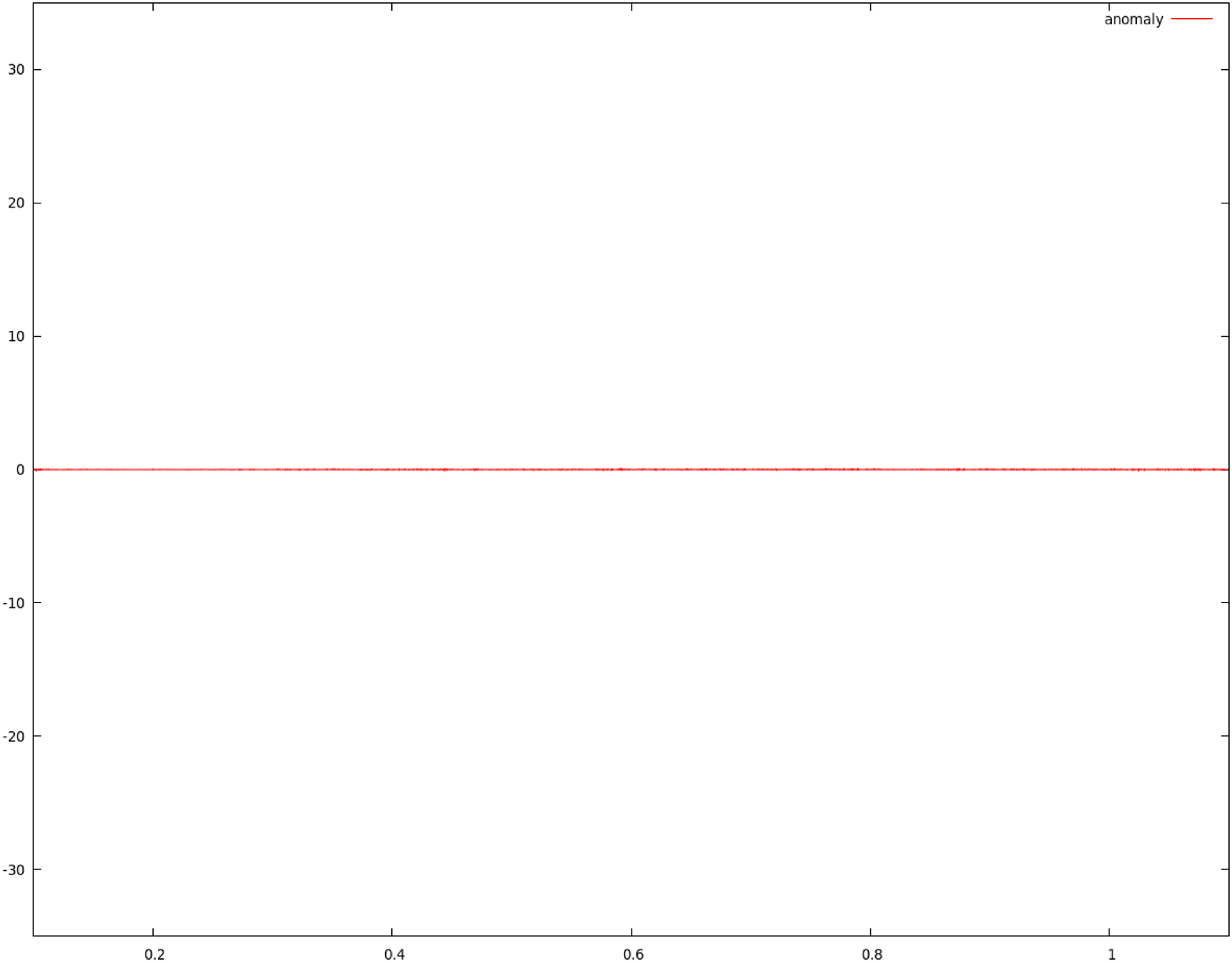}
\includegraphics[width=8cm,scale=5, angle=0, height=5cm]{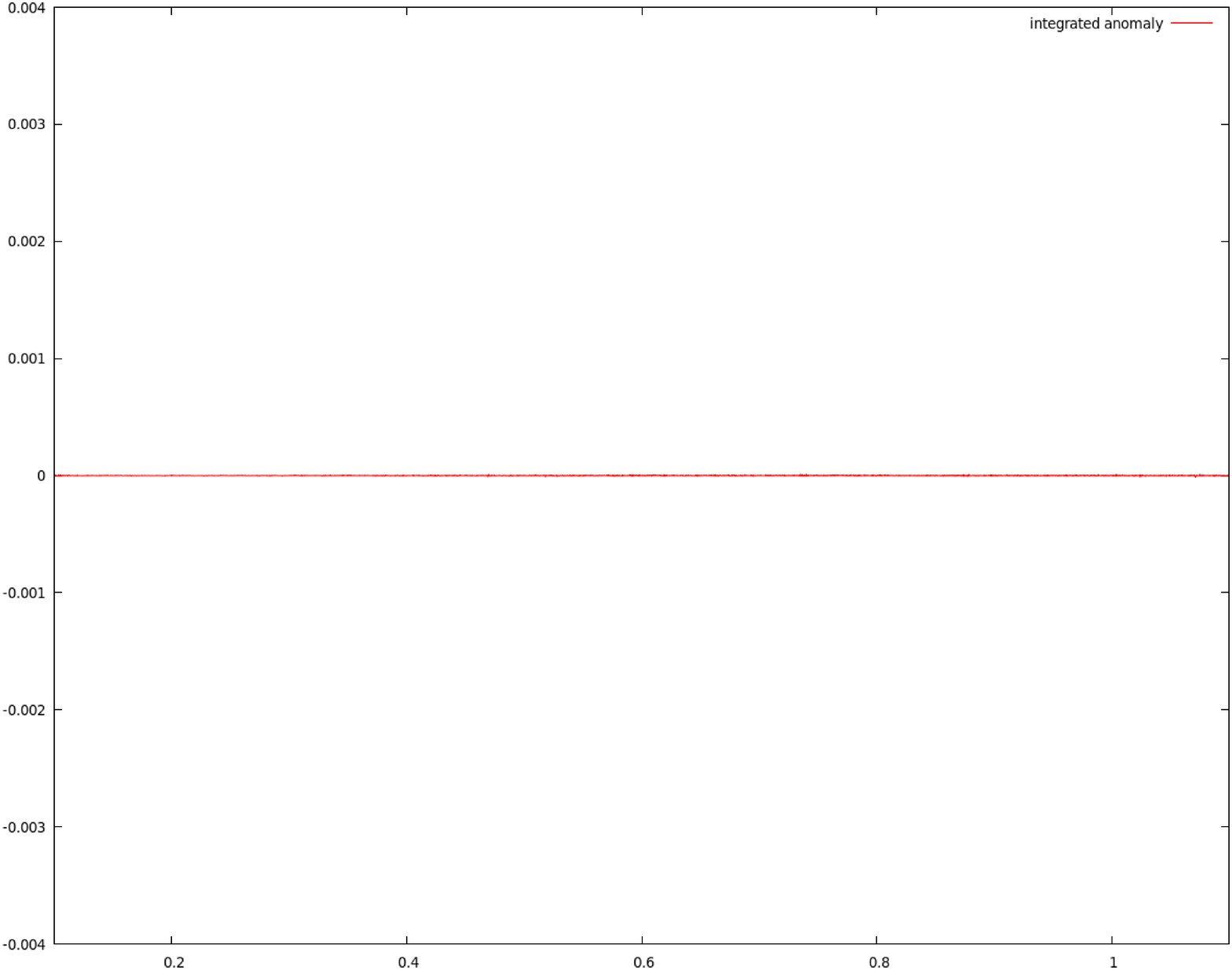}   
\parbox{5in}{\caption {(color online) Anomaly $\beta^{(4)}_r(t)$ (a) and time integrated anomaly $\int^t dt' \beta^{(4)}_r(t')$ (b) for the transmission  of two dark solitons of the cubic-quintic NLS model (\ref{cqsol2}) with different velocities (different amplitudes) are plotted for $\epsilon=0.04,\,|\psi_0|=6,\,\eta = 2.5$, sent at   $v_1 = -10.31 \sqrt{2}$ (right soliton),\,$v_2 = 10.77 \sqrt{2}$ (left soliton), $v_s = 11.32 \sqrt{2}$.}}
\end{figure}

Since the two-soliton solutions admit in general the symmetry (\ref{pari1}), as presented in section \ref{stp} through perturbation theory on the deformation parameter, we expect their anomalies to vary during collision and the charges to be asymptotically conserved. However, for a special soliton solutions with space-reflection symmetry, the anomaly $\b^{(4)}$ belongs to the sequence of even order charges and it must vanish during the whole collision process. In fact, this behaviour is observed through our numerical simulations for various soliton configurations and different values of the deformation parameter $\epsilon$. So, we can argue that the dynamics favours the soliton configurations with symmetry properties (\ref{xsym}) at all orders in perturbation theory, as presented in section  \ref{srs}, for the collision of two solitons with opposite and  equal velocities.     

Notice that, we have shown in general that the well separated individual solitons provide vanishing anomalies (\ref{vanish1}). Remarkably, for 2-soliton collisions with different velocities (different amplitudes) one notices that in Figs. 7, 8, 9 and 10 the anomalies vanish  during the  whole collision process. Even though the incident and outgoing solitons do not form a 2-soliton solution with definite even parity under space-reflection ${\cal P}_x$, the collision dynamics reproduces the vanishing of $\beta^{(4)}_r$ in (\ref{qcon11}). Therefore, the charges must be exactly  conserved, within numerical accuracy. So, our numerical simulations suggest that the space-reflection ${\cal P}_x$ symmetry is a sufficient but not a necessary condition for the vanishing of the anomaly $\beta^{(4)}_r$.  

The mechanism responsible for the (asymptotic) conservation of an infinite number of charges is not well understood yet, but in all examples presented in the literature \cite{jhep1, jhep2} the two-soliton type solutions present special properties under a
space-time parity transformation. Certainly, a better understanding of non-linear dynamics governing the behaviour of the charges and anomalies during  the soliton collisions deserve a further analyses.

\subsection{Second deformation: $V'[I] = 2 \eta I - \epsilon \frac{I^q}{1+ I^q}$}
\label{second}
Next, we consider the numerical solutions of the deformed NLS equation (\ref{nlsd}) defined for the potential (\ref{enns1}). Since we do not know a solitary wave solution in analytical form we search for a numerical solution  by considering a seed solitary wave provided by the CQNLS solitary wave solution (\ref{dark22})-(\ref{dark221}). So, one notices that as the solitary wave evolves in time it emits radiation on its both sides overlapping with the {\sl cwb}. Then, once the edge of the radiation waves are sufficiently far from the numerically generated dark soliton itself one can remove the radiation  completely from the both hand sides such that only the {\sl cwb} is allowed to oscillate in these regions. Then, two of these numerically generated dark solitons  are taken far apart and stitched together at the middle point in order to provide an initial condition for the two-dark soliton collision of the deformed model.

In the Figs. 11-20  the simulations of two soliton collisions in the second deformed NLS model (\ref{enns1}) are presented. In the Figs.  11-14 we show  the collision of equal velocity (equal amplitude) solitons (top left figures). The relevant anomaly $\beta^{(4)}_r(t)$  (\ref{beta11}) and the integrated anomaly  $\int^t dt' \beta^{(4)}_r(t')$ (\ref{tint}) as functions of time, are plotted (bottom line figures). Moreover, the integrand functions $\g(x,t)$ in (\ref{func11}) have been plotted as functions of $x$ for successive times (top right figures). These plots show qualitatively the behaviour of these functions such that its integration in the whole space furnishes vanishing anomalies.  The two-soliton reflection results are shown in Figs. 11 and 13 with parameters  $\epsilon = 0.06,\, q=2$ and $\epsilon = -0.06,\, q=2$, respectively, for  slow solitons. The two-soliton transmission are considered in Figs. 12 and 14 with parameters  $\epsilon = 0.1, q=3$ and $\epsilon = -0.06, q=2$, respectively,  for fast solitons. 

In the Figs.  15-20 we show  the collision of different velocity (different amplitude) solitons with parameters $(\epsilon, q) \equiv \{(0.1, 3),(-0.06, 2),(0.06, 4),(-0.06, 3), (0.06,3), (-0.1, 5)\}$, respectively. In the Figs. 15 and 16 we show the soliton profiles for three successive times (top left figures), and the integrand function $\g(x,t)$ in (\ref{func11}) is plotted as function of $x$ for successive times (top right figures). These plots show qualitatively the behaviour of these functions such that its integration in the whole $x-$space furnishes vanishing anomalies for different amplitude soliton collisions. The relevant anomaly $\beta^{(4)}_r(t)$  (\ref{beta11}) and the integrated anomaly  $\int^t dt' \beta^{(4)}_r(t')$ (\ref{tint}) as functions of time, are plotted in all the figures 15-20 (bottom line figures in Figs. 15 and 16).   

Therefore, in the Figs. 15-20 we have shown numerically, with higher precision, that the  charge  $Q^{(4)}$ is indeed conserved when
the configuration does not possess reflection symmetry (solitons with different velocities).

Notice that the charge $Q^{(4)}$ in (\ref{qcon11})-(\ref{charge}) is conserved to within an accuracy of $ \approx 10^{-11}$ in the second modified NLS  model even for deformation parameters of the order of $|\epsilon| \approx  0.1$, whereas in the first modification the CQNLS model this error is of the order of $ \approx 10^{-5}$ for $|\epsilon| \approx 0.01$. In our extensive simulations we noticed that in general the vanishing of the anomalies occur with small numerical errors provided that  the parameter value $\epsilon$ is small and the solitons are fast. Then, we are quite confident of our simulations in the second modified NLS model in which the anomalies vanish within an error of less than $ 10^{-10}$ for $|\epsilon| <  1.5 $ and $n \geq 2$ in the whole process of the collisions of two fast and/or slow solitons. The main conclusion that we can make from our simulations for several two-soliton collisions is that the charge $Q^{(4)}$ is not only  asymptotically conserved, but exactly conserved during the whole  collision process, within the numerical accuracy.

The results of our simulations of the second deformation of the NLS model (\ref{enns1}) and the fact that it possesses a submodel (with $\epsilon = 1,\, q\in \IZ_{+}$) which passes the Painlev\'e test \cite{enns} suggest that it might be an exactly integrable model for some set of the parameters $\{|\psi_0|, \eta, \epsilon,\,q \}$. To the best of our knowledge its Lax pair formulation and analytic soliton solutions are not known in the literature and the search for them are worth attempting. However, the existence of two-solitons is not indicative of integrability. In fact, some non-linear models present elastic scattering of two solitons but they do not posses three-soliton solutions. Nevertheless, according to the ref. \cite{hietarinta} the above deformed NLS model (\ref{enns1}) may be regarded as a partially integrable model in the sense that it possesses two-soliton solutions, even though it has been obtained numerically. Moreover, the existence of three-soliton solutions in partially integrable models is quite demanding and it is very rare. In fact, in the Hirota's bilinear approach the existence of three-solitons is a strict integrability criterion \cite{hietarinta}. So, it would be interesting to tackle this problem in the lines discussed above for some deformations of the NLS model, since the presence of three-soliton combinations of any type may indicate integrability of the model.

\begin{figure}
\centering
\label{fig8}
\includegraphics[width=8cm,scale=4.5, angle=0, height=4.5cm]{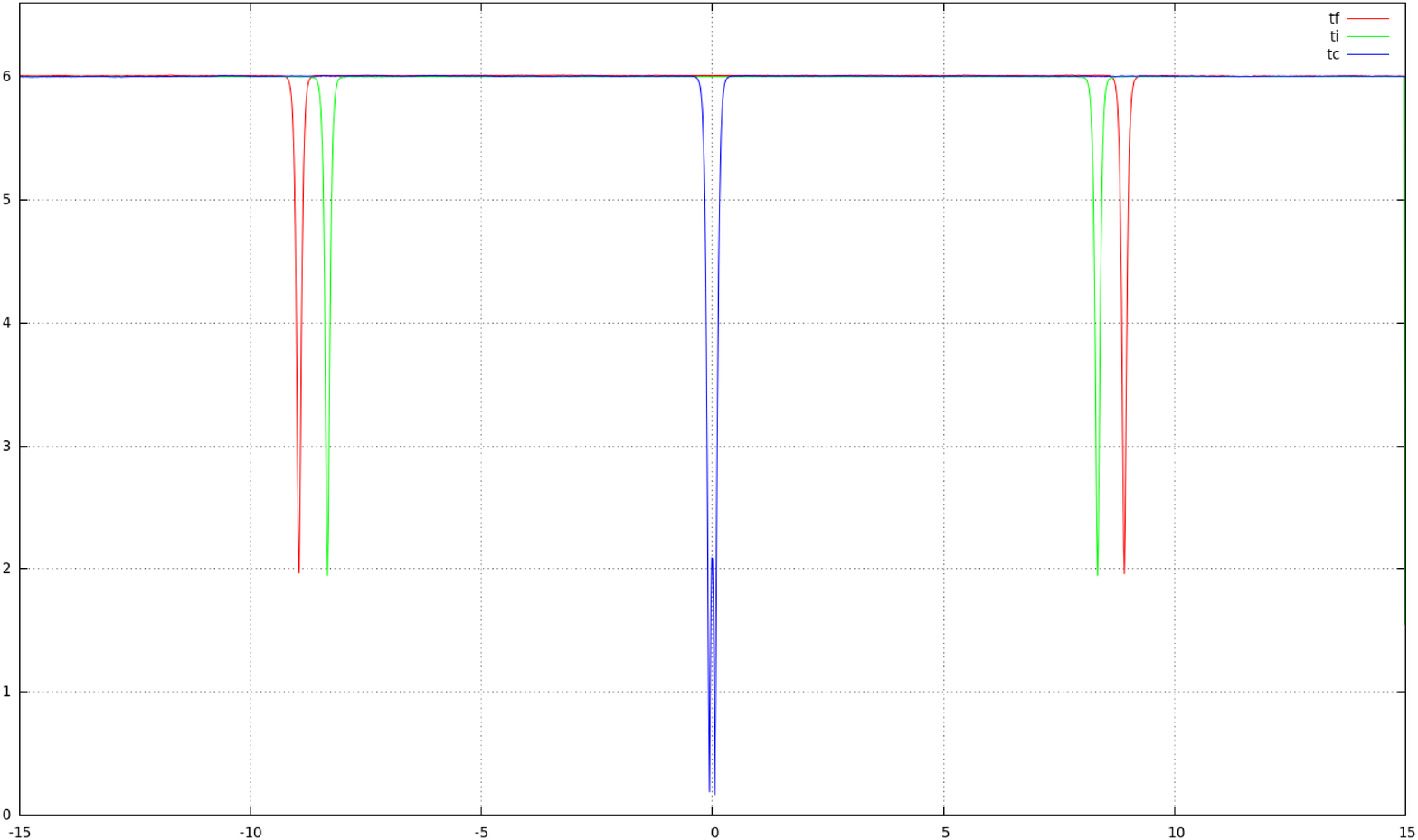} 
\includegraphics[width=8cm,scale=4.5, angle=0, height=4.5cm]{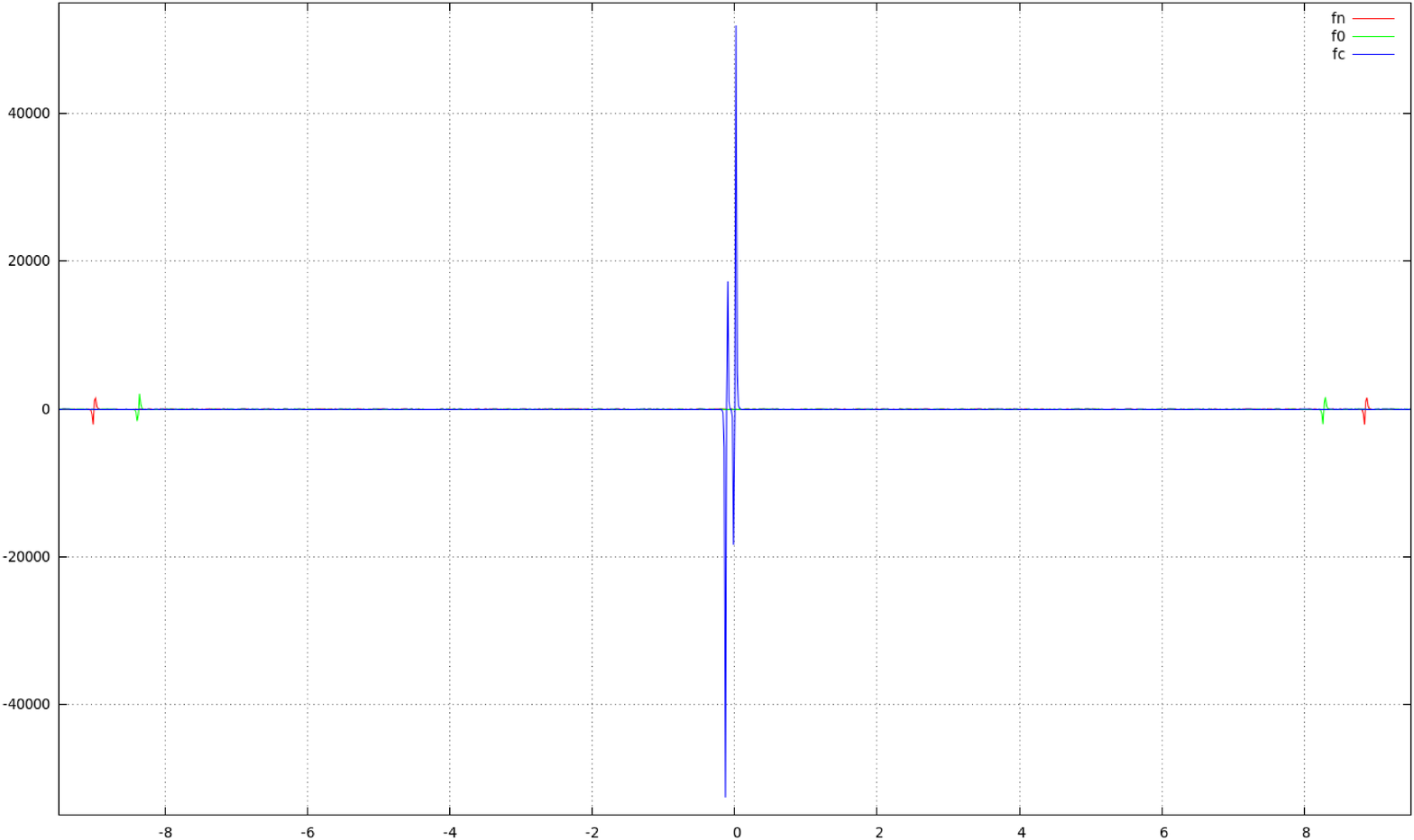}
\includegraphics[width=8cm,scale=4.5, angle=0, height=4.5cm]{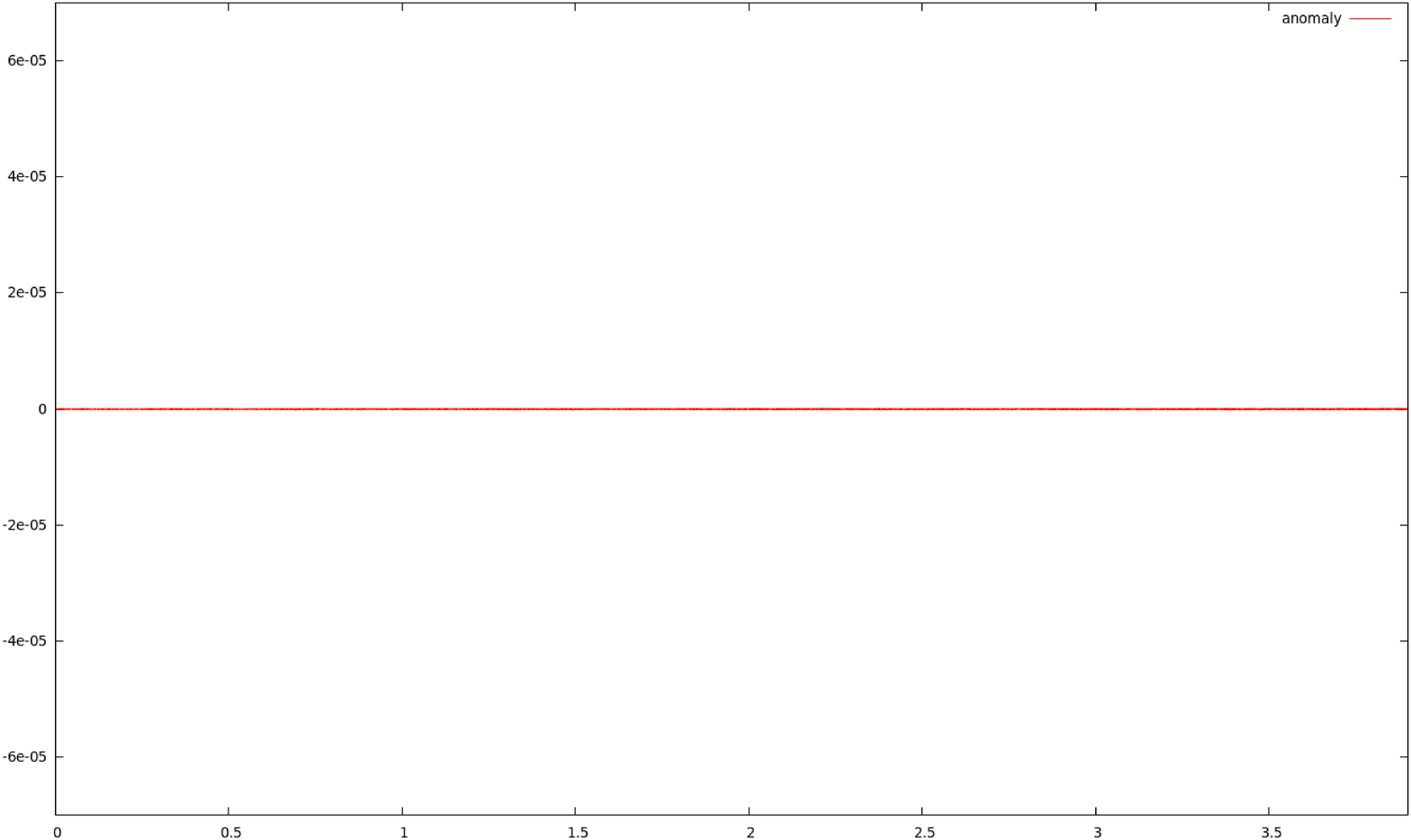}
\includegraphics[width=8cm,scale=4.5, angle=0, height=4.5cm]{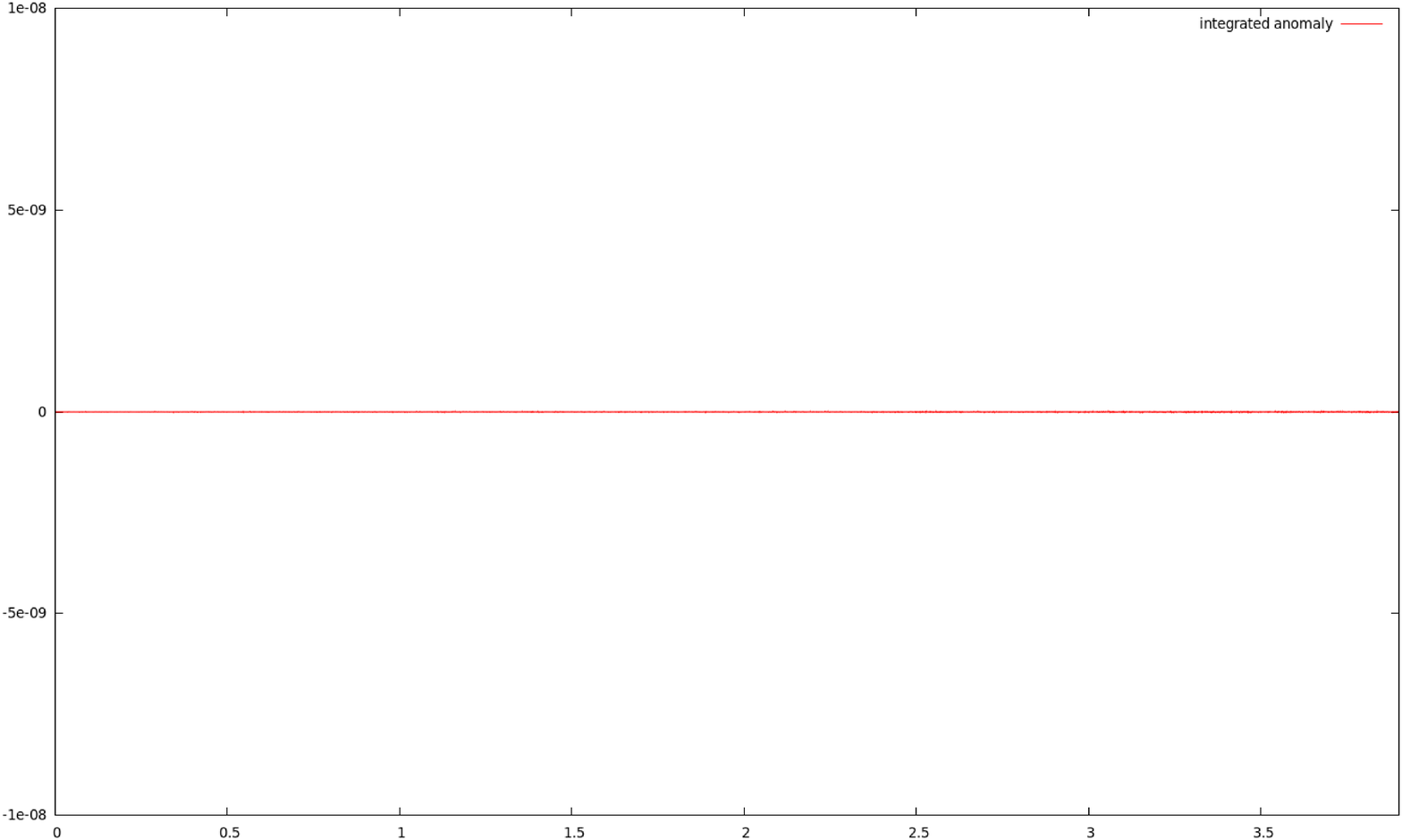}   
\parbox{5in}{\caption {(color online) Top left: reflection  of two dark solitons of the model  (\ref{enns1}) with $q=2$ and equal velocities is plotted for $\epsilon=0.06,\,|\psi_0|=6,\,\eta = 2.5$. The initial gray solitons ($t_i$=green line) travel in opposite direction with velocities $v_1 = -3.7 \sqrt{2}$ (right soliton),\,$v_2 = 3.7 \sqrt{2}$ (left soliton). They partially  overlap ($t_c$=blue line) and then reflect to each other. The gray solitons after collision are plotted as a red line ($t_f$). Note that  $|v_1| + v_2  < v_s$ ($v_s = 10.1 \sqrt{2}$). Top right: the integrand $\g(x,t)$ of the anomaly plotted for three successive times ($t_i$, $t_c$ and $t_f$). Bottom: the anomaly $\beta^{(4)}_r(t)$ and time integrated anomaly $\int^t dt' \beta^{(4)}_r(t')$, respectively.}}
\end{figure}

\begin{figure}
\centering
\label{fig9}
\includegraphics[width=8cm,scale=4.5, angle=0, height=4.5cm]{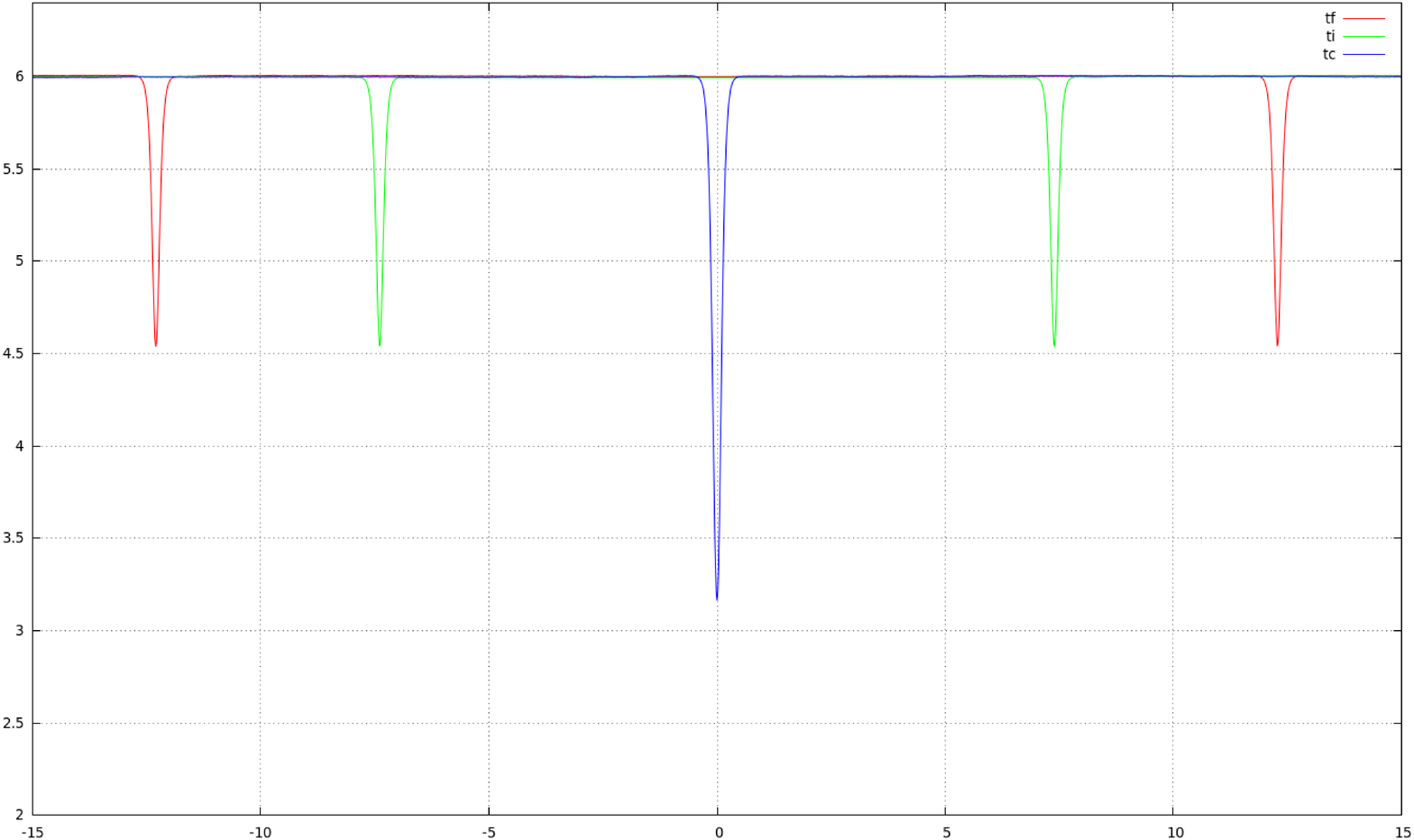} 
\includegraphics[width=8cm,scale=4.5, angle=0, height=4.5cm]{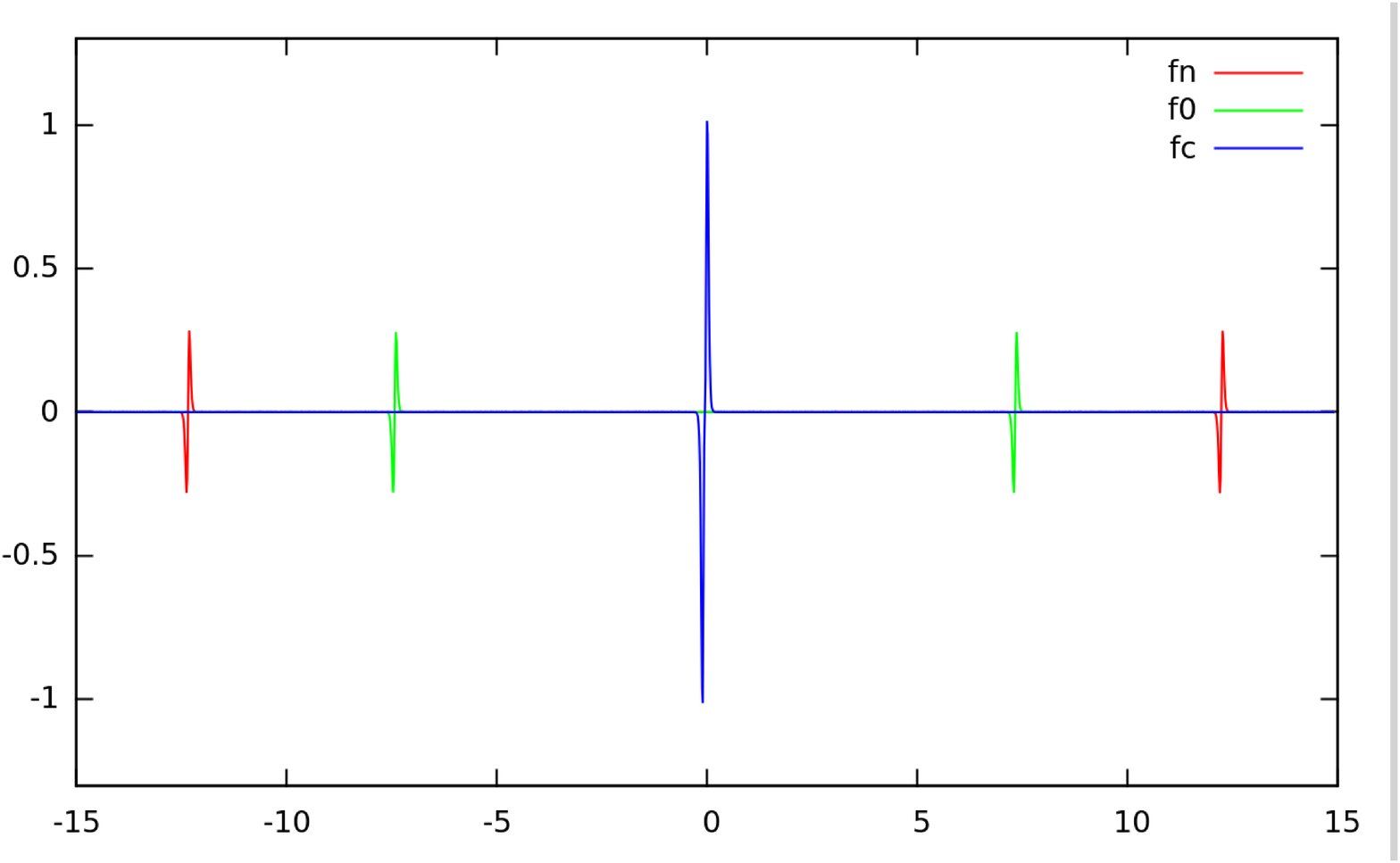}
\includegraphics[width=8cm,scale=4.5, angle=0, height=4.5cm]{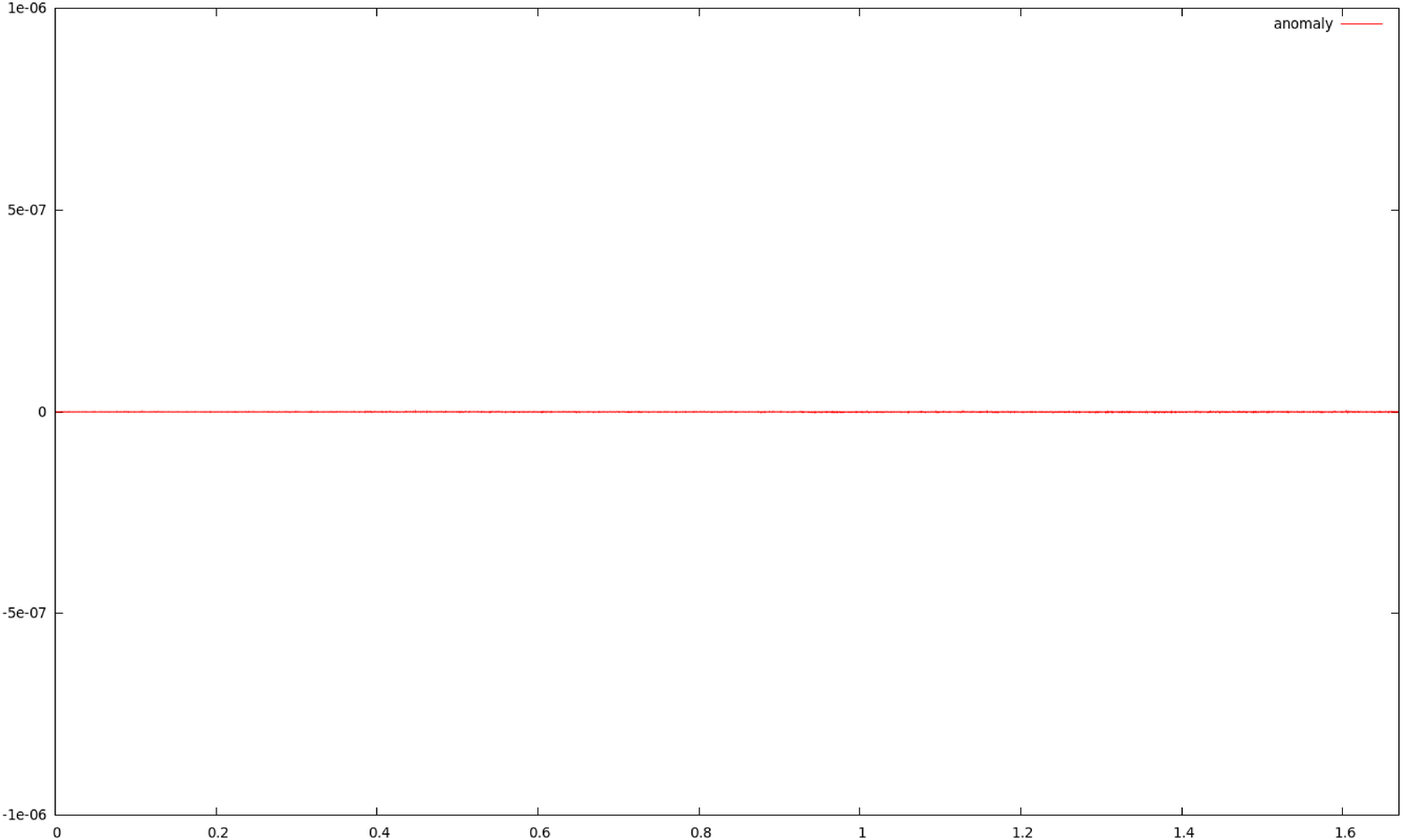}
\includegraphics[width=8cm,scale=4.5, angle=0, height=4.5cm]{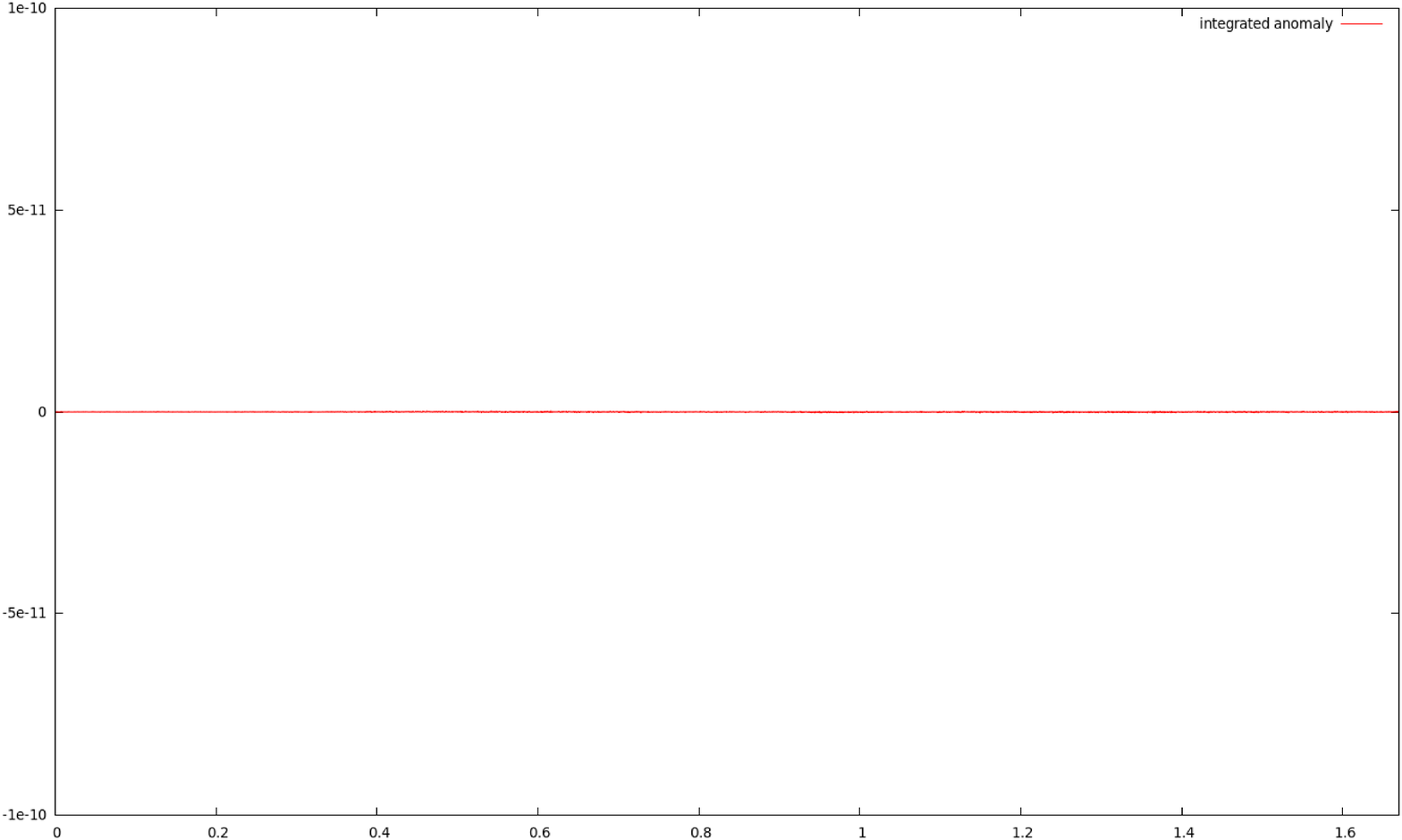}   
\parbox{5in}{\caption {(color online) Top left: transmission   of two dark solitons of the model  (\ref{enns1}) with $q=3$ and equal velocities are plotted for $\epsilon=0.1,\,|\psi_0|=6,\,\eta = 2.5$. The initial gray solitons ($t_i$=green line) travel in opposite direction with velocities $v_1 =-6.42\sqrt{2}$ (right soliton),\,$v_2 = 6.42 \sqrt{2}$ (left soliton). They completely  overlap ($t_c$=blue line) and then transmit to each other. The gray solitons after collision are plotted as a red line ($t_f$). Note that  $|v_1| + v_2  > v_s$ ($v_s = 7.1 \sqrt{2}$). Top right: the integrand $\g(x,t)$ of the anomaly plotted for three successive times ($t_i$, $t_c$ and $t_f$). Bottom:the anomaly $\beta^{(4)}_r(t)$ and time integrated anomaly $\int^t dt' \beta^{(4)}_r(t')$, respectively.}}
\end{figure}

\begin{figure}
\centering
\label{fig10}
\includegraphics[width=8cm,scale=5, angle=0, height=5cm]{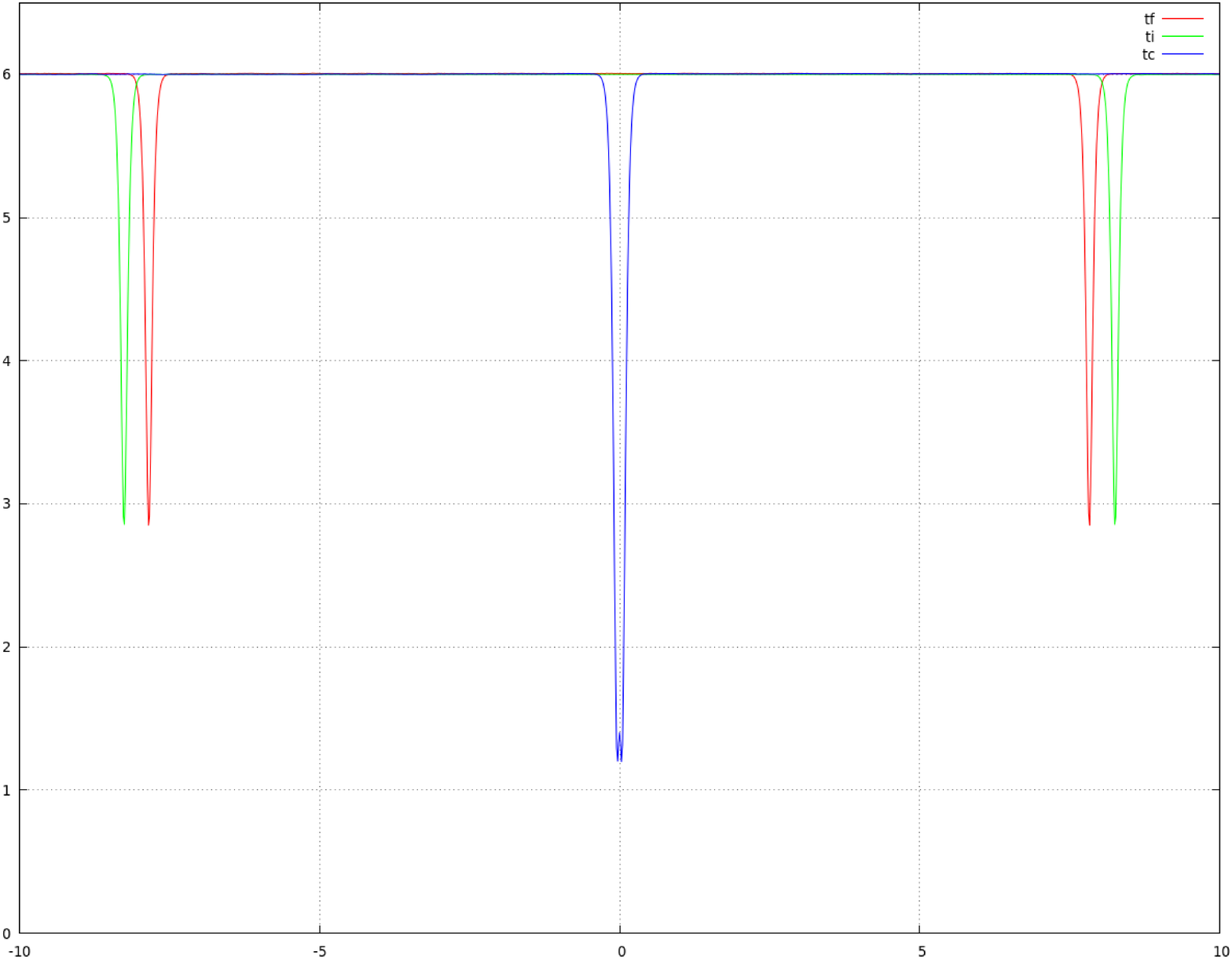} 
\includegraphics[width=8cm,scale=5, angle=0, height=5cm]{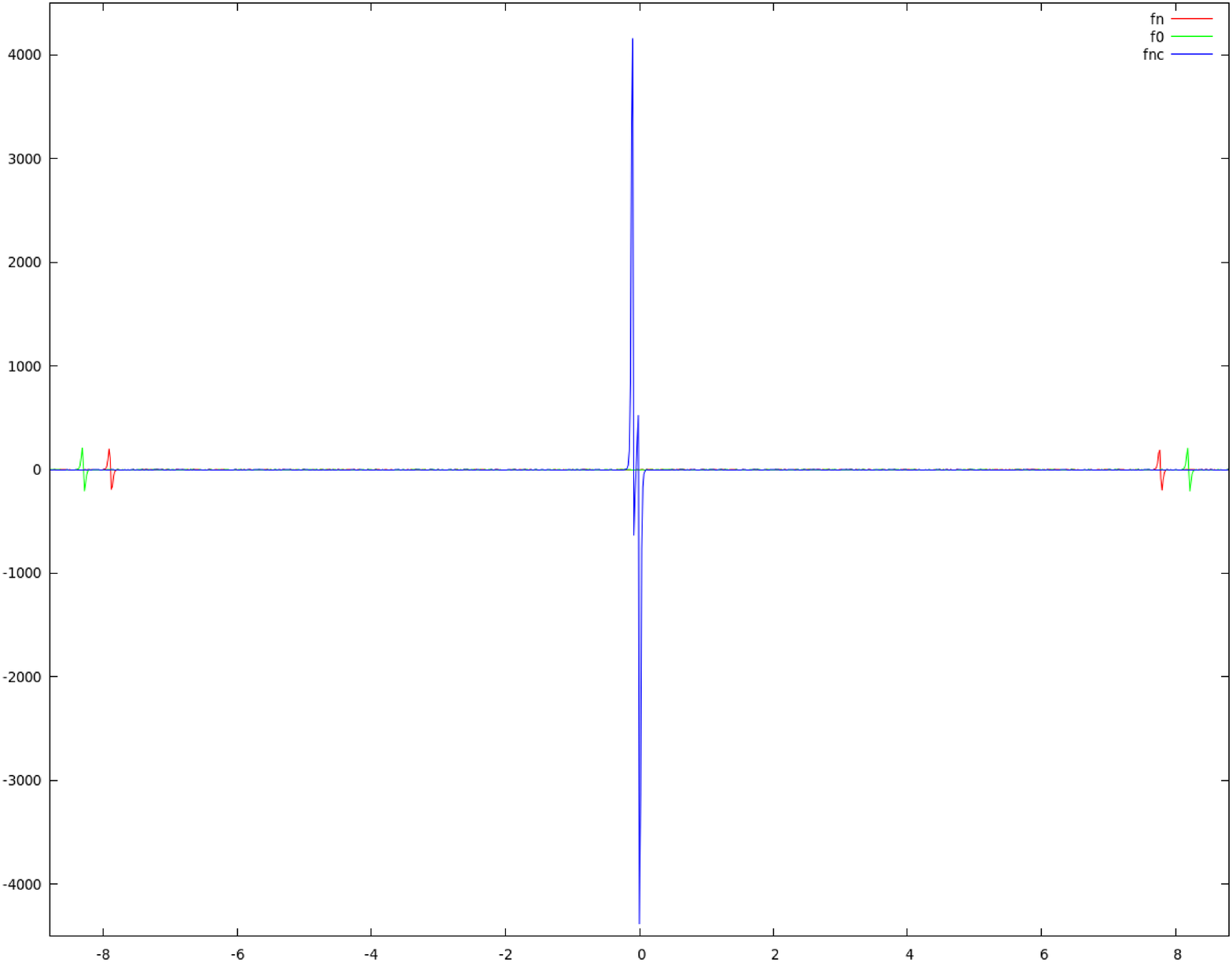}
\includegraphics[width=8cm,scale=5, angle=0, height=5cm]{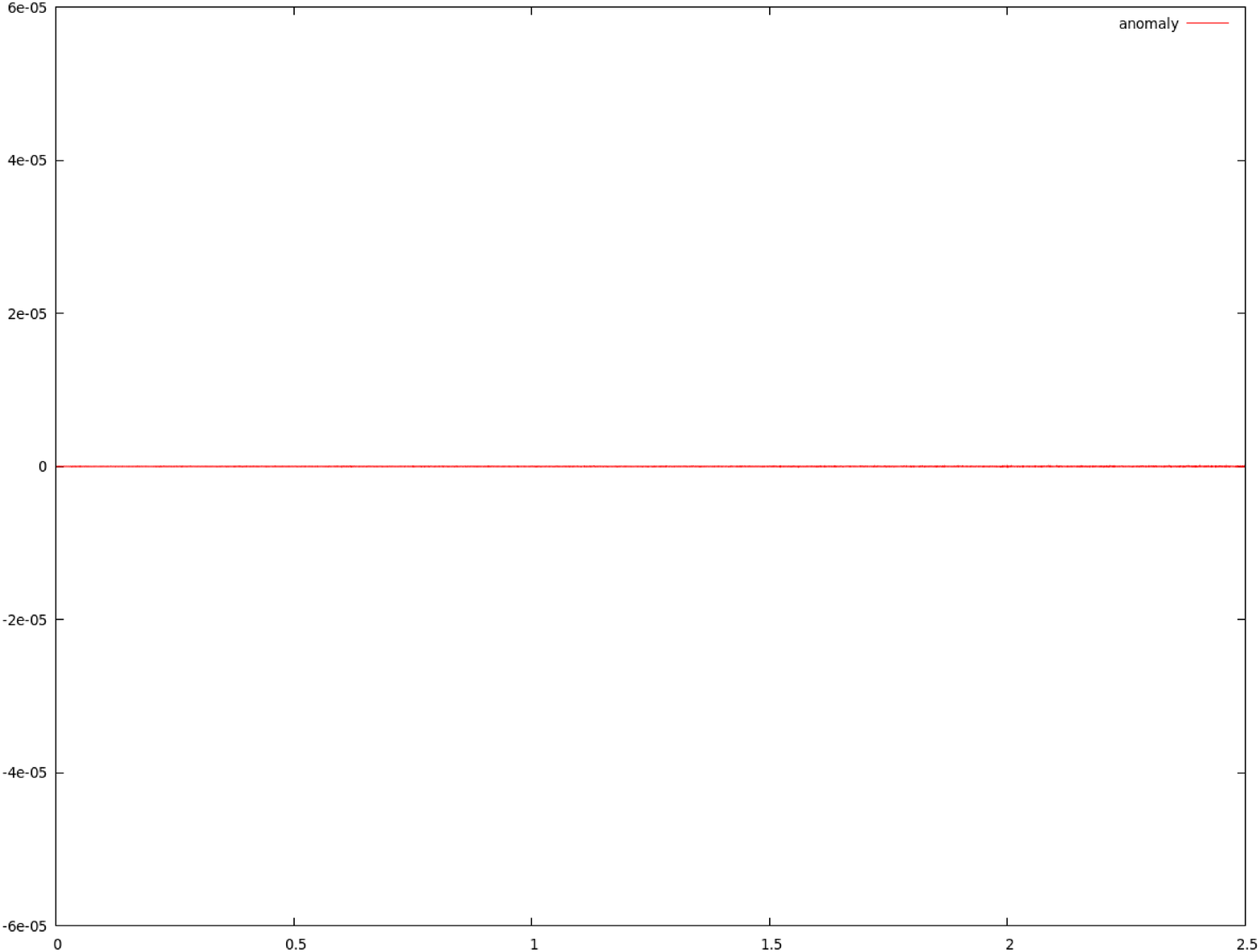}
\includegraphics[width=8cm,scale=5, angle=0, height=5cm]{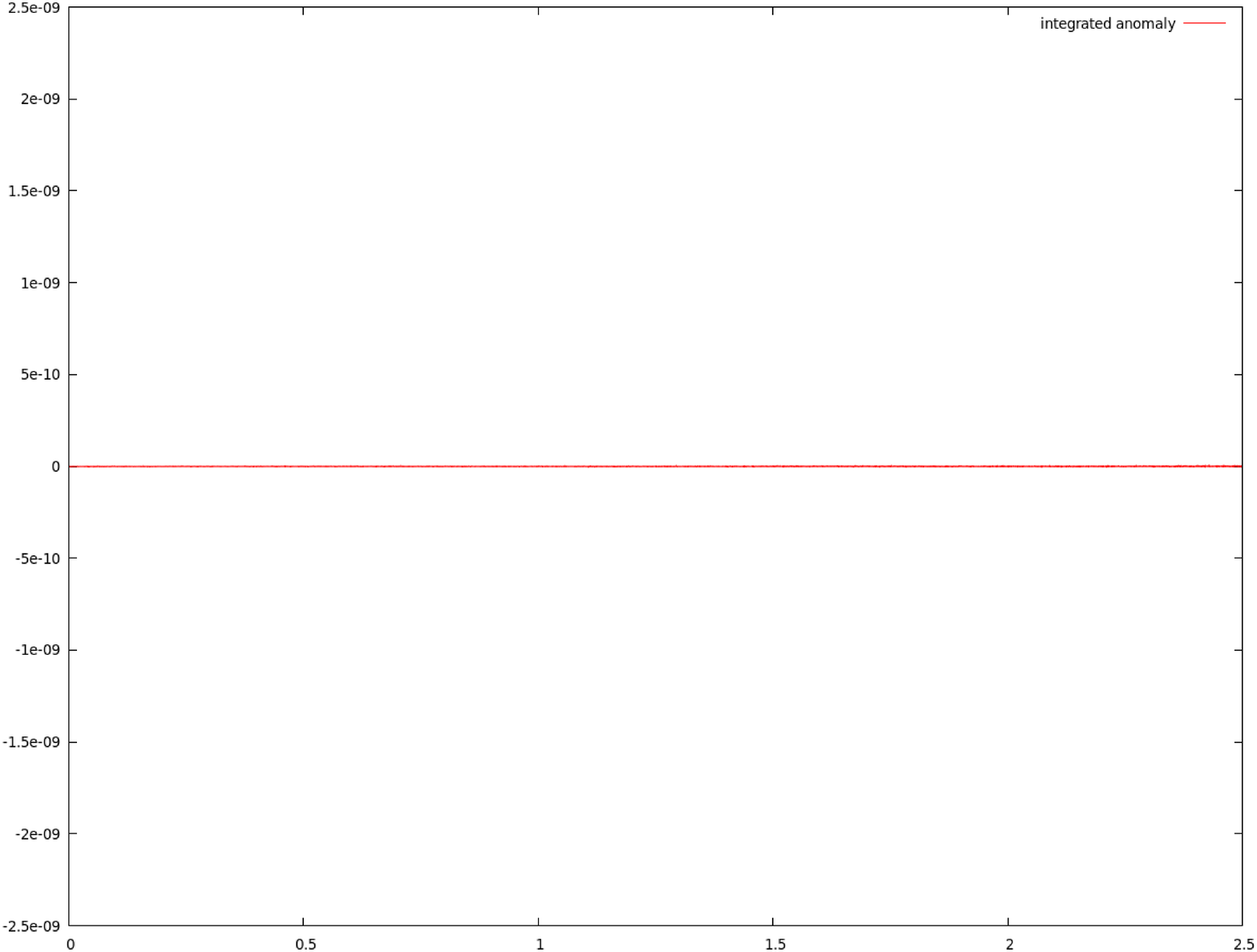}   
\parbox{5in}{\caption {(color online) Top left: reflection  of two dark solitons of the model  (\ref{enns1}) with $q=2$ and equal velocities is plotted for $\epsilon=-0.06,\,|\psi_0|=6,\,\eta = 2.5$. The initial gray solitons ($t_i$=green line) travel in opposite direction with velocities $v_1 =-7.16\sqrt{2}$ (right soliton),\,$v_2 = 7.16 \sqrt{2}$ (left soliton). They partially  overlap ($t_c$=blue line) and then reflect to each other. The gray solitons after collision are plotted as a red line ($t_f$). Note that  $|v_1| + v_2  < v_s$ ($v_s = 16 \sqrt{2}$). Top right: the integrand $\g(x,t)$ of the anomaly plotted for three successive times ($t_i$, $t_c$ and $t_f$). Bottom: the anomaly $\beta^{(4)}_r(t)$ and time integrated anomaly $\int^t dt' \beta^{(4)}_r(t')$, respectively.}}
\end{figure}

\begin{figure}
\centering
\label{fig11}
\includegraphics[width=8cm,scale=5, angle=0, height=5cm]{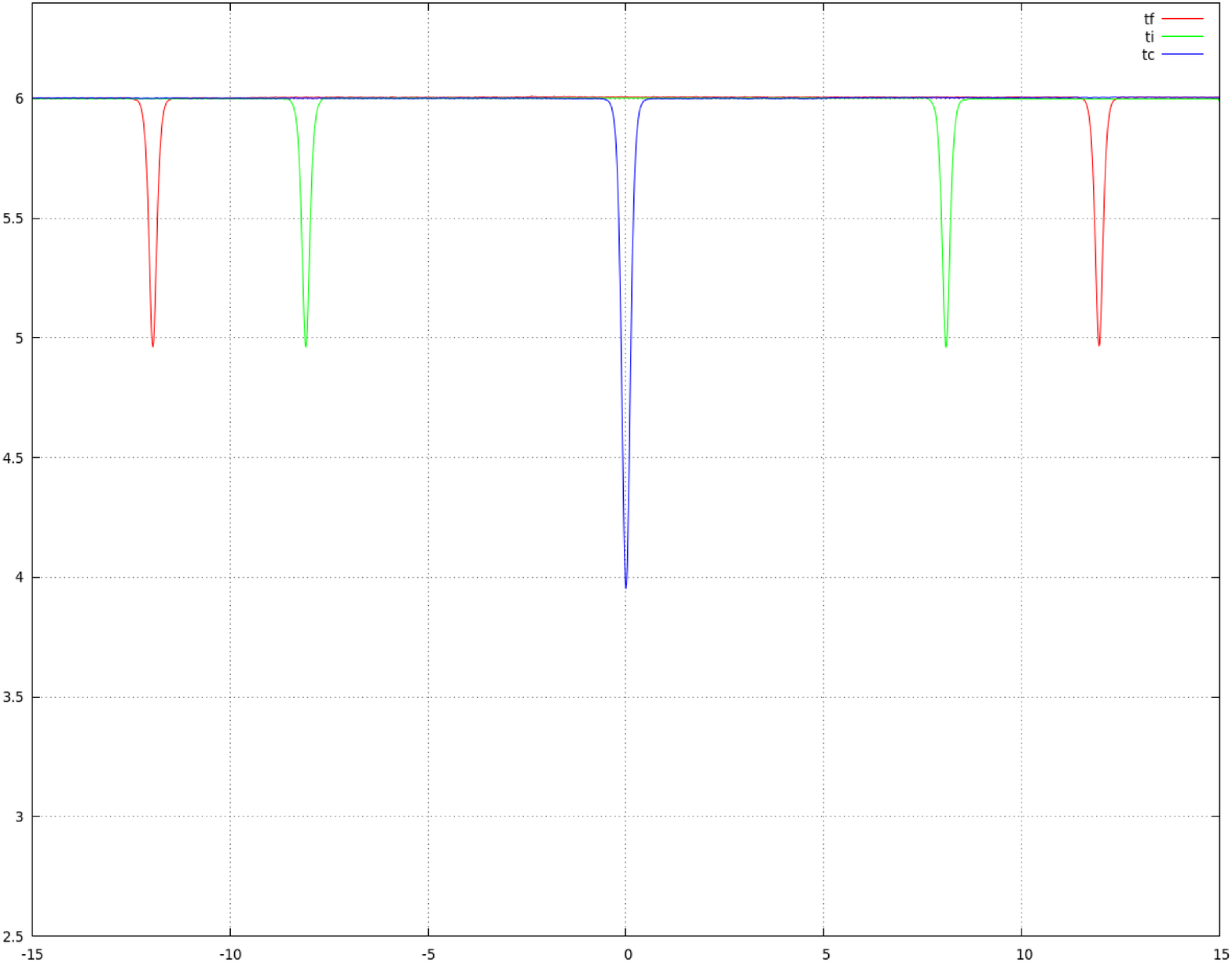} 
\includegraphics[width=8cm,scale=5, angle=0, height=5cm]{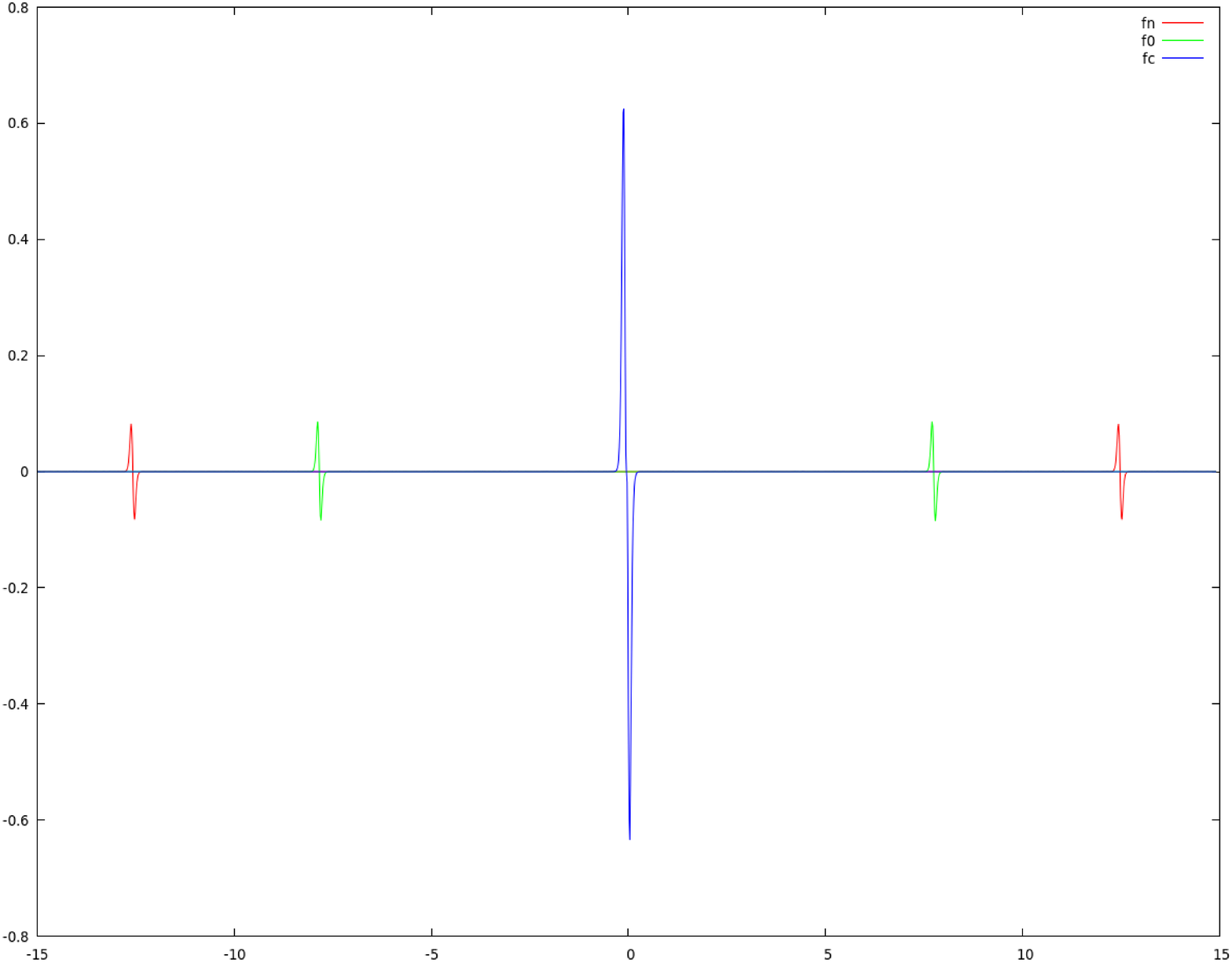}
\includegraphics[width=8cm,scale=5, angle=0, height=5cm]{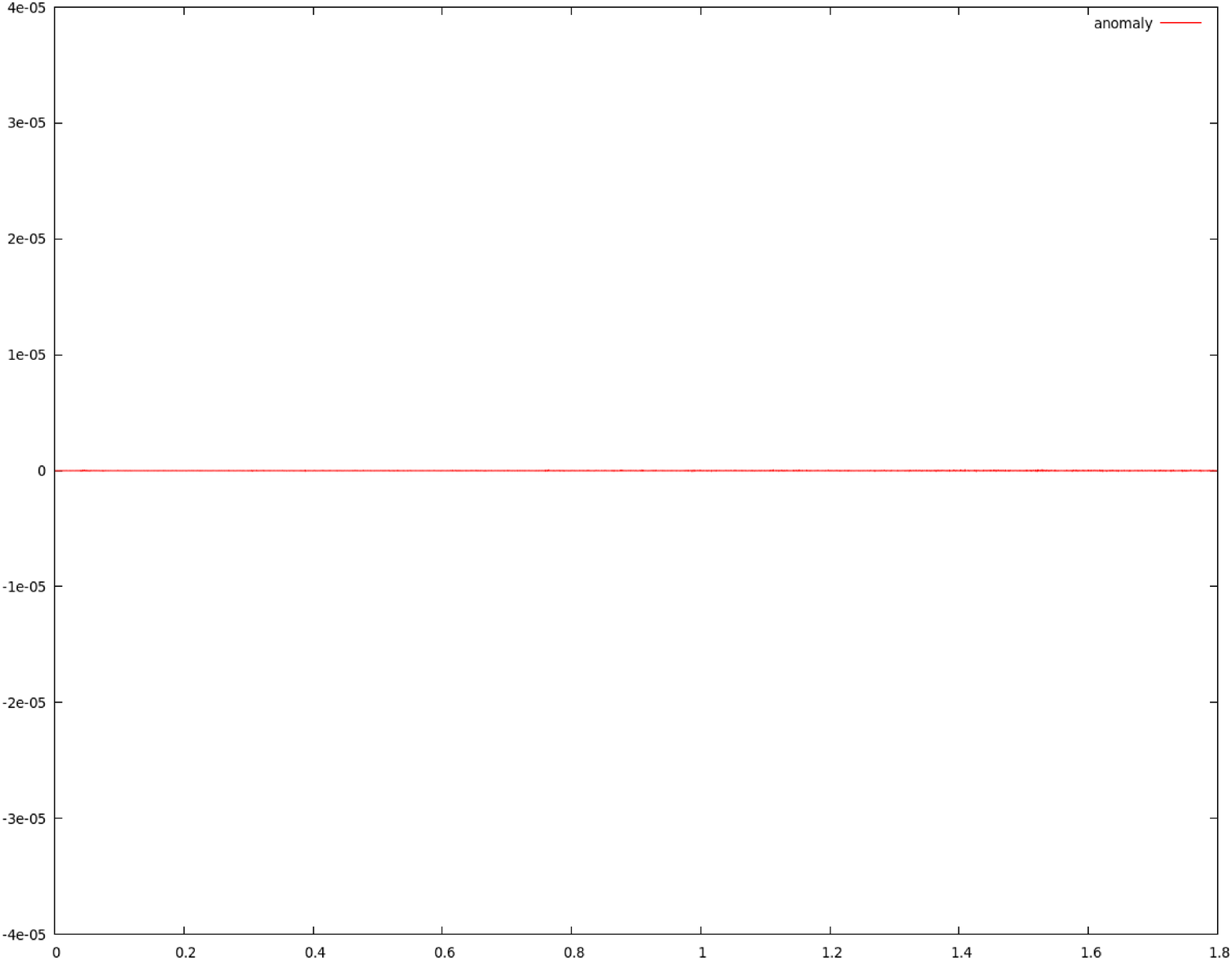}
\includegraphics[width=8cm,scale=5, angle=0, height=5cm]{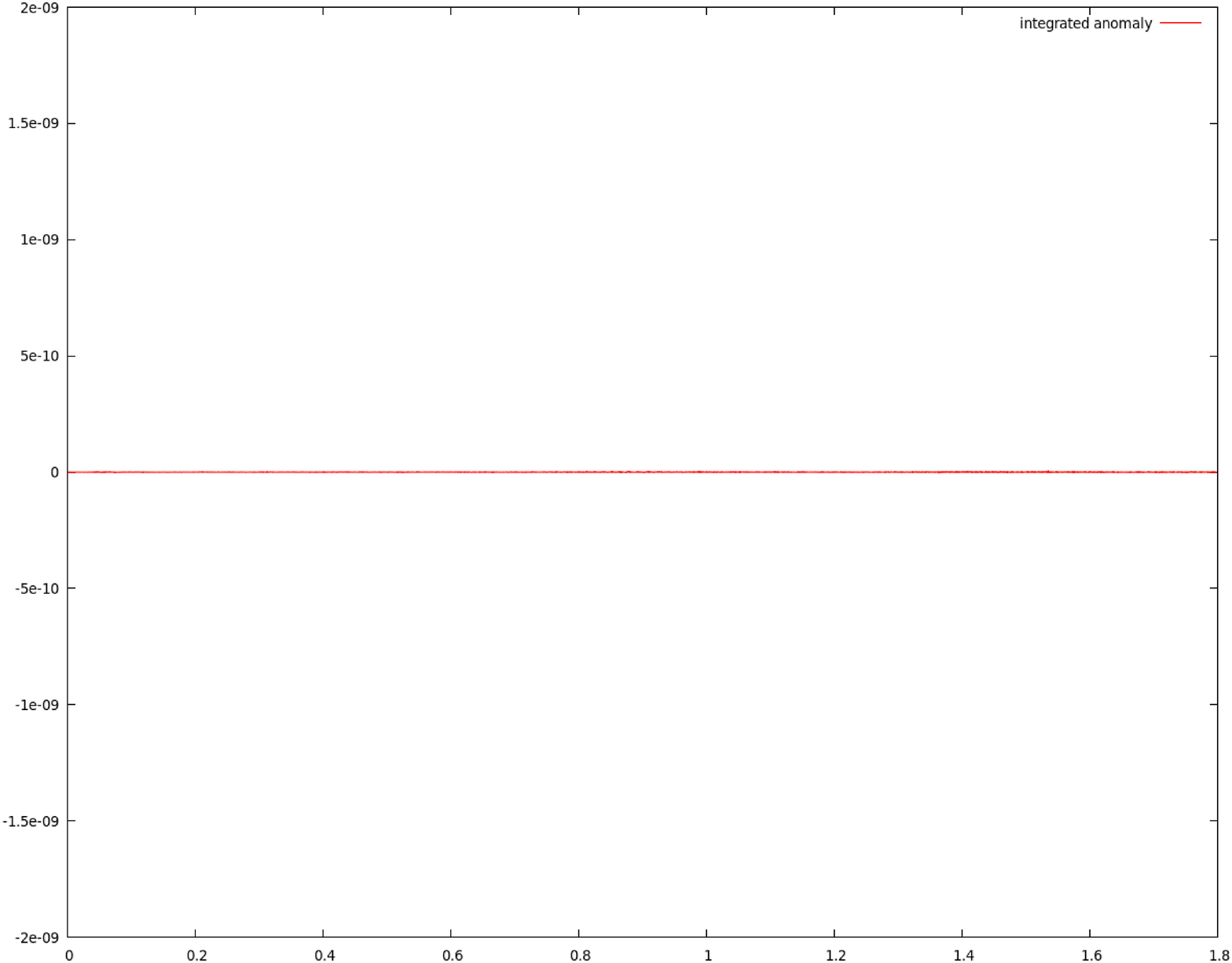}   
\parbox{5in}{\caption {(color online) Top left: transmission  of two dark solitons of the model (\ref{enns1}) with $q=2$ and equal velocities is plotted for $\epsilon=-0.06,\,|\psi_0|=6,\,\eta = 2.5$. The initial gray solitons ($t_i$=green line) travel in opposite direction with velocities $v_1 =-12.97\sqrt{2}$ (right soliton),\,$v_2 = 12.97 \sqrt{2}$ (left soliton). They completely  overlap ($t_c$=blue line) and then transmit to each other. The gray solitons after collision are plotted as a red line ($t_f$). Note that  $|v_1| + v_2  > v_s$ ($v_s = 16 \sqrt{2}$). Top right: the integrand $\g(x,t)$ of the anomaly plotted for three successive times ($t_i$, $t_c$ and $t_f$). Bottom: the anomaly $\beta^{(4)}_r(t)$ and time integrated anomaly $\int^t dt' \beta^{(4)}_r(t')$, respectively.}}
\end{figure}

\begin{figure}
\centering
\label{fig12}
\includegraphics[width=6cm,scale=4, angle=0, height=4cm]{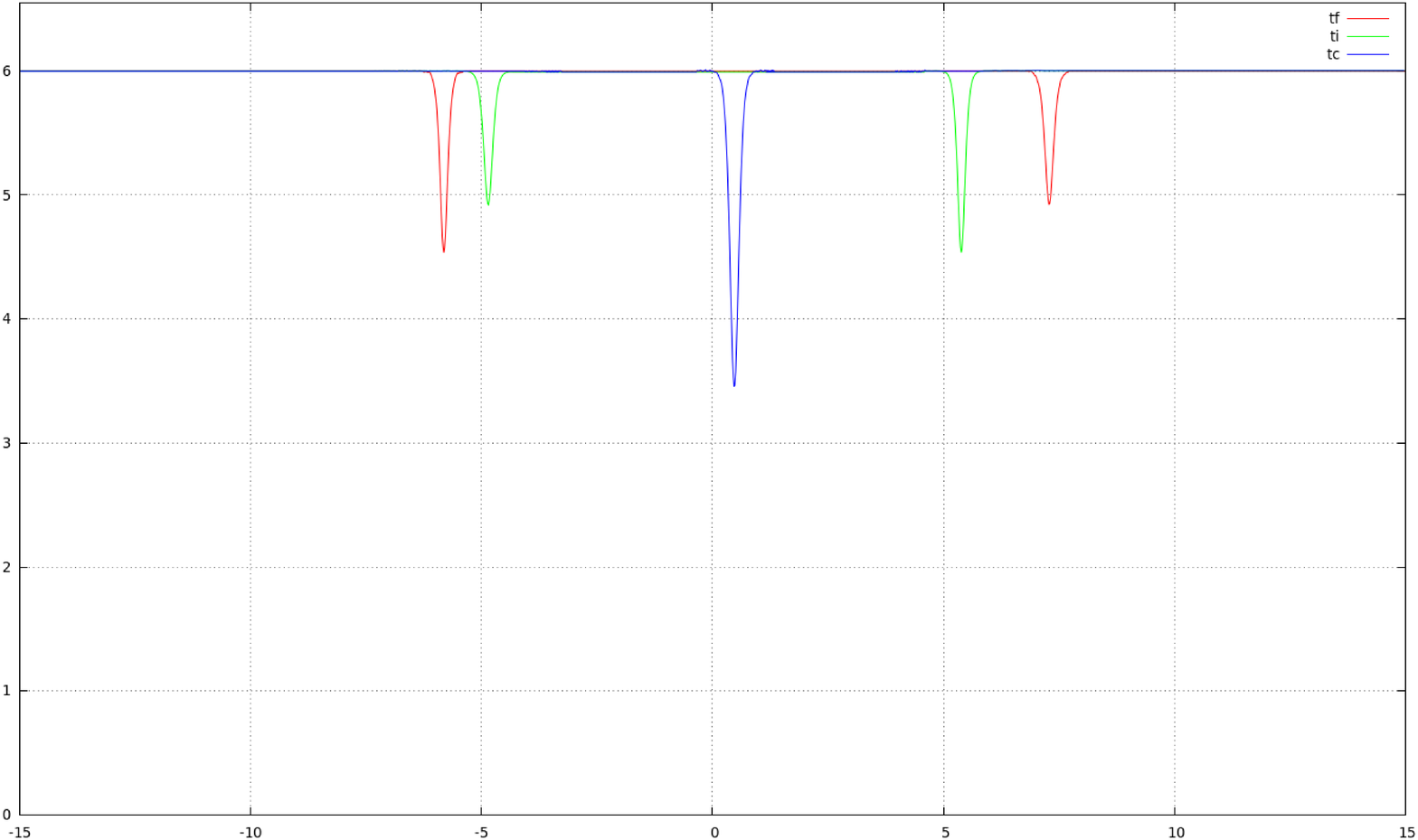} 
\includegraphics[width=6cm,scale=4, angle=0, height=4cm]{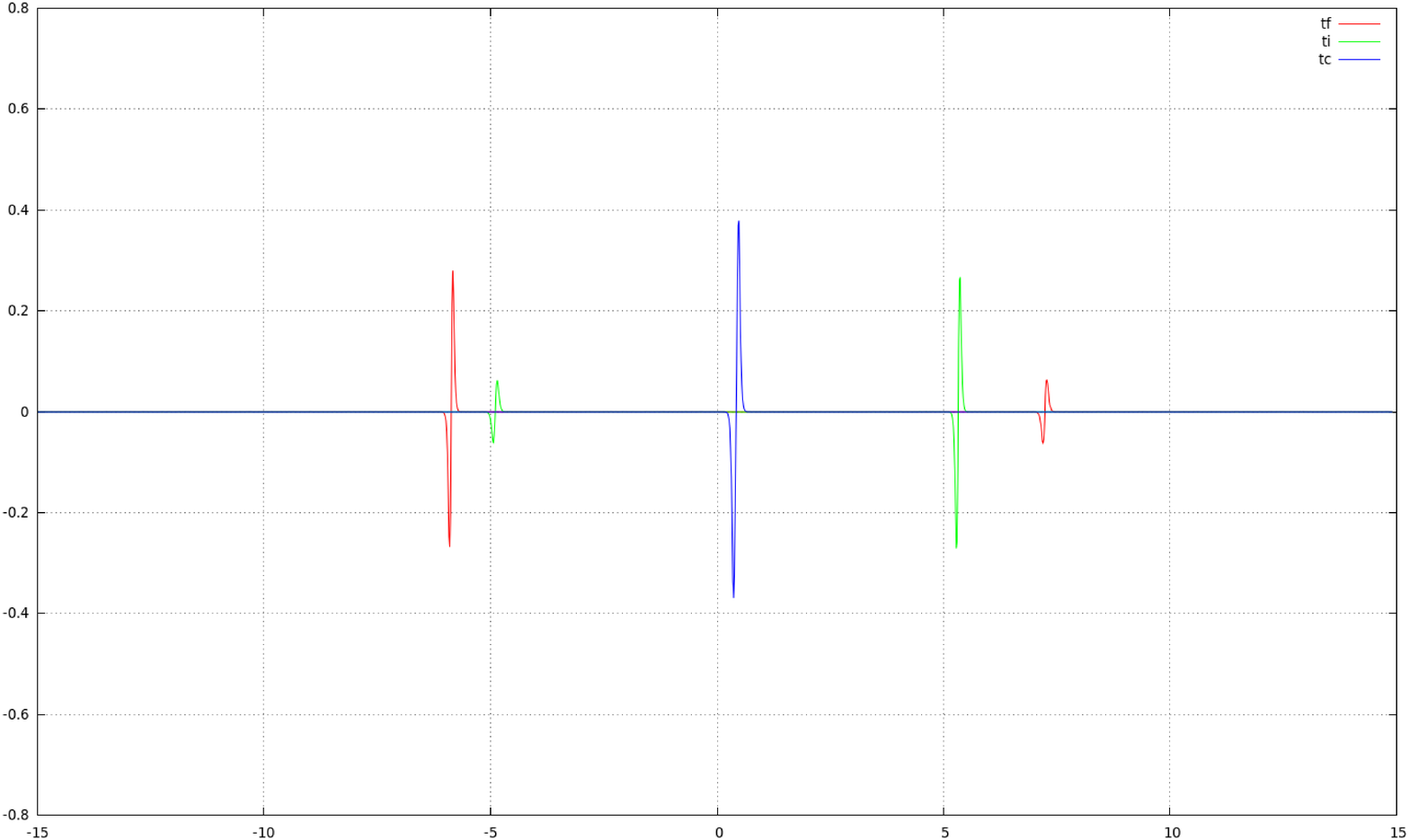}
\includegraphics[width=8cm,scale=5, angle=0, height=4cm]{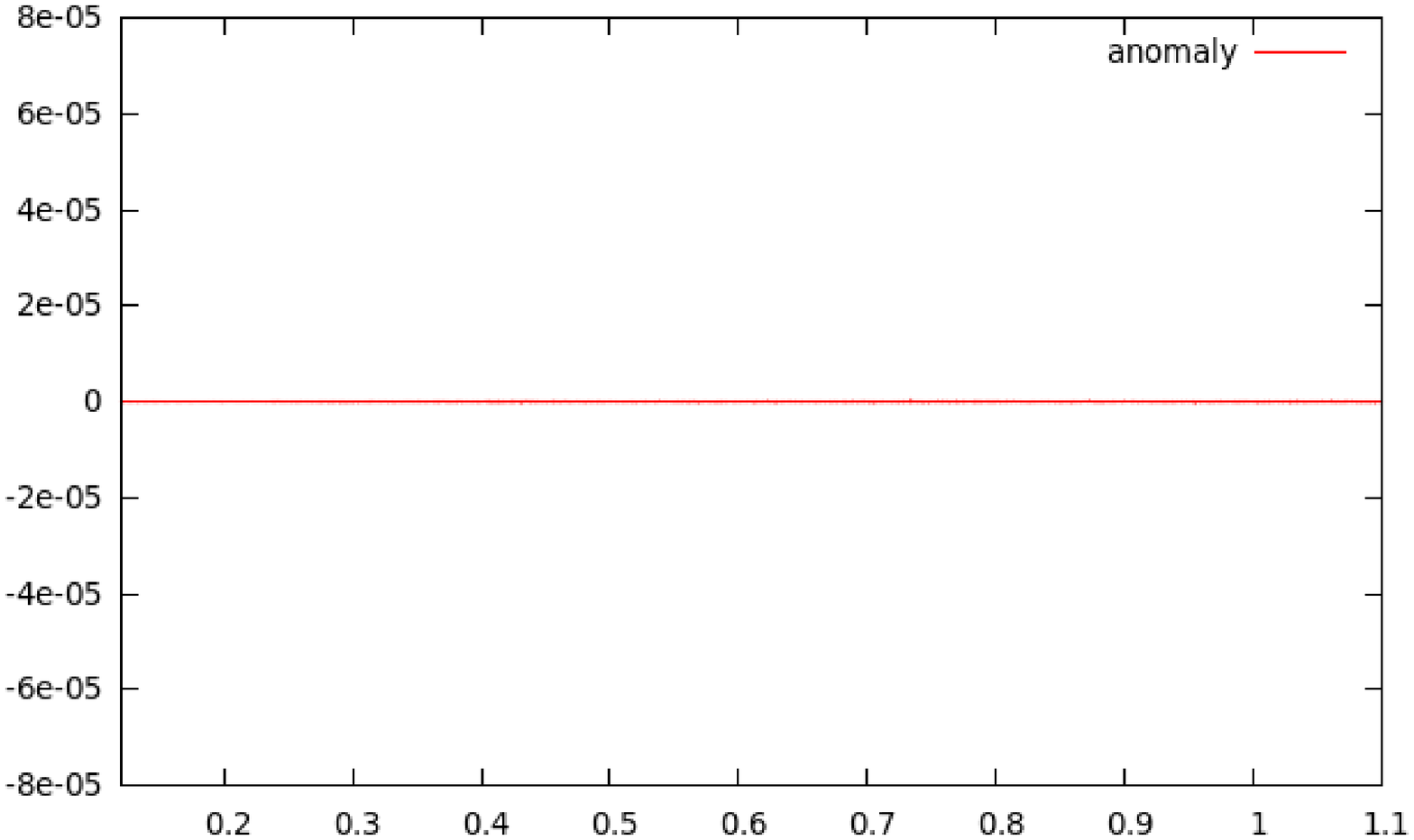}
\includegraphics[width=8cm,scale=5, angle=0, height=4cm]{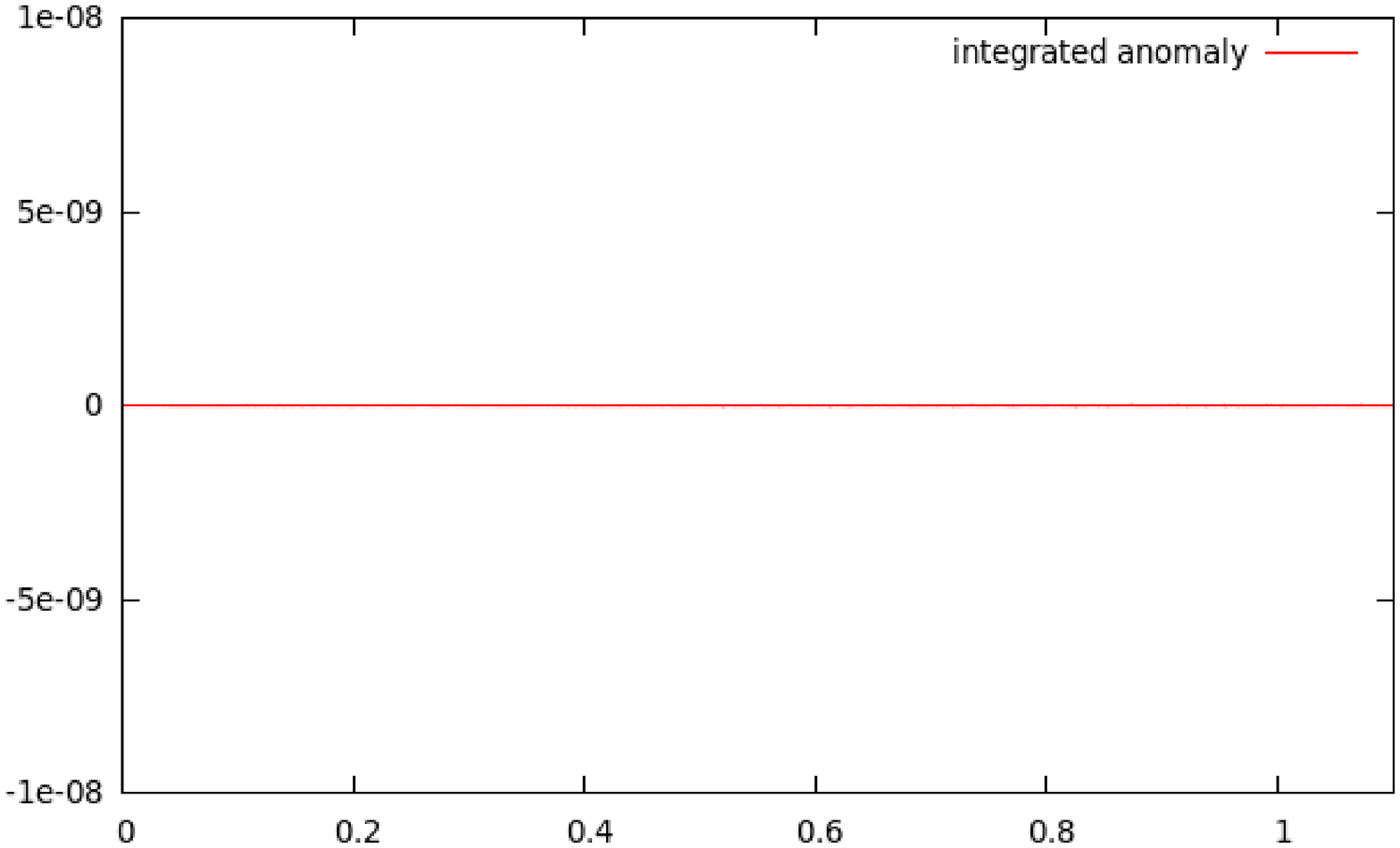}   
\parbox{5in}{\caption {(color online) Top left: transmission  of two dark solitons of the model (\ref{enns1}) with $q=3$ and different velocities (different amplitudes) is plotted for $\epsilon=0.1,\,|\psi_0|=6,\,\eta = 2.5$. The initial gray solitons ($t_i$=green line) travel in opposite direction with velocities $v_1 = -6.42 \sqrt{2}$ (right soliton),\,$v_2 = 6.7 \sqrt{2}$ (left soliton). They completely  overlap ($t_c$=blue line) and then transmit to each other. The gray solitons after collision are plotted as a red line ($t_f$). Note that  $|v_1| + v_2 >  v_s$ ($v_s = 7.1\sqrt{2}$). Top right: the integrand $\g(x,t)$ of the anomaly plotted for three successive times ($t_i$, $t_c$ and $t_f$). Bottom: the anomaly $\beta^{(4)}_r(t)$ and time integrated anomaly $\int^t dt' \beta^{(4)}_r(t')$, respectively.}}
\end{figure}

\begin{figure}
\centering
\label{fig13}
\includegraphics[width=8cm,scale=4, angle=0, height=5cm]{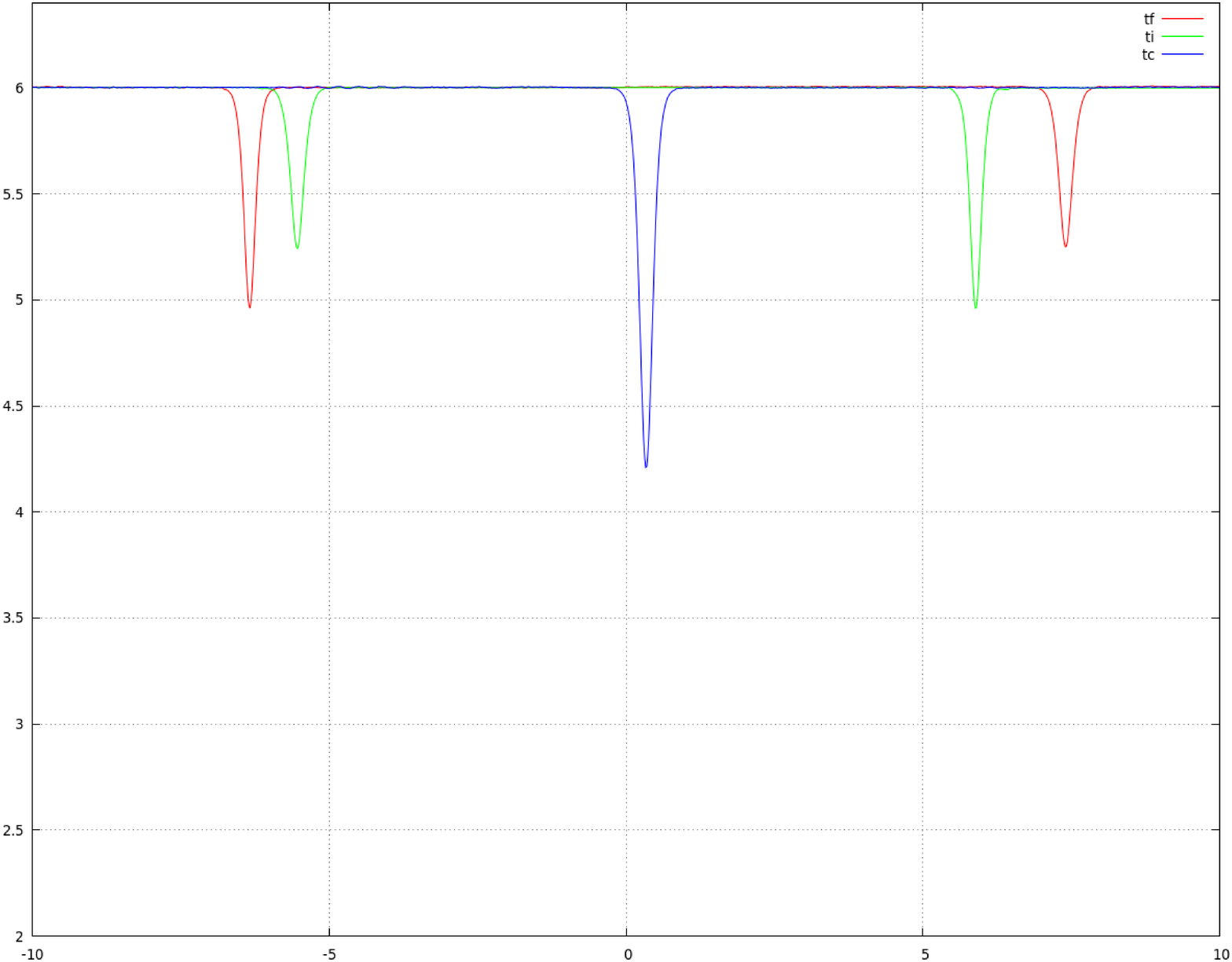} 
\includegraphics[width=8cm,scale=4, angle=0, height=5cm]{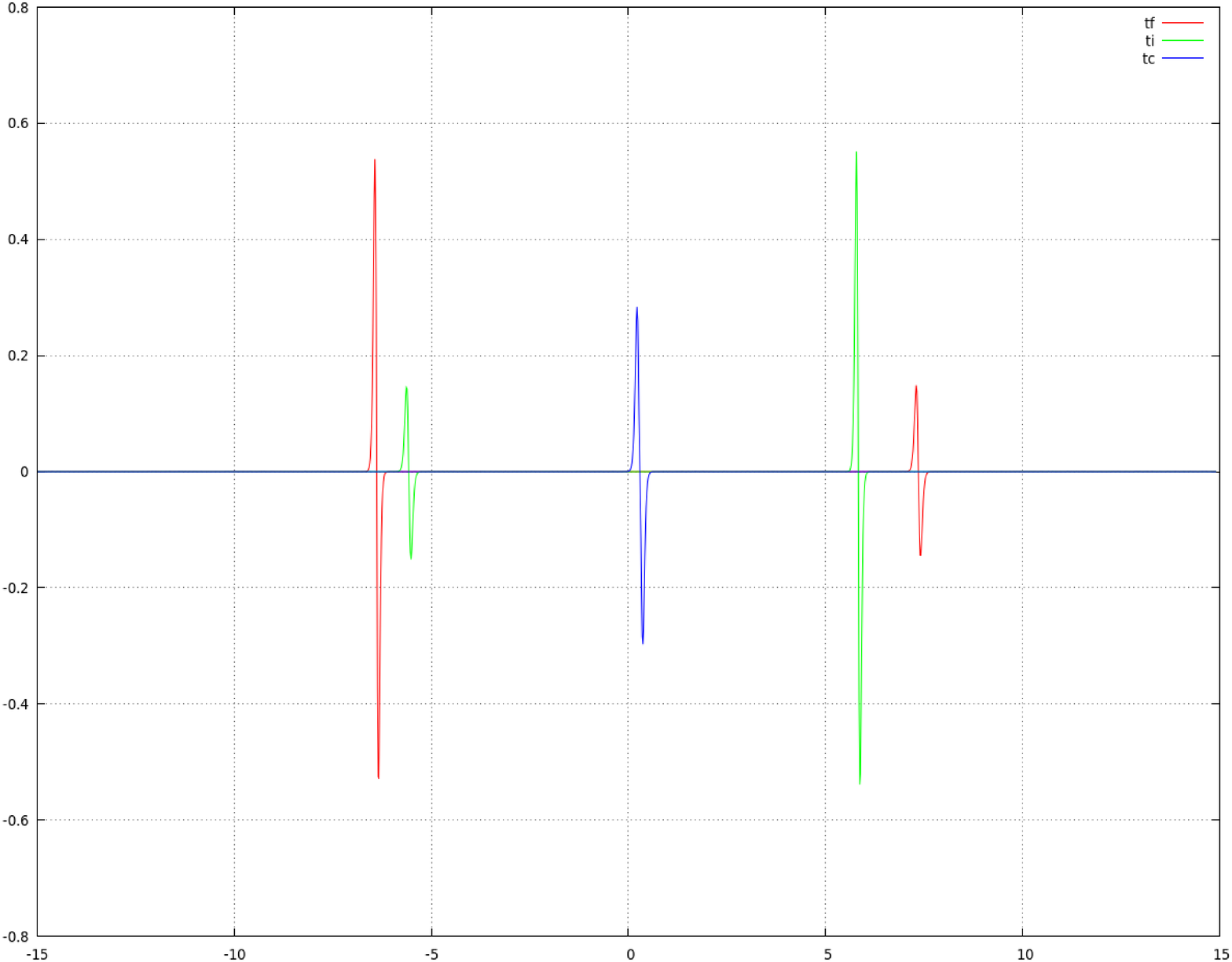}
\includegraphics[width=8cm,scale=5, angle=0, height=5cm]{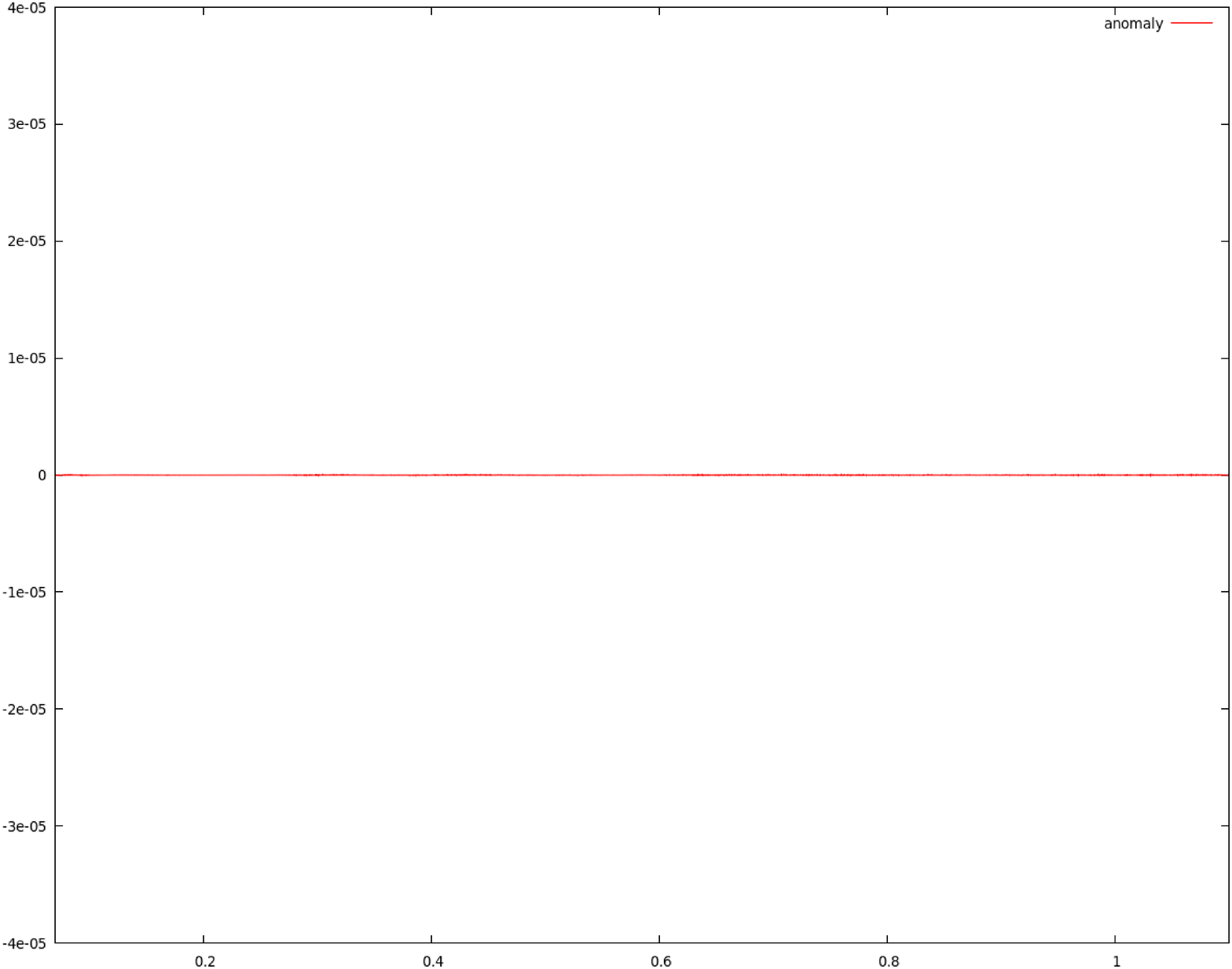}
\includegraphics[width=8cm,scale=5, angle=0, height=5cm]{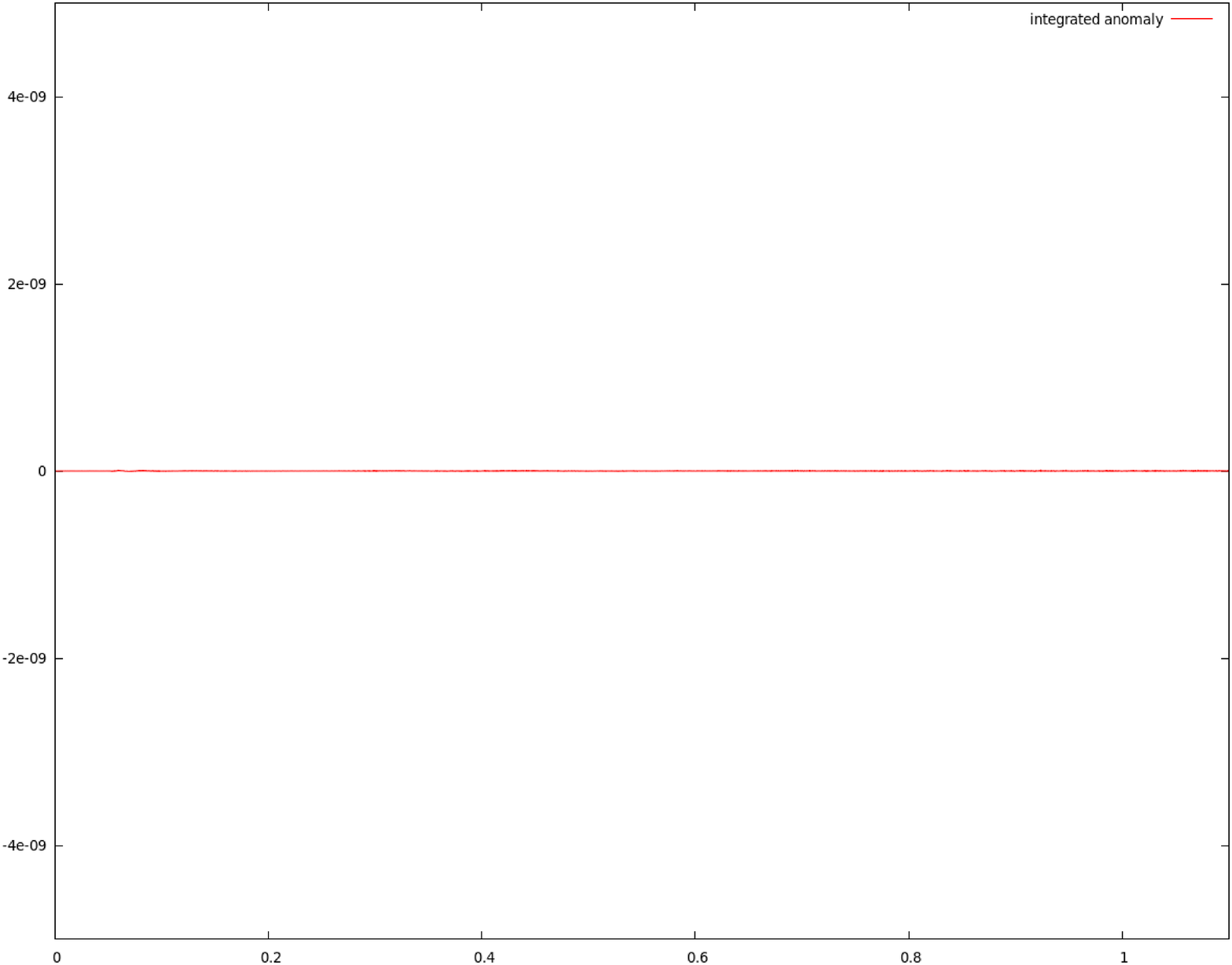}   
\parbox{5in}{\caption {(color online) Top left: transmission  of two dark solitons of the model (\ref{enns1}) with $q=2$ and different velocities (different amplitudes) is plotted for $\epsilon=-0.06,\,|\psi_0|=6,\,\eta = 2.5$. The initial gray solitons ($t_i$=green line) travel in opposite direction with velocities $v_1 = -12.97 \sqrt{2}$ (right soliton),\,$v_2 = 13.8 \sqrt{2}$ (left soliton). They completely  overlap ($t_c$=blue line) and then transmit to each other. The gray solitons after collision are plotted as a red line ($t_f$). Note that  $|v_1| + v_2 >  v_s$ ($v_s=16.0\sqrt{2}$). Top right: the integrand $\g(x,t)$ of the anomaly plotted for three successive times ($t_i$, $t_c$ and $t_f$). Bottom:the anomaly $\beta^{(4)}_r(t)$ and time integrated anomaly $\int^t dt' \beta^{(4)}_r(t')$, respectively.}}
\end{figure}

\begin{figure}
\centering
\label{fig14}
\includegraphics[width=8cm,scale=5, angle=0, height=4cm]{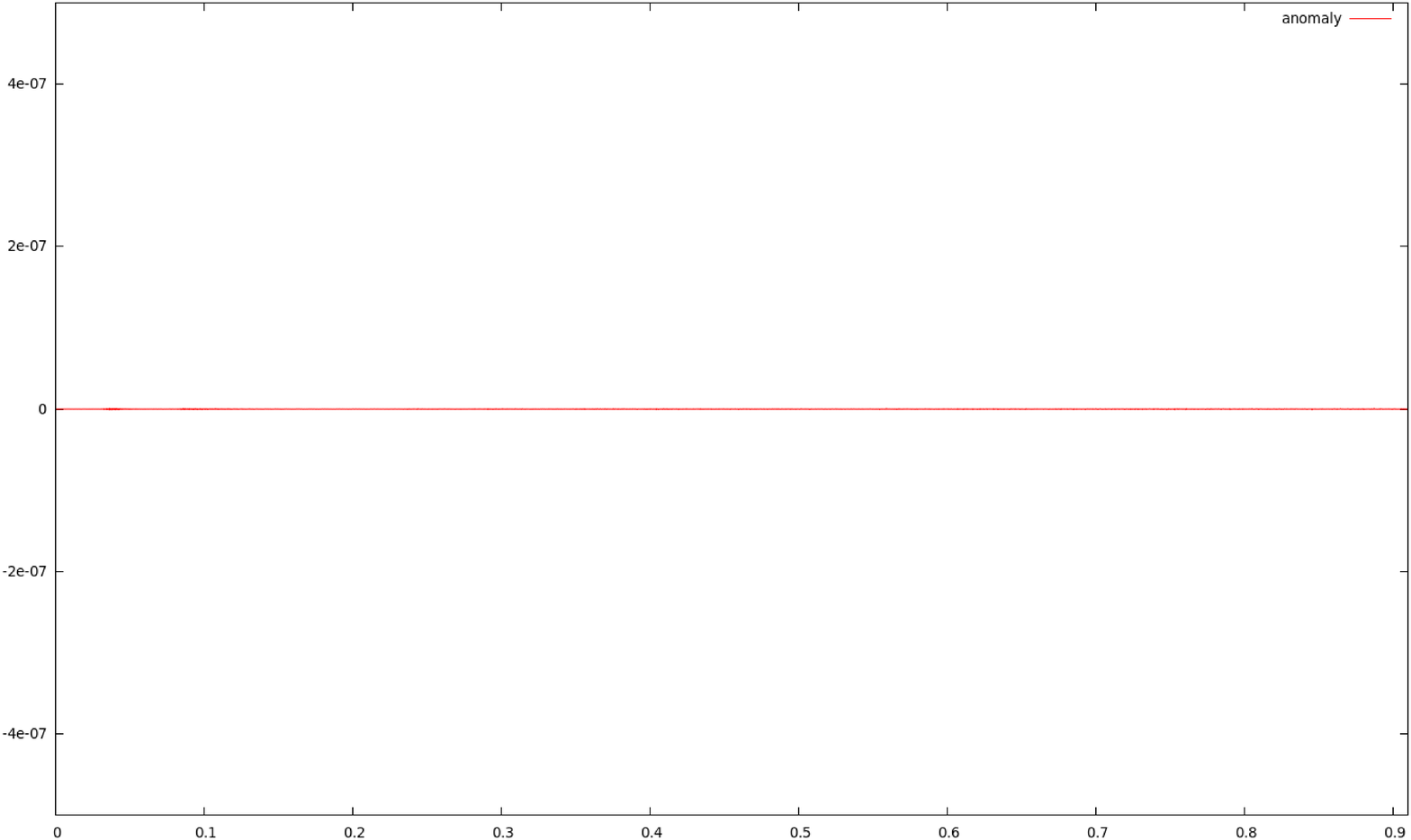}
\includegraphics[width=8cm,scale=5, angle=0, height=4cm]{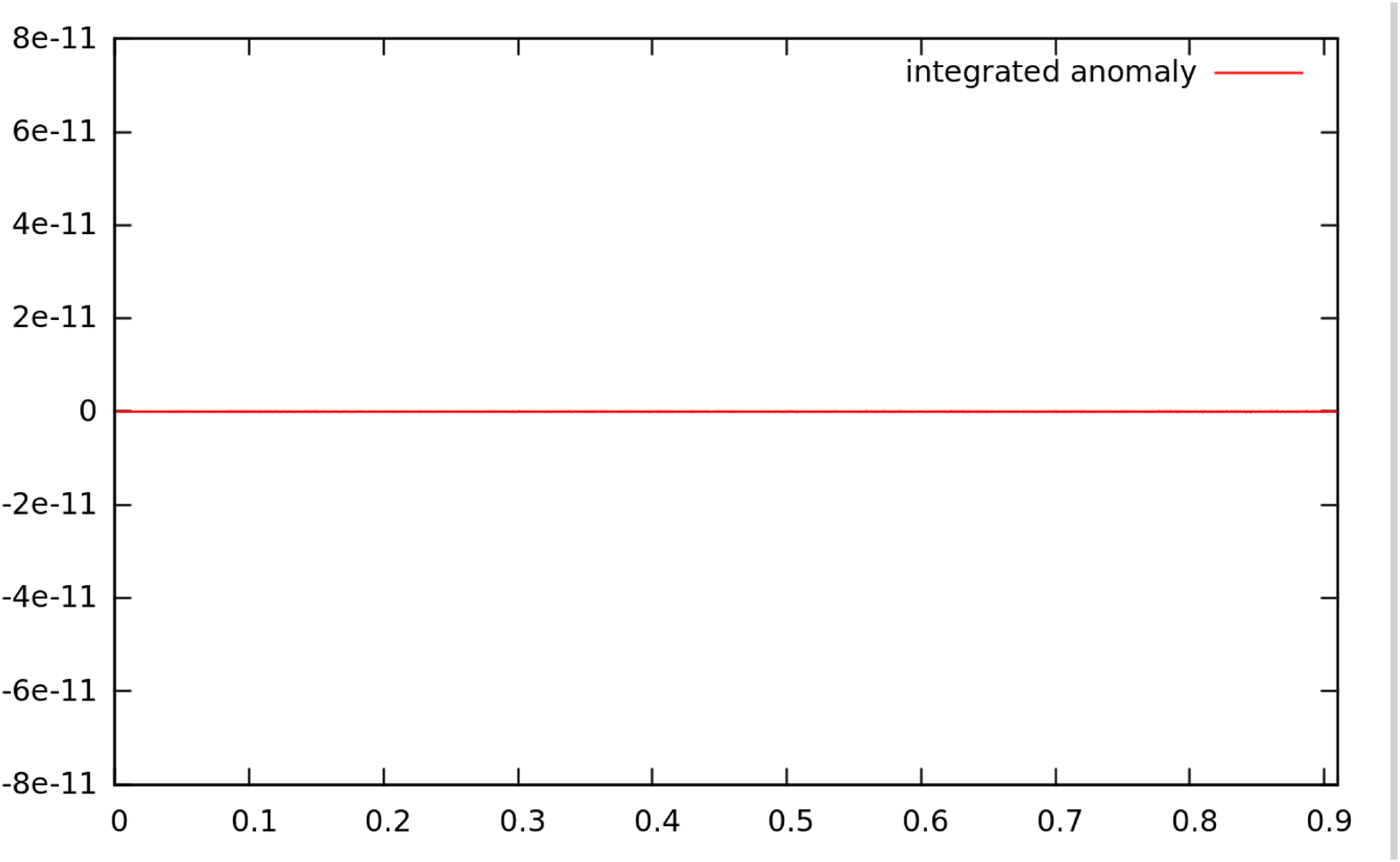}   
\parbox{5in}{\caption {(color online)  
Anomaly $\beta^{(4)}_r(t)$ (a) and time integrated anomaly $\int^t dt' \beta^{(4)}_r(t')$ (b) for the transmission  of two dark solitons of the model (\ref{enns1}) with $q=4$ and different velocities (different amplitudes) is plotted for $\epsilon=0.06,\,|\psi_0|=6,\,\eta = 2.5$, sent at   $v_1 = -8.38 \sqrt{2}$ (right soliton),\,$v_2 = 9.02 \sqrt{2}$ (left soliton), $v_s = 10.11\sqrt{2}$.}}
\end{figure}

\begin{figure}
\centering
\label{fig15}
\includegraphics[width=8cm,scale=5, angle=0, height=4cm]{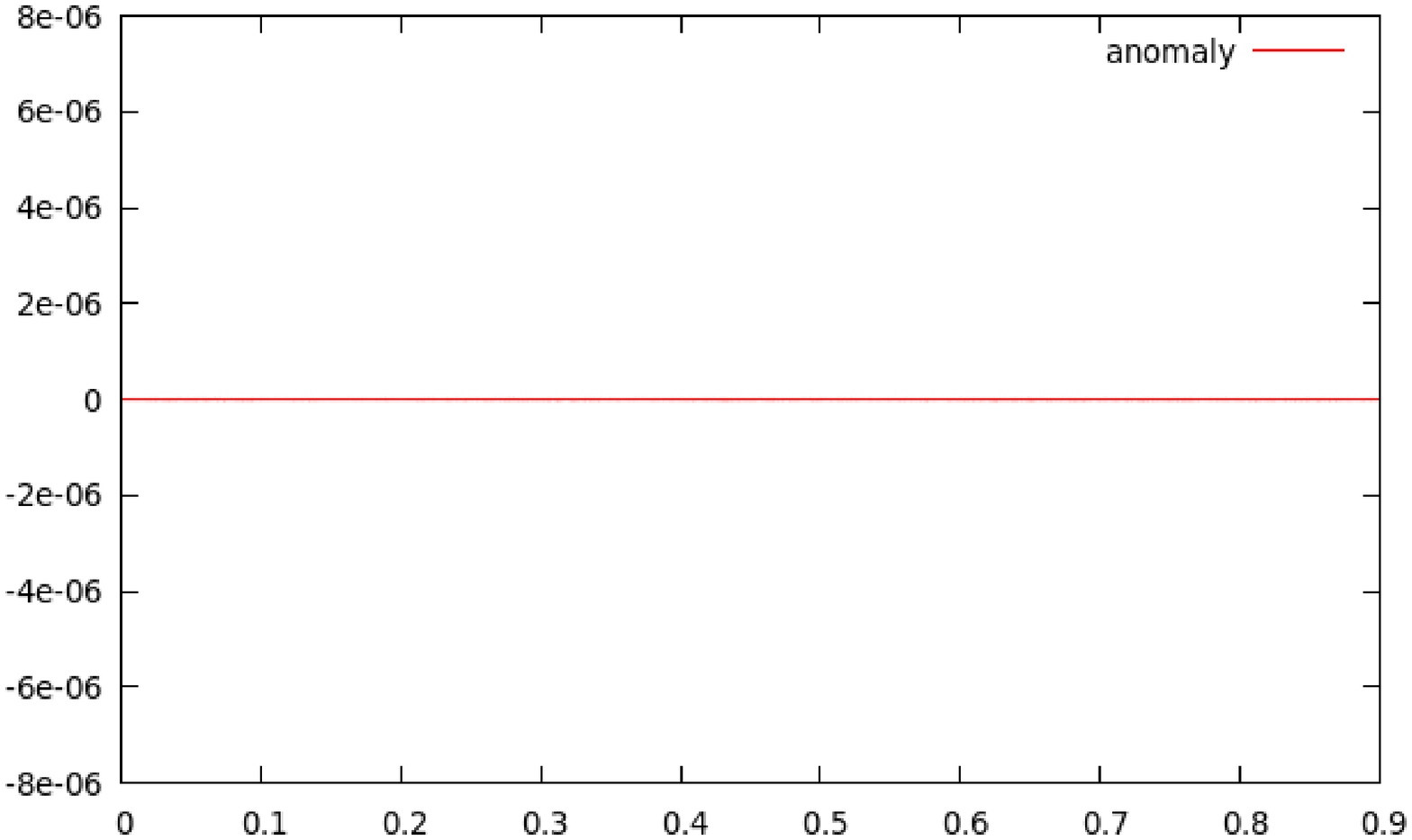}
\includegraphics[width=8cm,scale=5, angle=0, height=4cm]{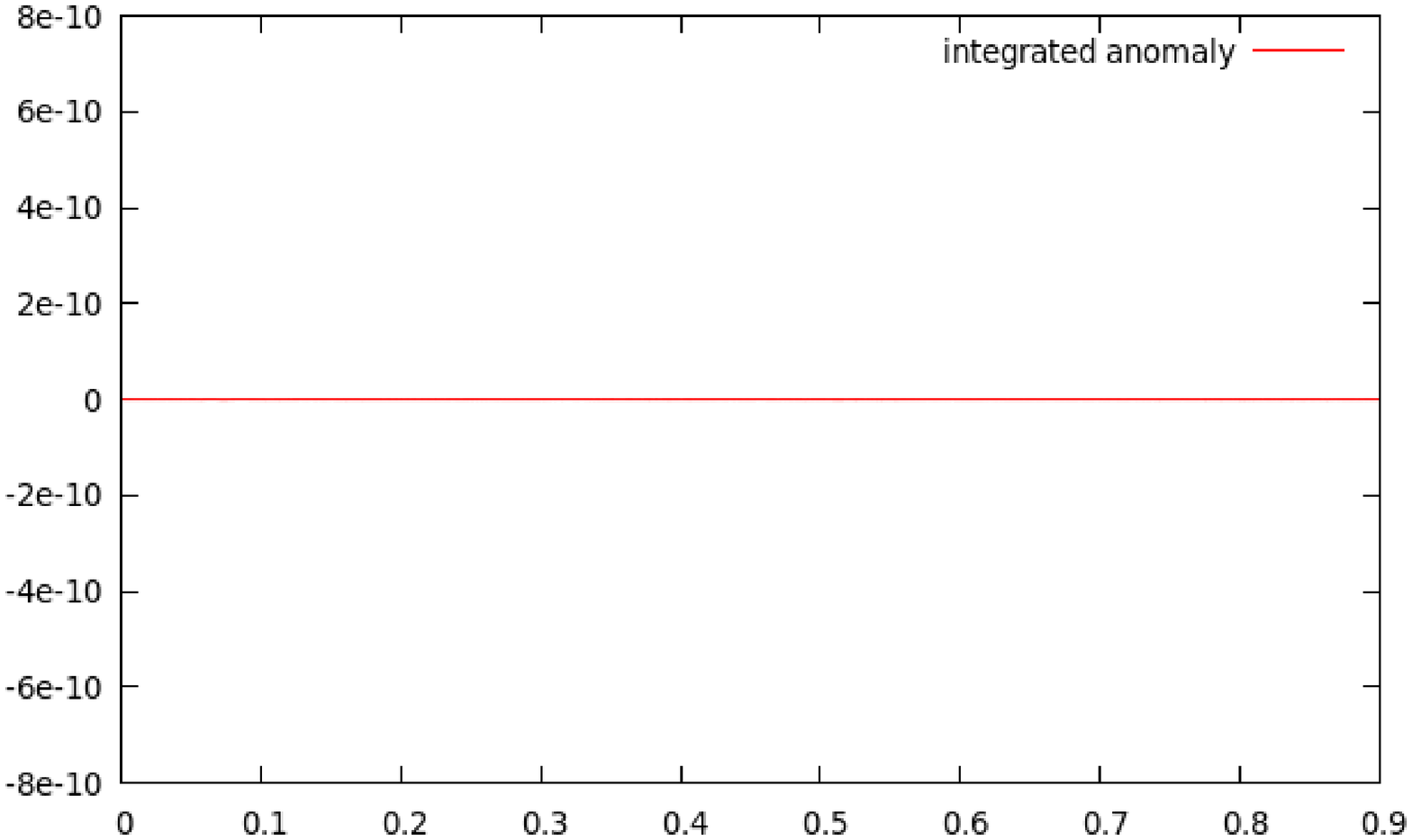}   
\parbox{5in}{\caption {(color online)  
Anomaly $\beta^{(4)}_r(t)$ (a) and time integrated anomaly $\int^t dt' \beta^{(4)}_r(t')$ (b) for the transmission  of two dark solitons of the model (\ref{enns1}) with $q=3$ and different velocities (different amplitudes) is plotted for $\epsilon=-0.06,\,|\psi_0|=6,\,\eta = 2.5$, sent at   $v_1 = -12.97 \sqrt{2}$ (right soliton),\,$v_2 = 13.8 \sqrt{2}$ (left soliton), $v_s = 16.1\sqrt{2}$.}}
\end{figure}

\begin{figure}
\centering
\label{fig16}
\includegraphics[width=8cm,scale=5, angle=0, height=5cm]{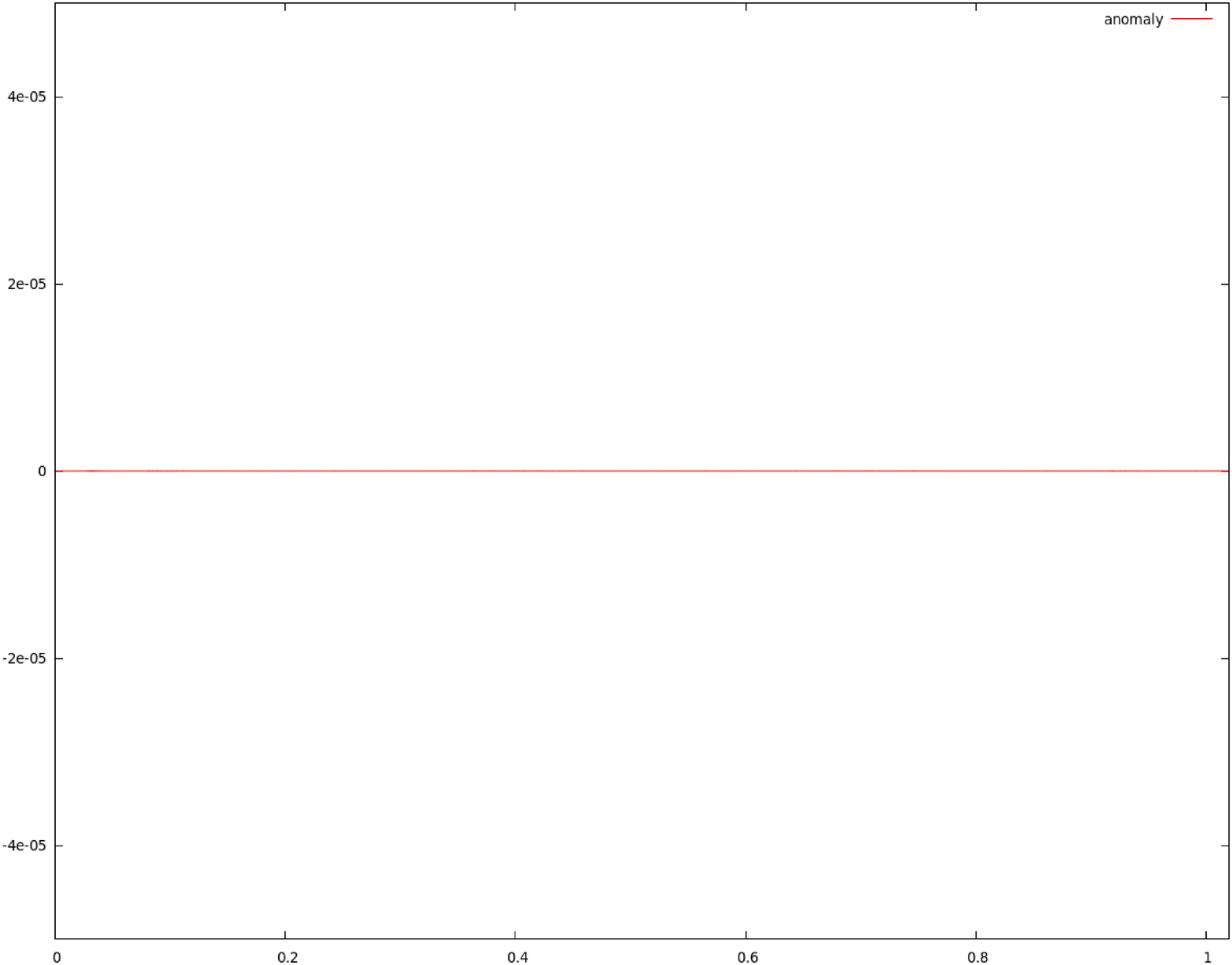}
\includegraphics[width=8cm,scale=5, angle=0, height=5cm]{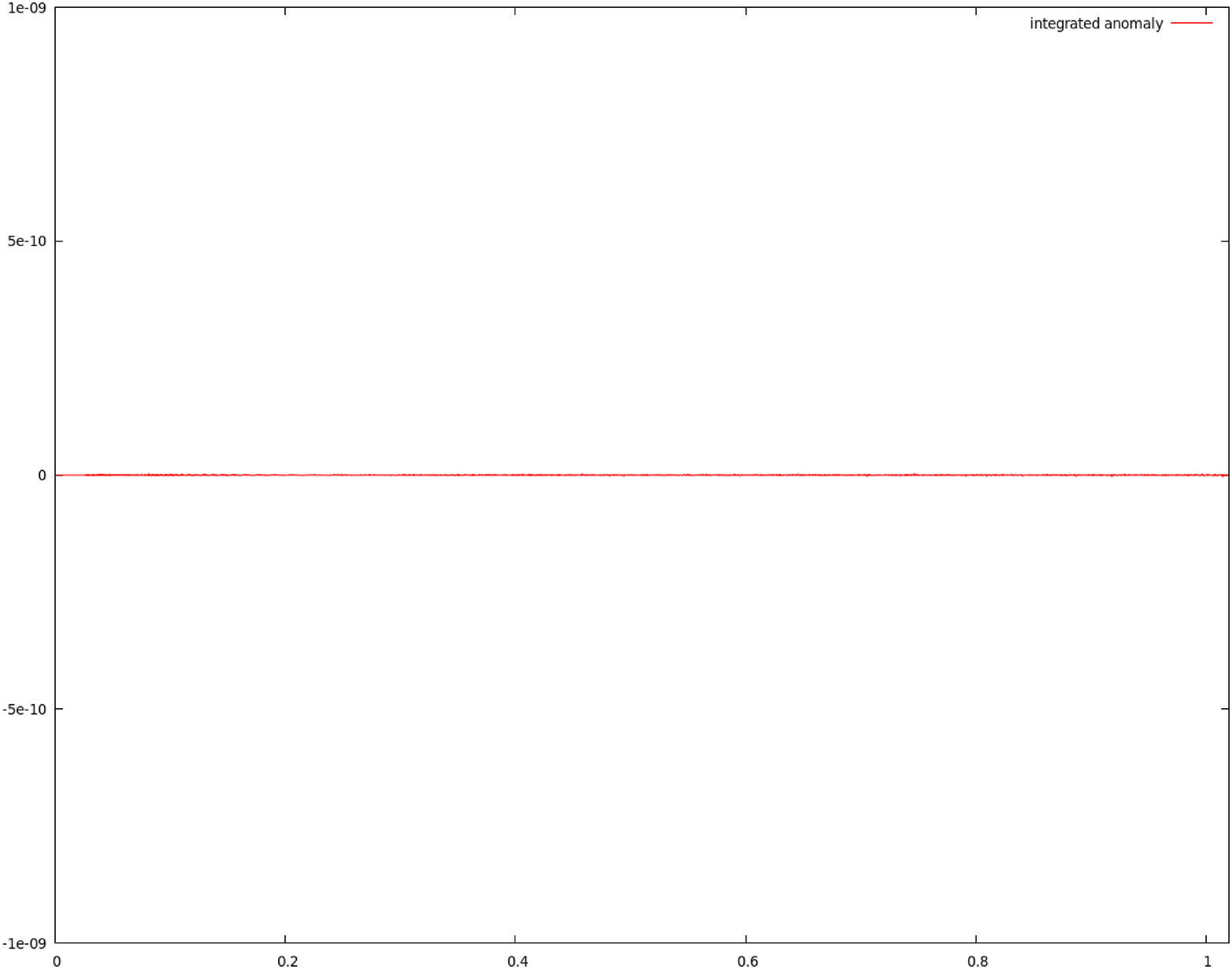}   
\parbox{5in}{\caption {(color online)  
Anomaly $\beta^{(4)}_r(t)$ (a) and time integrated anomaly $\int^t dt' \beta^{(4)}_r(t')$ (b) for the transmission  of two dark solitons of the model (\ref{enns1}) with $q=3$ and different velocities (different amplitudes) is plotted for $\epsilon=0.06,\,|\psi_0|=6,\,\eta = 2.5$, sent at   $v_1 = -8.38 \sqrt{2}$ (right soliton),\,$v_2 = 9.02 \sqrt{2}$ (left soliton), $v_s = 10.11\sqrt{2}$.}}
\end{figure}

\begin{figure}
\centering
\label{fig17}
\includegraphics[width=8cm,scale=5, angle=0, height=5cm]{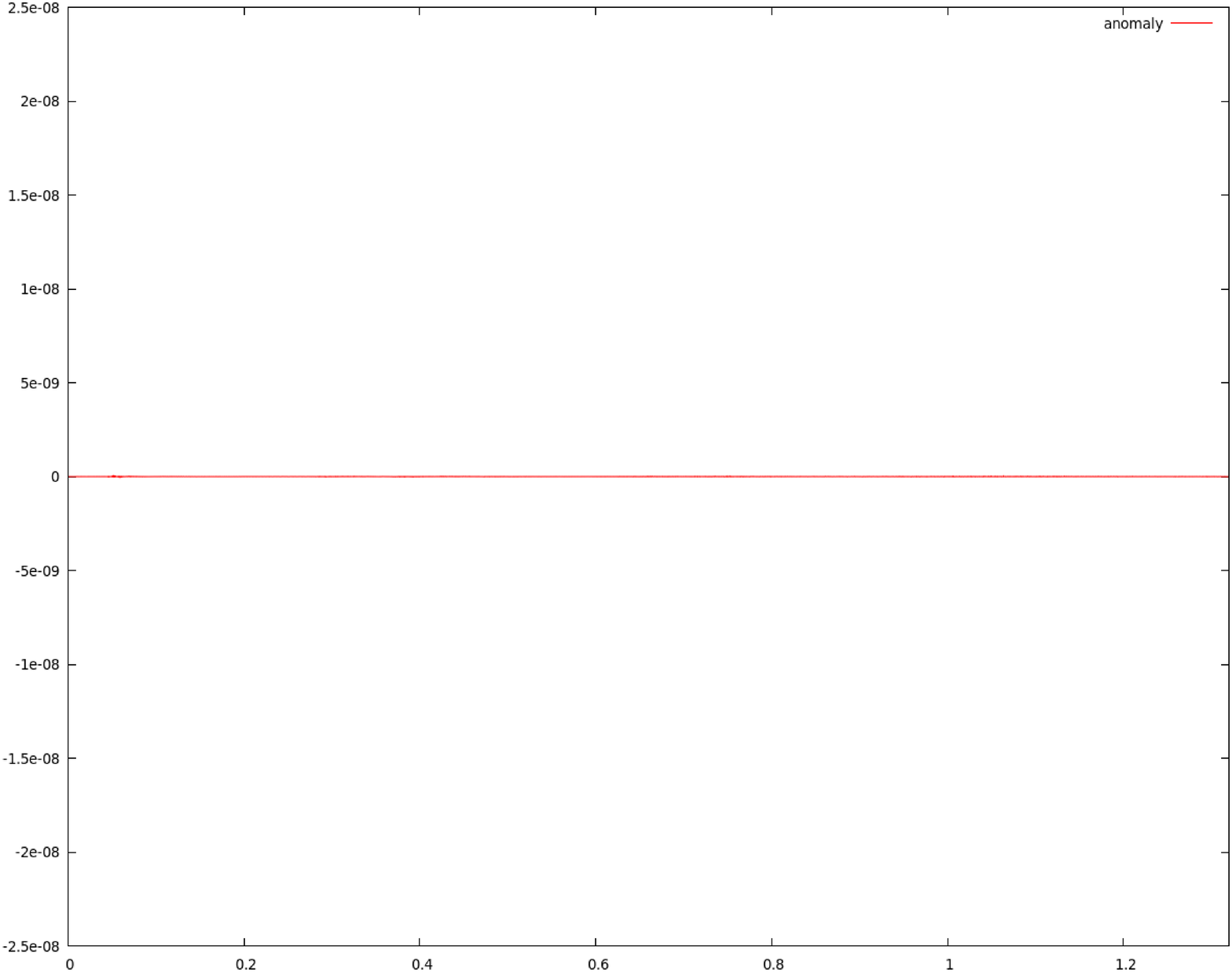}
\includegraphics[width=8cm,scale=5, angle=0, height=5cm]{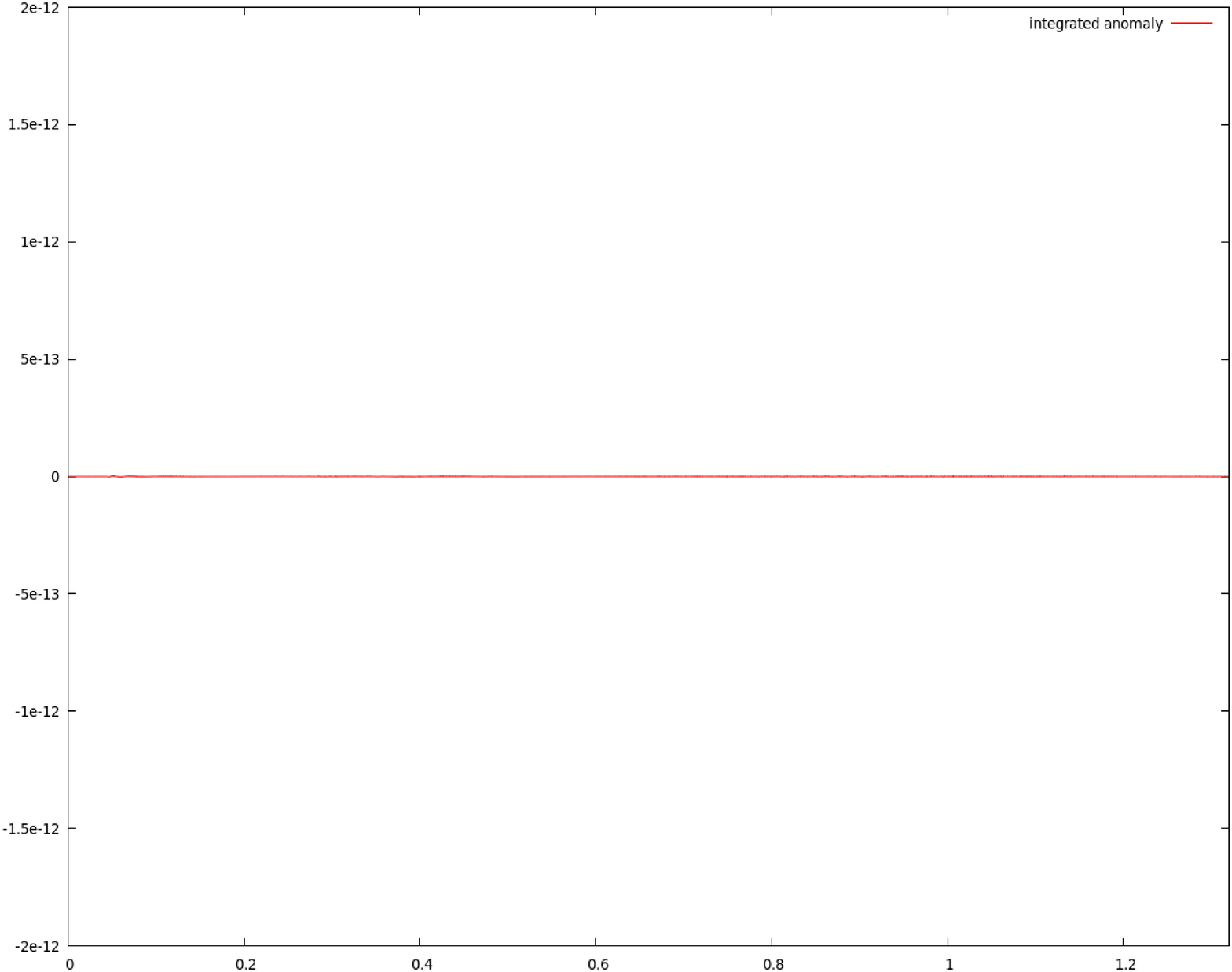}   
\parbox{5in}{\caption {(color online)  
Anomaly $\beta^{(4)}_r(t)$ (a) and time integrated anomaly $\int^t dt' \beta^{(4)}_r(t')$ (b) for the transmission  of two dark solitons of the model (\ref{enns1}) with $q=5$ and different velocities (different amplitudes) is plotted for $\epsilon=-0.1,\,|\psi_0|=6,\,\eta = 2.5$, sent at   $v_1 = -13.74 \sqrt{2}$ (right soliton),\,$v_2 = 14.74 \sqrt{2}$ (left soliton), $v_s = 17.60\sqrt{2}$.}}
\end{figure}
\newpage

\section{Discussions and some conclusions}
\label{conclu}

In this paper we have studied some properties of the deformed non-linear Schr\"odinger model defined by the eq. of motion (\ref{nlsd}), with the potential involving a deformation parameter $\epsilon$ and the non-vanishing boundary condition (\ref{vac1}). We have shown that such deformed NLS theories possess some sectors where the solitary wave  solutions exhibit
properties very similar to solitons in exactly integrable field theories. They possess an
infinite number of exactly conserved quantities for one-soliton type solutions,
i.e. for dark solitary waves travelling with a constant speed. For solutions possessing a special space-time parity symmetry, using analytical methods borrowed from integrable theories, it can be shown that those quantities are asymptotically conserved for two-soliton type solutions. This means that these quantities can vary in time during the collision process of two one-solitons but return, in the distant future (after the collision), to the values they had in the distant past (before the collision). Furthermore, for solutions possessing the space-time parity symmetry and in addition, a special space-reflection parity symmetry we can say even more.  Using an algebraic technique which combines the space-reflection and  an order two $\IZ_2$ automorphism of the  affine $sl(2)$ loop algebra, in the context of  an anomalous zero-curvature (Lax) equation, where the
Lax potentials live on the loop $sl(2)$ affine Kac-Moody algebra, we have shown that those infinite number of charges split into two sets, one of them contains a sequence of even order exactly conserved charges and the other one comprises asymptotically conserved charges. We have verified this property, analytically and numerically,  for the head-on collision of two solitons travelling in opposite directions with equal velocities.  

The mechanism behind the exact conservation of the sequence of even order charges is not well understood yet. As in all of the examples where the asymptotically conserved charges have been
observed so far, associated to a space-time parity symmetry  \cite{jhep1, jhep2}, the two-soliton type solutions associated to a sequence of exactly conserved charges present special properties under a
space-reflection parity transformation. The only explanation we have found, so far, for such behaviour of the even order charges, is that those special soliton-like solutions transform in a special way under a space-reflection parity transformation for a fixed time, where the point in $x-$coordinate around which space is reversed depends upon the parameters of the particular solution under consideration. 

In order to compute the vanishing of the anomaly for a general solitary wave solution we have used the properties of the solutions under the Galilei transformation and space-time 
transformations in the co-moving coordinate of the soliton to prove that the quantity $\a^{(3, -n)}$ satisfies the property (\ref{factor1}) under the parity transformation for $\widetilde{t}=0$. In addition, we have
shown that by considering the zeroth order solution in the deformation parameter $\epsilon$ to be a two-soliton solution of the integrable NLS model,
satisfying (\ref{parxx}), one can always construct by a power series expansion on $\epsilon$, a solution of the deformed model satisfying the parity property (\ref{parxx}), which implies the exact conservation of the even order charges.

We have confirmed the predictions of our analytical calculations through an efficient and accurate numerical method, the so-called time-splitting cosine pseudo-spectral finite difference method, which is appropriate for the case in which the cwb is at rest. The numerical code was devised  for two important deformations of the NLS. The first deformation is related to the cubic-quintic non-integrable NLS model and its numerical simulation considers as an initial condition two of its solitary waves located some distance apart.  The second deformation considers a saturable type potential and its solitary wave is generated numerically by taking  as a seed solution the analytic solitary wave of the  cubic-quintic NLS model. 

For the above two deformations we have computed the first non-trivial anomaly $\b^{(4)}$ of the $Q^{(4)}_r$ charge's non-conservation law. We have verified that this anomaly vanishes, and consequently the exact conservation of the charge $Q^{(4)}_r$ holds for various two-soliton configurations, within numerical accuracy. 

Remarkably, we have found that even for two-soliton solutions with different velocities (different amplitudes) which do not satisfy the space-reflection symmetry,  the anomaly $\b^{(4)}$ vanishes (see Figs. 7-10 for the cubic-quintic NLS model (\ref{cqnls2}) and Figs. 15-20 for the saturable NLS (\ref{enns1})), within the numerical accuracy. Then, we may argue that the parity property (\ref{spari11})-(\ref{xsym}) is not the cause of the exact conservation of the charges, but according to our analytical calculations it is a sufficient condition for these phenomena to happen. Further research work is necessary to settle such
questions which involve the non-linear dynamics of the scattering. So, the symmetries involved in the quasi-integrability phenomenon deserve further investigation and they may have relevant applications in many areas of non-linear sciences.  
 
\section{Acknowledgements}

The authors thank Prof. L.A. Ferreira for enlightening discussions and suggestions, and for hospitality at the IFSC-USP. The authors acknowledge FAPEMAT for financial support.

\appendix

\newpage
 
\section{Equations in the $R$ and $\vp$ parametrization}
\label{solitary}
  
In this appendix we discuss the deformed NLS equations of motion in the $R$ and $\vp$ parametrization and consider translationally invariant solitary wave solutions of the form 
\br
\label{darkap}
\psi(z,t) = \sqrt{R(z)}\,\, \mbox{exp}[i \vp(x,t)/2],\,\,\,\,\vp(x,t)\equiv 2 \Theta(z)+ 2 w t,\,\,\,z= x- v t,
\er
with $w= -V'[R]$ and satisfying the non-vanishing boundary condition
\br
\label{bc222}
R(z \rightarrow \pm \infty) = |\psi_0|^2,\,\,\,\,\, \pa_{z} R_{(z \rightarrow \pm \infty)} =  0, \,\,\,\,\,\pa_{z} \Theta_{(z \rightarrow \pm \infty)} = 0.   
\er   
Substituting (\ref{darkap}) into (\ref{eq111})-(\ref{eq222}) one gets the relationships
\br
\label{theta11}
\Theta'(z) - \frac{v}{2} (1- \frac{|\psi_0|^2}{R}) &=&0\\
2 w - 2 v \Theta'(z) + 2 [\Theta'(z)]^2 - \frac{\pa_z^2 R}{R} + \frac{1}{2} \frac{(\pa_z R)^2}{R^2}+2 V'[R] &=& 0
\er
Decoupling the last two equations one can write a differential equation for $R$ as 
\br
\label{sec1}
R^{-1}\Big\{\pa_z R \, \pa_z^2 R - \frac{[\pa_z R]^3}{2 R} \Big\}=   \pa_z R \, \{2 w - \frac{v^2}{2}(1-\frac{|\psi_0|^4}{R^2}) + 2 V'[R]\}
\er
Since the l.h.s. of the last equation can be written as 
\br
\label{lhs}
2 \pa_{z}[(\pa_z \sqrt{R})^2],\er
and the r.h.s. as
\br
\label{rhs}
\pa_z\{ \int_{|\psi_0|^2}^{R} [ V'[I]-V'[|\psi_0|^2] ] dI - \frac{v^2}{2}\frac{(R-|\psi_{0}|^2)^2}{R}\},
\er
one can integrate once the eq. (\ref{sec1}) to get
\br
\label{j11}
\pa_z \sqrt{R} &=& \frac{1}{\sqrt{2}} \Big\{ \int_{|\psi_0|^2}^{R} \( V'[I]-V'[|\psi_0|^2] \) dI - \frac{v^2}{2}\frac{(R-|\psi_0|^2)^2}{R}\Big\}^{1/2}
\\
&\equiv& J[R]\label{j12}\er

\section{Expressions corresponding to the gauge transformation}
\label{apen2}
In this appendix we present the expressions corresponding to the gauge transformation performed in section \ref{quasi}. Let us introduce the notation for the partial derivatives
\br
\star^{(m,n)} \equiv \pa_x^{m} \pa_{t}^n,
\er
which will be used in the r.h.s.'s of the expressions below.

The $\zeta_{i}^{(-n)}$ defined in (\ref{group}) are
\br
\zeta_{1}^{(-1)}&=&0,\\
\zeta_{2}^{(-1)}&=& 2 \sqrt{\eta} \sqrt{R},\\
\zeta_{1}^{(-2)}&=& \frac{i \sqrt{\eta} R^{(1,0)}}{\sqrt{R}},\\
\zeta_{2}^{(-2)}&=& \sqrt{\eta} \vp^{(1,0)} \sqrt{R},\\
\zeta_{1}^{(-3)}&=& \frac{i (\sqrt{\eta} R^{(1,0)} \vp^{(1,0)} + \sqrt{\eta} R  \vp^{(2,0)})}{\sqrt{R}},\\
\zeta_{2}^{(-3)}&=& \frac{ 16 \eta^{3/2} R^3 + 3 \sqrt{\eta} (\vp^{(1,0)})^2 R^2 - 6 \sqrt{\eta} R^{(2,0)} R + 3 \sqrt{\eta} (R^{(1,0)})^2}{6 R^{3/2}},\\
\zeta_{1}^{(-4)}&=& \frac{i}{12 R^{5/2}} \Big[ 64 \eta^{3/2} R^{(1,0)} R^3 + 9 \sqrt{\eta} (\vp^{(1,0)})^2 R^{(1,0)} R^2 + 18 \sqrt{\eta} \vp^{(1,0)} \vp^{(2,0)} R^3\\
&& -12 \sqrt{\eta} R^{(3,0)} R^2 + 18 \sqrt{\eta} R^{(1,0)}  R^{(2,0)} R - 9 \sqrt{\eta} ( R^{(1,0)} )^3 \Big],\\
\zeta_{2}^{(-4)}&=& \frac{1}{4 R^{3/2}} \Big[ 16 \eta^{3/2} R^3 \vp^{(1,0)} - 6 \sqrt{\eta} \vp^{(2,0)} R^{(1,0)} R - 6 \sqrt{\eta} \vp^{(1,0)} R^{(2,0)} R\\
&& +3 \sqrt{\eta} (R^{(1,0)})^2 \vp^{(1,0)} +  \sqrt{\eta} (\vp^{(1,0)})^3  R^2 - 4 \sqrt{\eta} R^2 \vp^{(3,0)} \Big]
\er
The $a_{x}^{(3,n)}$ introduced in (\ref{ax}) are
\br
a_{x}^{(3,0)} &=& \frac{i}{2} \vp^{(1,0)}\\\
a_{x}^{(3,-1)} &=& 2 i \eta R,\\  
a_{x}^{(3,-2)} &=& i \eta \vp^{(1,0)} R,\\
a_{x}^{(3,-3)} &=& \frac{i \eta \( 4 \eta R^3 + (\vp^{(1,0)})^2  R^2 - 2 R^{(2,0)} R + (R^{(1,0)})^2 \)}{2 R}, \\
a_{x}^{(3,-4)} &=& \frac{i \eta}{4 R} \Big[ 12 \eta \vp^{(1,0)} R^3 - 6 R \(\vp^{(2,0)}  R^{(1,0)} +\vp^{(1,0)}  R^{(2,0)} \) + 3 \vp^{(1,0)}  (R^{(1,0)})^2 \\
&& + \( (\vp^{(1,0)})^3 - 4 \vp^{(3,0)} \) R ^2\Big]
\er
The parameters $\a^{(j,-n)}$ defined in (\ref{a1n}) are
\br
\a^{(3,0)} &=& 1,\\
\a^{(3,-1)} &=& 0,\\
\a^{(3,-2)} &=& 2 \eta R,\\
\a^{(3,-3)} &=& 2 \eta R \vp^{(1,0)},\\
\a^{(3,-4)} &=& 6 \eta^2 R^2 + \frac{3}{2} \eta  (\vp^{(1,0)})^2 R - 2 \eta R^{(2,0)}+ \frac{3 \eta (\vp^{(1,0)})^2 }{2 R},
\er
and
\br
\a^{(1,0)} &=& 0,\\
\a^{(1,-1)} &=& -2 \sqrt{\eta} \sqrt{R},\\
\a^{(1,-2)} &=& - \sqrt{\eta} \sqrt{R} \vp^{(1,0)},\\
\a^{(1,-3)} &=& -4 \eta^{3/2} R^{3/2} - \frac{1}{2} \sqrt{\eta} (\vp^{(1,0)})^2 \sqrt{R} - \frac{\sqrt{\eta} (R^{(1,0)})^2}{2 R^{3/2}}  + \frac{\sqrt{\eta} R^{(2,0)}}{ \sqrt{R}} ,\\
\a^{(1,-4)} &=& -6 \eta^{3/2} \vp^{(1,0)} R^{3/2} - \frac{3 \sqrt{\eta} (R^{(1,0)})^2 \vp^{(1,0)}}{4 R^{3/2}} + \frac{3 \sqrt{\eta} \vp^{(1,0)} R^{(2,0)}}{2 \sqrt{R}} \\
&&+ \frac{3 \sqrt{\eta} R^{(1,0)} \vp^{(2,0)}}{2 \sqrt{R}} - \frac{1}{4} \sqrt{\eta} (\vp^{(1,0)})^3 \sqrt{R} + \sqrt{\eta} \vp^{(3,0)} \sqrt{R} .
\er
and 
\br
\a^{(2,0)} &=& 0,\\
\a^{(2,-1)} &=& 0,\\
\a^{(2,-2)} &=& - \frac{ \sqrt{\eta} R^{(1,0)}}{\sqrt{R}},\\
\a^{(2,-3)} &=& -4 \frac{\sqrt{\eta} \vp^{(1,0)} R^{(1,0)} }{\sqrt{R}} - i \sqrt{\eta} \vp^{(2,0)} \sqrt{R},\\
\a^{(2,-4)} &=& -6 i \eta^{3/2} R^{(1,0)} R^{1/2} - \frac{3 i \sqrt{\eta} R^{(1,0)} (\vp^{(1,0)})^2}{4 R^{1/2}} -  \frac{3}{2} i  \sqrt{\eta} \vp^{(1,0)} \vp^{(2,0)} \sqrt{R} \\
&& + \frac{3 i \sqrt{\eta} (R^{(1,0)})^3}{4 R^{5/2}} -  \frac{3 i\sqrt{\eta}  R^{(2,0)} R^{(1,0)}}{2 R^{3/2}} + \frac{i\sqrt{\eta} R^{(3,0)}}{ \sqrt{R}} .
\er

\end{document}